\documentclass[twocolumn]{aastex631}

\usepackage{newtxtext,newtxmath}
\usepackage{ae,aecompl}

\usepackage{graphicx}	% Including figure files
\usepackage{amsmath}	% Advanced maths controls

\newcommand\msun{{\,M_\odot}}

\newcommand{\unit}[1]{\ensuremath{\, \mathrm{#1}}}
\newcommand{\cmci}{~\mbox{cm}^{-3}}

\newcommand{\gadgetthree}{{\sc p-gadget3}}

\newcommand{\Msun}{~\mbox{M}_{\odot}}

\newcommand{\music}{{\sc music}}
\newcommand{\cloudy}{{\sc cloudy}}

\begin{document}

\title[Simulating UFDs: Metallicity and Size Challenges]{Understanding Stellar Mass-Metallicity and Size Relations in Simulated Ultra-Faint Dwarf Galaxies}

%\author[0009-0000-8108-6456]{Tae Bong Jeong}
%\affiliation{School of Space Research, Kyung Hee University, 1732 Deogyeong-daero, Giheung-gu, Yongin-si, Gyeonggi-do 17104, Republic of Korea}

\author[0009-0008-7277-3818]{Minsung Go}
\affiliation{School of Space Research, Kyung Hee University, 1732 Deogyeong-daero, Yongin-si, Gyeonggi-do 17104, Republic of Korea}

\author[0000-0001-6529-9777]{Myoungwon Jeon}
\affiliation{School of Space Research, Kyung Hee University, 1732 Deogyeong-daero, Yongin-si, Gyeonggi-do 17104, Republic of Korea}
\affiliation{Department of Astronomy \& Space Science, Kyung Hee University, 1732 Deogyeong-daero, Yongin-si, Gyeonggi-do 17104, Republic of Korea}
\correspondingauthor{Myoungwon Jeon}
\email{myjeon@khu.ac.kr}

\author[0000-0001-7863-2591]{Yumi Choi}
\affiliation{NSF National Optical-Infrared Astronomy Research Laboratory, 950 North Cherry Avenue, Tucson, AZ 85719, USA}

\author[0000-0002-3204-1742]{Nitya Kallivayalil}
\affiliation{Department of Astronomy, The University of Virginia, 530 McCormick Road, Charlottesville, VA 22904, USA}

\author[0000-0001-8368-0221]{Sangmo Tony Sohn}
\affiliation{Space Telescope Science Institute, 3700 San Martin Drive, Baltimore, MD 21218, USA}
\affiliation{Department of Astronomy \& Space Science, Kyung Hee University, 1732 Deogyeong-daero, Yongin-si, Gyeonggi-do 17104, Republic of Korea}

\author[0000-0003-0715-2173]{Gurtina Besla}
\affiliation{Department of Astronomy, University of Arizona, 933 N. Cherry Ave, Tucson, AZ 85721, USA}

\author[0000-0002-3188-2718]{Hannah Richstein}
\affiliation{Department of Astronomy, The University of Virginia, 530 McCormick Road, Charlottesville, VA 22904, USA}

\author[0000-0003-2990-0830]{Sal Wanying Fu}
\affiliation{Department of Astronomy, University of California, Berkeley, Berkeley, CA, 94720, USA}

\author[0009-0000-8108-6456]{Tae Bong Jeong}
\affiliation{School of Space Research, Kyung Hee University, 1732 Deogyeong-daero, Yongin-si, Gyeonggi-do 17104, Republic of Korea}

\author[0000-0001-5135-1693]{Jihye Shin}
\affiliation{Korea Astronomy and Space Science Institute, 776 Daedeokdae-ro, Yuseong-gu, Daejeon 34055, Republic of Korea}

\received{21 Nov 2024}
\revised{15 Apr 2025}
\accepted{29 Apr 2025}
\submitjournal{ApJ}

%\begin{keywords}
%galaxies: formation -- galaxies: dwarf -- galaxies: star formation -- methods: numerical -- cosmology: dark age, reionization, first stars
%\end{keywords}

%\keywords{Early universe (435), Galaxy formation (595), High-redshift galaxies (734), Hydrodynamical simulations (767), Population II stars (1284), Population III stars (1285), James Webb Space Telescope (2291)}


\section{Introduction}
Ultra-faint dwarf galaxies (UFDs) are identified as the faintest, least massive ($M_V\gtrsim-7.7$, $M_{\star} \leq 10^5\Msun$), and most metal-poor ($\rm \langle[Fe/H]\rangle\lesssim-2$) systems in the Universe \citep[see reviews by][]{Simon2019,McConaachie2012,Frebel2015}. Considering that some Milky Way (MW) UFDs likely formed over 70\% of their stellar mass before reionization and were subsequently quenched by the combined effects of supernovae (SNe) and the global heating from cosmic reionization, UFDs provide valuable insights into star formation and the underlying physical mechanisms that governed them in the early Universe (e.g., \citealp{Bovill2009, Tolstoy2009, Brown2014, Weisz2014}).

Recently, significant theoretical efforts have been made to explain the properties of observed UFDs through high-resolution cosmological simulations. These simulations have achieved stellar mass resolutions of $m_{\star}\approx500 M_{\odot}$ (e.g., \citealp{Jeon2017, Applebaum2021, Sanati2023, Kim2023}) and even higher resolutions of $m_{\star} \leq 60 M_{\odot}$ (e.g., \citealp{Wheeler2019, Agertz2020, Lee2024}). The enhanced resolution of these simulations allows the generation of theoretical UFD samples that are more comparable to observational data. Despite these advancements, challenges remain in accurately understanding certain observed properties of MW UFDs, such as their stellar metallicity (e.g., \citealp{Wheeler2019, Agertz2020, Prgomet2022, Sanati2023}) and galaxy size (e.g., \citealp{Rey2019,  Tarumi2021, Revaz2023, Goater2024}).

The stellar mass-metallicity relation (MZR) in the UFD regime exhibits unique trends that differ from those observed in more massive classical dwarf galaxies. Typically, there is a positive correlation between stellar mass and the average metallicity of galaxies, where more massive halos tend to form more generations of stars, leading to higher metallicities (e.g., \citealp{Zaritsky2008, Behroozi2010, Read2017}; \citealp{Dekel1986, Kirby2013}). However, this correlation is less evident in UFDs compared to relatively more massive dwarf galaxies (e.g., \citealp{Simon2019}). Notably, a plateau in average metallicity, with $\rm \langle[Fe/H]\rangle\sim -2.5$, has been observed for galaxies with $M_{\star}\lesssim10^5\Msun$, based on data from the Hubble Space Telescope (HST) for 13 MW UFDs (\citealp{Fu2023}). Meanwhile, cosmological simulations of UFDs tend to show a continued decline in metallicity with decreasing stellar masses, showing overall $\rm \langle[Fe/H]\rangle$ values that are 1-2 dex lower than observational results (e.g., \citealp{Wheeler2019, Agertz2020, Applebaum2020, Jeon2021a, Sanati2023, Azartash-Namin2024}).

There have been theoretical proposals to explain and narrow this gap. For instance, \citet{Agertz2020} suggested that the low-metallicity found in simulated UFDs could be due to excessively strong SN energy, which prevents the gas from retaining metals within star-forming regions. \citet{Sanati2023} proposed considering the contribution of the first generation of stars, known as Population III (Pop III) stars, which are expected to be relatively massive, releasing more metals into the surrounding medium upon their death as SNe. A more effective method to explain the discrepancy has been suggested by \cite{Prgomet2022}, who used a metallicity-dependent initial mass function (IMF). This approach assumes that the IMF in the early Universe might have been more top-heavy compared to those observed in the local Universe (e.g., \citealp{Geha2013, Sharda2022}). By employing a varying IMF, they achieved relatively high average metallicity in simulated UFDs compared to other works, as more metals are released from more frequent SNe events compared to the commonly used IMFs, such as the Chabrier and Salpeter IMFs. However, it remains a topic of debate whether the star-forming environment, especially at low-metallicity in the early Universe, would favor a different IMF (e.g., \citealp{Bate2019, Guszejnov2022}).

Most theoretical works so far have focused on the differences in the average [Fe/H] values between UFD analogs and observations. Meanwhile, the metallicity distribution function (MDF) could potentially provide a more detailed understanding of the origin of this discrepancy. The composite MDFs of observed UFDs tend to show overall similar distributions across the UFDs (e.g., \citealp{Simon2019, Fu2023}). These distributions highlight the limited presence of low-metallicity stars ($\rm [Fe/H] < -4$), while relatively high-metallicity stars ($\rm [Fe/H] \geq -2.0$) constitute about 20\% of the total stars in the composite MDFs. Such a distribution could be a key to understanding the chemical enrichment history of UFDs, as producing high-metallicity stars ($\rm [Fe/H] \geq -2.0$) is expected to be challenging due to their vulnerability to stellar feedback and the relatively short duration of star formation episode prior to reionization. Therefore, it is crucial to understand the star-forming environments of UFDs and to assess whether it is theoretically possible to explain the observed metallicity distribution of the stars.

The star-forming environments of UFD galaxy analogs likely differ from those of more massive dwarf galaxies (e.g., \citealp{Jeon2017, Revaz2023}). In the initial stages of UFDs' hierarchical growth, when halo masses are still within the minihalo scale ($M_{\rm vir}\sim10^{5-6}\Msun$) and multiple progenitor\footnote{Multiple progenitor halos can be seen as synonymous with subhalos, which eventually merge into a single UFD analog.} halos remain distinct prior to merging, star formation commences in these progenitor halos once their virial mass reaches about $M_{\rm vir}\sim10^{6}\Msun$ (e.g., \citealp{Bromm2004}). This is followed by the early quenching of UFDs due to stellar feedback and cosmic reionization, after which the progenitor halos merge into a UFD analog. During these early assembly stages, the most massive progenitor does not necessarily host the majority of the stars; instead, stars form across multiple small progenitor halos. This contrasts with more massive dwarf galaxies, where most stars predominantly form within a single, most massive halo with a deep potential well (e.g., \citealp{Fitts2018}).

As stars form in diverse environments within different progenitor halos of a UFD analog, each halo develops distinct metallicity distributions shaped by local conditions. It means that the resulting MDF of the galaxy at $z=0$ is determined not only by a single progenitor halo but also by multiple progenitor halos of diverse environments. Nevertheless, the MDFs of galaxies are typically interpreted using a one-zone chemical evolution model, which assumes most stars form within a single main progenitor halo (e.g., \citealp{Weinberg2017, Escala2018, Sandford2024}). However, due to the varied star-forming environments across multiple progenitor halos in UFD analogs, employing such a model to UFDs requires caution, as it may oversimplify the complexity observed in their MDFs.

Another challenge in explaining the properties of observed UFDs through theoretical work is accurately predicting their size. Current high-resolution simulations tend to predict larger values for the half-light radius, $r_{\rm h}$, which serves as a proxy for galaxy size, than those observed (e.g., \citealp{Wheeler2019, Rey2019, Prgomet2022, Revaz2023}). Specifically, these theoretical works struggle to achieve the very small $r_{\rm h} \lesssim 100$ pc values indicated by observational data (e.g., \citealp{McConaachie2012, Simon2019, Richstein2022, Richstein2024, Cerny2023, Cerny2024}). Although it was expected that enhanced simulation resolution would allow for producing smaller UFD analogs, the opposite trend is observed (e.g., \citealp{Wheeler2019}). Recent work by \citet{Revaz2023}, using dark-matter (DM)-only simulations, suggested the conditions necessary for the formation of small-size UFDs. They proposed that if all member stars originate from the same initial progenitor halo in a compact distribution ($r_{\rm h}\lesssim15$ pc) before reionization, smaller UFDs could form. Additionally, they highlighted the role of mergers between progenitor halos in producing extended stellar distributions.

Discovering new member stars at the outskirts of UFDs could potentially reconcile the discrepancy between observations and theoretical models. For instance, \citet{Wheeler2019} speculated that current observations might be sensitive only to the bright core of dwarf galaxies, missing the outer regions, thus predicting smaller galaxy sizes ($r_{\rm h} \lesssim 100$ pc). Recent Gaia data (\citealp{Abdurrouf2022}) have facilitated the probing of extended stellar structures in individual UFDs (e.g., \citealp{Chiti2021, Yang2022, Longeard2022, Sestito2023, Waller2023, Hansen2024}) and statistically in multiple UFDs (e.g., \citealp{Tau2024, Jensen2024}). Specifically, \citet{Jensen2024} identified seven UFDs with $M_V \gtrsim -7.7$ (Bootes I, Bootes III, Draco II, Grus II, Segue I, Tucana II, and Tucana III) out of 60 MW UFDs, exhibiting extended stellar distributions with a low-density outer profile. Interestingly, \citet{Chiti2021} additionally reported on the metallicity of distant stars found at $\gtrsim 9 r_{\rm h}$ in Tucana II, showing lower metallicities of $\rm [Fe/H] \approx - 3.02$ compared to the average value of Tucana II ($\rm \langle[Fe/H]\rangle \ = \ -2.7 $). Therefore, the presence of distant stars with low metallicities could possibly result in a lowered average metallicity and an increased size of UFDs.

In addition to the conventional expectation that such extended structures of MW UFDs might arise from tidal interactions with the MW, it has been suggested that mergers between progenitor halos during the assembly process could also cause the UFDs to be extended (e.g., \citealp{Rey2019, Tarumi2021, Revaz2023, Goater2024}). As mentioned earlier, if member stars of UFDs indeed form in multiple progenitor halos rather than predominantly in a single halo, it could be challenging to expect any relationship between the inner and outer populations, such as metallicity gradient, due to the mixing caused by mergers. Contrary to this expectation, \cite{Chiti2021} argued that Tucana II seems to exhibit a metallicity gradient, indicating a possible relation between the inner and outer stellar populations. They attributed this to stellar feedback, which can puff older, metal-poor stars to the outskirts of a galaxy. This is a similar explanation for the metallicity gradient observed in more massive dwarf galaxies (e.g., \citealp{Mercado2021, Fu2024}). 
However, this explanation assumes that most member stars originate from a single main progenitor halo and that differences in their stellar properties are primarily a function of cosmic time rather than being determined by different progenitor environments and their mergers.

Given the short duration of star formation of UFDs in the early Universe, understanding the origins of stellar metallicities and their structural distribution in the observed UFDs requires examining the environments within the progenitor halos during their formation. It is also crucial to explore how these progenitor halos have assembled over time into the systems observed today. To achieve this, we perform cosmological simulations on six UFD analogs with masses ranging from $\rm M_{vir} \sim 10^{8}\Msun$ to $\rm M_{vir}\sim 10^{9}\Msun$, tracing the era of Pop III star formation up to $z=0$. Moreover, based on the possibility that discrepancies between simulations and observations in the UFD regime could arise from different methodologies for deriving physical quantities, we analyze the physical properties of simulated UFD analogs, while accounting for observational limitations and employing standard observational techniques. This approach aims to examine potential differences between simulation outcomes and observational data, therefore providing valuable insights into the interpretation of observed values. 

In this work, we also implement an improved method for star particle sampling compared to those used in \citet{Jeon2017, Jeon2021a}, by individually sampling massive stars eligible for SN explosions from a given IMF. Unlike the traditional single stellar population (SSP) method, which is suitable for low-resolution simulations where SN energy is released all at once to regulate subsequent star formation but may be too intense for UFD analogs (e.g., \citealp{Agertz2020, Applebaum2020}), our method ensures the discrete injection of SN energy. This approach allows for a more realistic realization of SN feedback effects.

The paper is structured as follows. Section 2 details the various physical models employed in our cosmological simulations, including the methodology for individual star sampling. Section 3 presents the simulation results, discussing mass evolution, SFH, stellar metallicities, and structural distribution. Finally, we discuss and conclude this work in Section 4. Unless otherwise noted, all distances are provided in physical (proper) units for consistency.

\section{Numerical methodology}

\begin{table}
\centering
\begin{tabular}{ccccc}
\hline
UFD analog & $M_{\rm vir}$ & $R_{\rm vir}$ & $M_{\star}$ & $\rm \langle[Fe/H]\rangle$ \\
 & $\rm [10^{8} \ M_{\odot}]$ & [kpc] & $\rm [10^{4} \ M_{\odot}]$ & \\
\hline
{\sc Halo1} & 1.8 & 14.6 & 0.24 & -2.19 \\
{\sc Halo2} & 4.2 & 19.5 & 1.22 & -2.78 \\
{\sc Halo3} & 6.0 & 22.0 & 0.52 & -2.98 \\
{\sc Halo4} & 7.9 & 24.1 & 1.58 & -2.65 \\
{\sc Halo5} & 11.7 & 27.5 & 2.14 & -2.71 \\
{\sc Halo6} & 12.7 & 28.2 & 0.89 & -2.98 \\
\hline
\end{tabular}
\caption{Summary of the simulations. Column (1): Set name. Column (2): Virial mass at $z=0$. Column (3): Virial radius. Column (4): Stellar mass. Column (5): Average stellar iron-to-hydrogen ratios.}
\label{tab:insitu}
\end{table}

\begin{figure*}
  \centering
  \includegraphics[width=175mm]{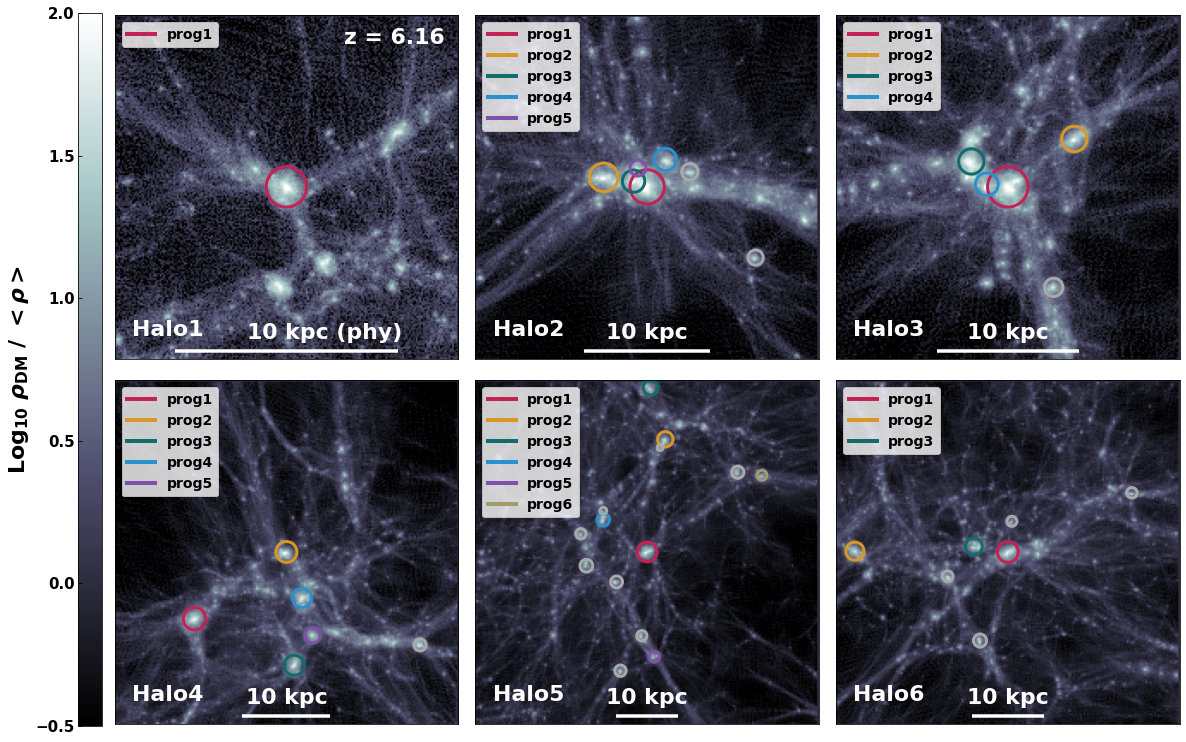}  
  \caption{
  DM projections along the z-direction at $z\sim6$, a point when star formation in all progenitor halos is quenched due to the combined effects of reionization and SNe feedback. The length scale is differently applied to best present the structure of progenitors, but for comparison purposes, a 10 kpc scale is indicated at the bottom of each panel. Progenitors identified as to-be-accreted are marked with circles, with their radii corresponding to their virial radii. Progenitors contributing more than 5\% of the total stellar mass of each halo are numbered according to their virial mass at $z\sim6$ and are color-coded, whereas progenitors contributing less than 5\% are uniformly colored in grey.}
  \label{fig:morph_all.png}
\end{figure*}

\subsection{Simulation Set Up}
In this work, we conduct a cosmological hydrodynamic simulation to explore the formation of UFD analogs. We utilize a modified version of parallel N-body/smoothed particle hydrodynamics (SPH) code \gadgetthree \ (\citealp{Springel2001, Springel2005, Schaye2010}). The initial conditions are generated using the cosmological initial condition generator, \music \ (\citealp{Hahn2011}). To select a target halo at $z=0$, we initially perform a low-resolution DM-only simulation within a box of $L_{\rm box} = 3.125 \ h^{-1}$ comoving Mpc from $z = 128$ to $z = 0$. After identifying isolated UFD analogs at $z=0$, we apply a zoom-in technique to the region surrounding the identified UFD analog. This region is determined by particles within $3 \ R_{\rm vir}$ at $z=0$, ensuring that boundary effects in the Lagrangian region do not impact the properties of our target UFD analogs. As a result, we generate high-resolution initial conditions, achieving a resolution where $M_{\rm{gas}} \approx 60 \ {\rm{M_{\odot}}}, $ and $M_{\rm{DM}} \approx 300 \ \Msun$.

We create six initial conditions for isolated UFDs with varying virial masses in the range of $M_{\rm vir} = \rm 10^{8} \sim 10^{9} \ M_{\odot}$. In this setup, the total mass of the most massive halo, {\sc Halo6}, is approximately ten times that of the least massive halo, {\sc Halo1}. This mass range enables us to investigate how virial mass influences galactic properties such as stellar mass, average metallicity, SFH, and more. The detailed properties are provided in Table~\ref{tab:insitu}. We set the cosmological parameters as follows, a matter density parameter of $\Omega_{\rm m}=1-\Omega_{\Lambda}=0.265$, a baryon density of $\Omega_{\rm b}=0.0448$, a present-day Hubble expansion rate of $\rm H_{\rm 0}=71\unit{km, s^{-1} Mpc^{-1}}$, a spectral index of $n_{\rm s}=0.963$, and a normalization of $\sigma_8=0.8$ (\citealp{Komatsu2011,planck2016}). The adopted softening length in our simulations is $\epsilon_{\rm soft} \sim 20$ physical pc, which remains constant across all redshifts for both DM and star particles. In addition, we use adaptive softening lengths for the gas particles, where the softening length is proportional to the SPH kernel length, with a minimum value of $\epsilon_{\rm gas, min} = 2.8 \ \rm pc$.

We update the abundances of primordial species ($\rm H, H^{+}, H^{-}, H_{2}, H_{2}^{+}, He, He^{+}, He^{++}, e^{-}, D, D^{+}$ and HD) by solving the non-equilibrium rate equations at every time step, incorporating all relevant cooling processes. These processes include collisional ionization of H and He, excitation and recombination cooling, bremsstrahlung, inverse Compton cooling, and collisional excitation cooling of $\rm H_{2}$ and HD. In addition, we account for metal line cooling involving seven species, C, O, Si, Mg, Ne, N, and Fe. The cooling rates for these metals are interpolated based on density, temperature, and metallicity using tables generated by the photoionization package \cloudy \ \citep{Ferland1998} (see \cite{Wiersma2009} for more details). Furthermore, we adopt the UV background values from \cite{Haardt2012}, initiating with zero intensity at $z = 7$ and linearly increasing it to full strength by $z = 6$, under the assumption of complete reionization by $z = 6$ (e.g., \citealp{Gunn1965, Fan2006}). Self-shielding of dense gas is implemented by attenuating the UV background according to $\exp(-N_{\rm HI} \ \overline{\sigma}_{\rm ion})$, where the neutral hydrogen column density is given by $N_{\rm H\ I} = {\rm x} \ n_{\rm HI}$. Here, $\rm x = h$ represents the SPH kernel size, $n_{\rm H I}$ is the number density of neutral hydrogen, and $\overline{\sigma}_{\rm ion}$ denotes the frequency-averaged photoionization cross section for $\rm HI$. For more details, we refer readers to \citet{Jeon2017}

\subsection{Star formation}
We employ a stochastic star formation method based on the Schmidt law (\citealp{Schmidt1959}), where the star formation rate (SFR) is defined as 
$\dot{\rho_{\ast}}=\rho/\tau_{\ast}$. In this equation, $\rho$ represents the gas density for star formation, and $\tau_{\ast}$ is given by $\tau_{\rm{ff}} / \epsilon_{\rm{ff}}$, where $\tau_{\rm{ff}}$ is the free-fall time at  $\rho$ and $\epsilon_{\rm{ff}}$ is the star formation efficiency. The free-fall time is expressed as $\tau_{\rm ff}=[3\pi/(32G\rho)]^{1/2}$ and $\epsilon_{\rm ff} = 0.01$, a value calibrated from observations in the local Universe (e.g., \citealp{Leroy2008}). Consequently, the star formation timescale can be normalized as follows
\begin{equation}
\tau_{\ast}=\frac{\tau_{\rm ff} (n_{\rm H})}{\epsilon_{\rm ff}}\sim400 {\rm Myr} \left(\frac{n_{\rm H}}{100 \cmci}\right)^{-1/2}.
\end{equation}

When the number density of a gas particle surpasses a threshold of $n_{\rm H, \ th}=100$ $\rm cm^{-3}$, it transforms into a star particle, retaining the metallicity of the gas. To assess the dependence of stellar mass on different parameters, $\epsilon_{\rm ff}$ and $n_{\rm H,th}$, we used the same initial conditions for {\sc Halo4} and performed two additional test runs with $\epsilon_{\rm ff} = 0.3$ and $\epsilon_{\rm ff} = 1.0$, respectively. Furthermore, we performed another run with $n_{\rm H,th} = 500~\mathrm{cm^{-3}}$, keeping $\epsilon_{\rm ff}$ fixed at 0.01 to isolate the effect of the increased density threshold. The resulting stellar masses varied by a factor of two compared to the default setup ($\epsilon_{\rm ff} = 0.01, \ n_{\rm H,th} = 100~\mathrm{cm^{-3}}$) and no clear trend is observed---for example, increasing $\epsilon_{\rm ff}$ did not lead to a corresponding increase in stellar mass. Previous studies have shown that the stellar mass is insensitive to the adopted star formation parameters (e.g., \citealp{hopkins2014, Gutcke2021}), and similar insensitivity might be expected in UFD analogs. However, due to their extremely low stellar masses, UFD analogs are more significantly affected by stochastic processes, which likely hinders the convergence of stellar mass across different runs.

The type of stellar population is then determined by the metallicity of the gas. We utilize a critical metallicity value of $Z_{\rm crit}=10^{-5.5} Z_{\odot}$. Gas with metallicity above this threshold gives rise to Pop II stars, while gas with $Z_{\ast} \leq Z_{\rm crit}$ leads to the formation of Pop III stars (e.g., \citealp{Omukai2000, Schneider2010, Safranek2016}). If a star particle is identified as a Pop III star, it is treated as an individual star and its mass is sampled from the assumed top-heavy IMF over the mass range $[\rm 1, 260] \ M_{\odot}$ according to the following equation,
\begin{equation}
\phi = \frac{dN_{\rm PopIII}}{d \ln m} \propto m^{-1.3} \exp{\left[ -\left(\frac{m_{\rm char}}{m}\right)^{1.6}\right]}\mbox{\ ,}
\end{equation}
where the characteristic mass is $m_{\rm char}=30\Msun$ (e.g., \citealp{Wise2012}).

On the other hand, if a star particle is classified as a Pop II star, we perform individual IMF sampling from a star particle with a mass of $m_{\star} = 60 \Msun$, thus forming a SSP. We sample masses from the Salpeter IMF (\citealp{Salpeter1955}), defined by the equation $dN/d\log m \approx m^{-\alpha}$, with a slope $\alpha = 1.35$ over the mass range $[0.1 - 100] \Msun$, until the total mass reaches $60 \Msun$. If, during this sampling, a mass falls within the range of $[8 - 40] \Msun$, the eligible range for a Type II SN explosion, we designate this particle as a Type II SN progenitor with the sampled mass. 

Since the individually sampled star particles have masses within $8\Msun-100 \Msun$, we use the following method to conserve the original gas-particle mass of $60\Msun$. If a sampled star has a mass $m_{\star}\in[8 - 60]\Msun$, we assign a lost mass of $60-m_{\star}$ to the next SSP particle. Conversely, if a star with $m_{\star}\in[60,100]\Msun$, we assign a reduced mass of $120-m_{\star}$ to the next SSP particle. This approach of individual star sampling, where a single stellar particle can result in a discrete SN explosion, allows us to accurately capture the discrete nature of SN explosions. If the star particle does not contain a mass within the eligible range for a SN explosion, it consists of low-mass stars totaling $60 \Msun$ and does not trigger any SN explosion.

\citet{Revaz2016} highlighted that incomplete sampling of the IMF using high-resolution SSP particles can increase scatter in alpha abundance ratios, especially for low-metallicity stars, and as well as MDF. However, our study shows that star formation primarily occurs in starbursts over short timescales, minimizing the impact of Type Ia SNe and AGB star losses on stellar metallicity since our SSP is a 60 $\Msun$ particle with stars ranging from 0.1 to 8 $\Msun$, and the tendency for stars to form in groups helps average out the effects of incomplete IMF sampling.

\begin{figure*}
  \centering
  \includegraphics[width=175mm]{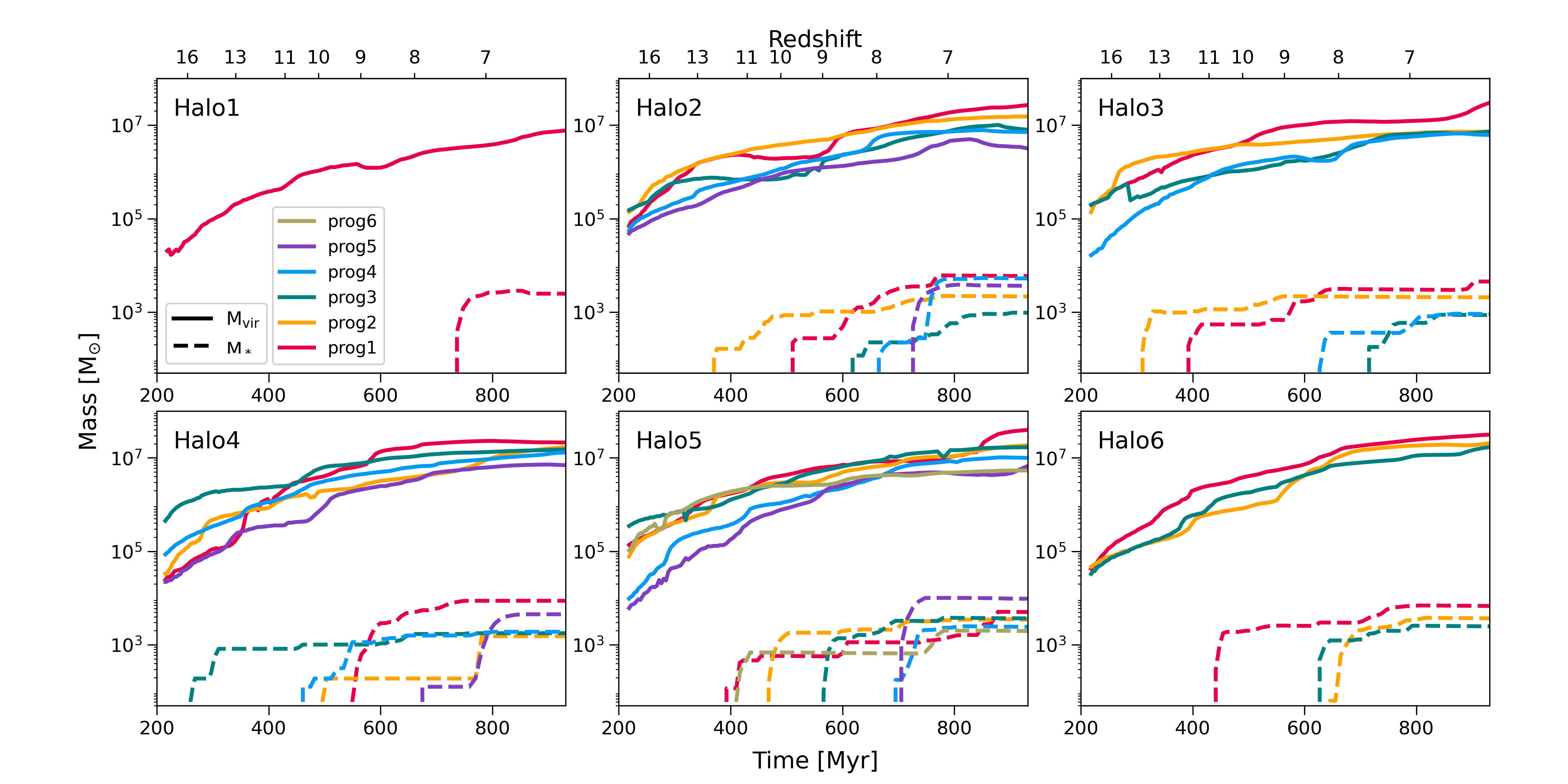}
  \caption{The virial (solid) and stellar (dashed) mass evolution of progenitor halos for each UFD analog as a function of cosmic time. Since all progenitors cease their star formation prior to $z\sim6$, we track their evolution up to that point and consider only Pop II SSP stars for stellar mass, as they are the only ones to survive to $z=0$. Each progenitor begins Pop II star formation when its virial mass reaches $M_{\rm vir}\sim 10^{6} \Msun$, leading to varying stellar mass growth histories. For instance, there is a trend where halos with lower masses at high redshift tend to begin star formation later.}
  \label{fig:Mass_evolution.png}
\end{figure*}

\begin{figure}
  \centering
  \includegraphics[width=85mm]{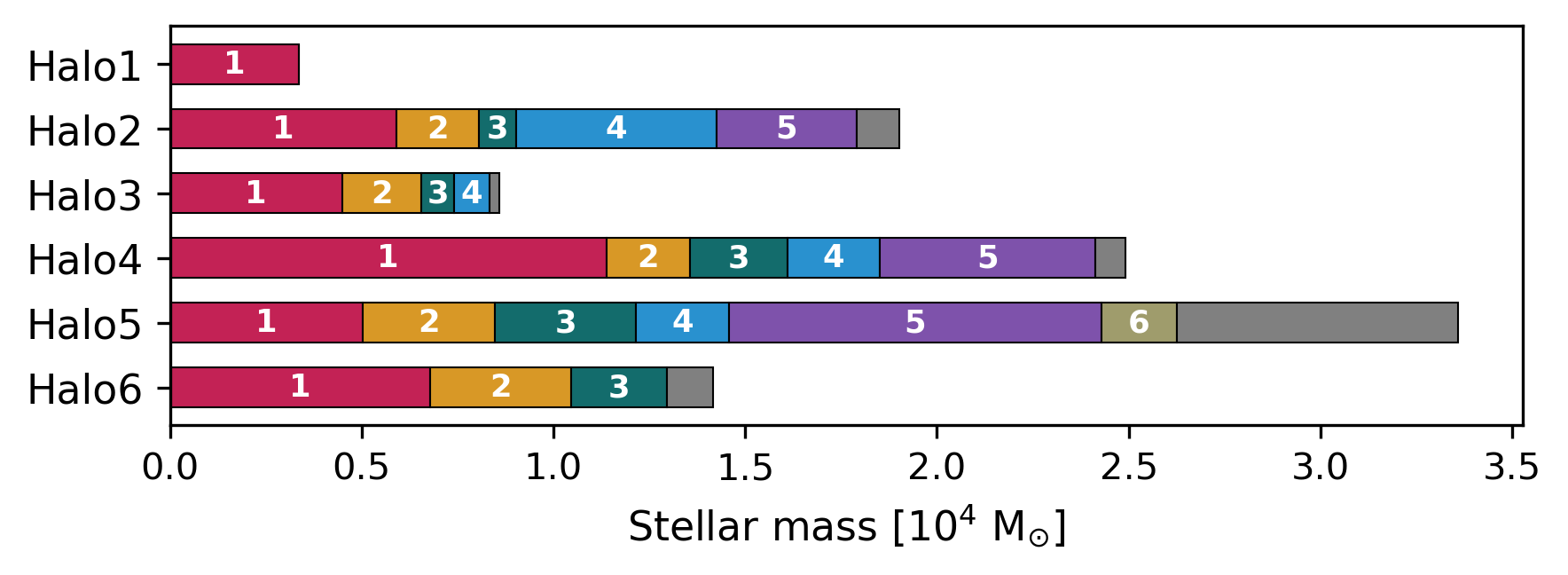}
  \caption{The stellar mass contribution of each progenitor, showing the total stellar mass of halos and distinguishing the stellar mass of each progenitor. The numbering and color-coding are consistent with Fig.~\ref{fig:Mass_evolution.png}. Note that the stellar mass is calculated at $z=6$, so mass loss due to AGB stars up to $z=0$ is not considered. Notably, across all halos, the most massive progenitor ({\sc prog1}) at $z=6$ contributes less than 50\% of the total stellar mass, indicating that star formation likely occurs in multiple progenitor halos rather than being dominated by a single progenitor.}\label{fig:stellar_mass_ratio.png}
\end{figure}

\subsection{Stellar feedback}
Pop III stars end their lives in three different ways depending on their initial mass (e.g., \citealp{Heger2003, Yoon2012}). Stars with an initial mass between $\rm 10 \ M_{\odot}$ and $\rm 40 \ M_{\odot}$ conclude their lives as core-collapse supernovae (CCSNe) with an energy of $E_{\rm SN}=10^{51}$ erg. In contrast, more massive stars with an initial mass between $\rm \ 140 \ M_{\odot}$ and $\rm \ 260 \ M_{\odot}$ end their lives as pair-instability supernovae (PISNe) with ten times the energy of CCSNe, $E_{\rm SN}=10^{52}$ erg. In this work, SN energy is injected as thermal energy into the surrounding medium. To prevent the commonly known over-cooling problem often associated with the thermal energy approach, we adopt the method suggested by \cite{Vecchia2012}. This method ensures that the temperature increase of gas particles receiving SN energy reaches above $\Delta T \sim 10^{7.5}$ K, a threshold necessary to avoid the rapid cooling of thermal energy, by reducing the number of gas particles that receive the SN energy. Specifically, for CCSNe with energy of $E_{\rm SN}=10^{51}$ erg, we inject energy into a single gas particle. For PISNe with energy of $E_{\rm SN}=10^{52}$ erg, we distribute the energy across ten gas particles. Since we do not consider the radiative transfer (RT) of photons emitted during the lifetime of stars in this work, all SN feedback is assumed to be triggered instantaneously. This approach prevents artificial star formation that might otherwise occur during the lifetimes of stars if SN explosions are delayed.

Since we treat Type II SN explosions individually for both Pop II and Pop III stars, metal enrichment by Type II SN is also achieved on an individual basis. For Pop III stars that end their lives as CCSNe, the amount of ejected mass and metal yields are provided by \cite{Heger2010}. For those that die as PISNe, the values are given by \cite{Heger2002}. Pop II stars yield their metals through three processes: AGB stars, CCSNe, and Type Ia SNe. When CCSNe occur, the metals are ejected based on values from \cite{Portinari1998}. Stars with masses in the range of $0.8 \ \Msun-8 \Msun$ undergo AGB mass loss, with yields provided by \cite{Marigo2001} as a function of their stellar age, metallicity, and mass. Type Ia SNe yields are determined based on the delay time distribution for stars within the mass range of $3 \ \Msun - 8 \ \Msun$. Given the uncertainties in the detailed evolution of Type Ia SN, we adopt an empirical delay-time function characterized by e-folding times (\citealp{Forster2006}). Notably, AGB and Type Ia SN yields are not ejected from individual stars but from a SSP star with a total mass of 60 $\rm M_{\odot}$, thus yielding IMF-averaged values at a given time. The expelled metals are initially transported among neighboring gas particles, approximately $N_{\rm Ngb}\approx48$, and then diffused onto the ambient medium using the method described by \cite{Greif2009a}. This diffusion scheme determines the degree of mixing on a sub-resolution scale based on physical properties such as velocity dispersion and density within the SPH kernel of gas particles.

\section{Results}
In this section, we present the results of our simulations. In Section~\ref{sec:3.1}, we explore the global properties, including morphology, mass growth history, and the stellar mass-halo mass relation. In Section~\ref{sec:3.2}, we delve into metallicity-related properties, such as the mass-metallicity relation and the MDF, and discuss the formation of high-metallicity stars. In Section~\ref{sec:3.3}, we analyze the sizes of the simulated galaxies by comparing them with observational data, and we also present other observables, including velocity dispersion and mass-to-light ratio.

\subsection{Morphology and mass growth history} \label{sec:3.1}
In this section, our analysis primarily focuses on the early epoch at $z\approx6$. During this period, the progenitor halos that will eventually merge into the UFD analog by $z=0$ remain distinct. Due to the physical separation of progenitor halos from one another, we can infer that the stellar properties in each progenitor halo are influenced solely by its own environment, without influence from other progenitor halos. Also, this epoch is appropriate because our target UFDs had already finished their star formation before $z\approx6$. We identify the progenitor halos at this epoch using the halo-finder code {\sc rockstar} (\citealp{Behroozi2013}). To determine which progenitor halos will eventually merge by $z = 0$, we compare the IDs of DM particles in each progenitor halo at $z \approx 6$ with those of particles within a UFD analog at $z = 0$. A progenitor halo is classified as a to-be-accreted progenitor if more than 90\% of its DM particles are found within the halo at $z = 0$. This results in the number of contributing progenitors being 1, 5, 4, 5, 6, and 3 for each UFD analog from {\sc Halo1} to {\sc Halo6}. These progenitors are selected based on their contribution of more than 5\% to the total stellar mass.

Fig.~\ref{fig:morph_all.png} illustrates the DM projection along the z-direction at $z \approx 6$, showing identified progenitor halos distributed at local density peaks. Different length scales are applied to each set for optimal illustration. The progenitor halos are depicted as circles with radii corresponding to their virial radii and are color-coded for distinction. The progenitors are numbered according to their virial mass at $z \approx 6$, with the most massive progenitor in each set labeled as {\sc prog1} (red). Progenitors with a stellar mass contribution below 5\% of the total stellar mass are uniformly colored grey in the plot. 

\begin{figure}
  \centering  \includegraphics[width=75mm]{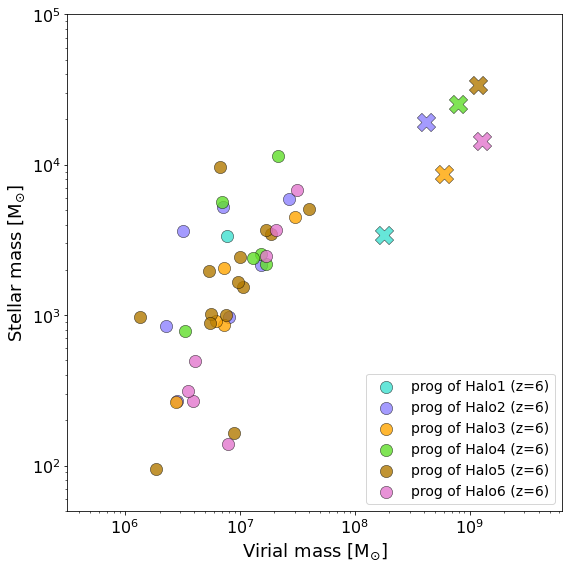}
  \caption{The stellar mass-halo mass relation for all progenitor halos at $z=6$ (circles) and the final UFD analogs at $z=0$ (crosses). Progenitor halos with stellar masses contributing less than 5\% are also included to explore the lower mass range. Progenitor halos that constitute the same UFD analog are represented by the same color. The relation shows a linear trend where stellar masses increase with rising halo masses. However, for progenitor halos with $M_{\rm vir} \sim 10^{7}\Msun$, there is a significant spread in stellar mass, ranging from $M_{\star}\approx10^{2}-10^{4}\Msun$, reflecting their sensitivity to environmental effects of progenitor halos.}
  \label{fig:SMHM_minihalo.png}
\end{figure}

We find that the number of star-forming progenitors is associated with the final halo mass at $z=0$. For example, as shown in Fig.~\ref{fig:morph_all.png}, {\sc Halo1} has only one star-forming progenitor halo, {\sc prog1}, likely due to {\sc Halo1}'s relatively small virial mass of $M_{\rm vir, \ z = 0} \approx 1.8 \times 10^{8} \Msun$. In contrast, {\sc Halo5}, which is approximately ten times more massive than {\sc Halo1} with a mass of $M_{\rm vir, \ z = 0} \approx 1.17 \times 10^{9} \Msun$, has a total of 14 star-forming progenitor halos, including those with stellar mass less than 5\% of the total. However, this trend is not strictly linear, as the most massive {\sc Halo6}, showing a greater virial mass than {\sc Halo5} at $M_{\rm vir, \ z=0} \approx 1.27 \times 10^{9} \Msun$, has 7 star-forming progenitors, which is half the number of progenitors in {\sc Halo5}.

\begin{figure}
  \centering  \includegraphics[width=77mm]{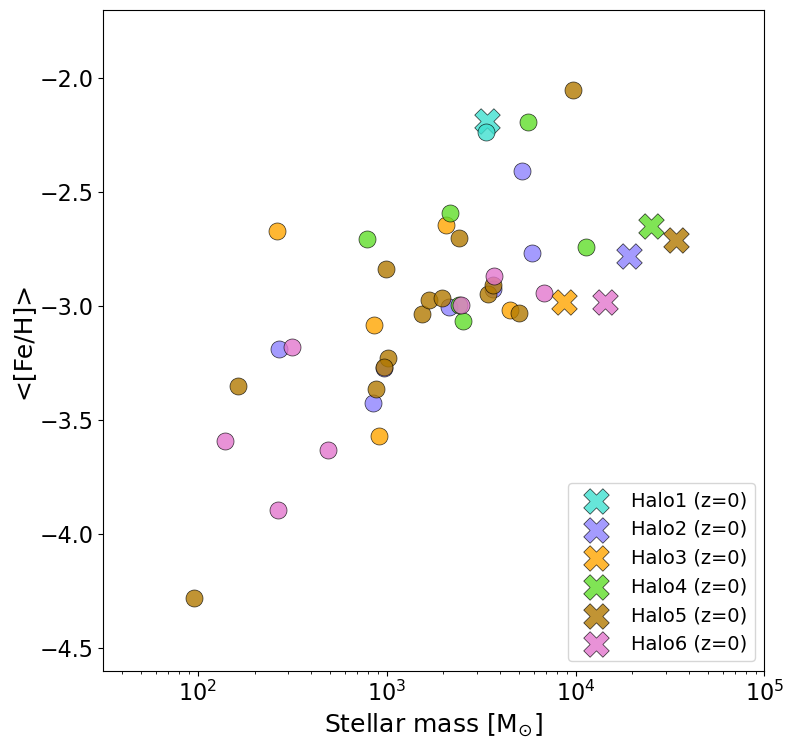}
  \caption{The mass-metallicity relation for progenitor halos, represented by the same symbols as in Fig.~\ref{fig:SMHM_minihalo.png}, exhibiting a positive correlation at $z=6$. When compared with the relation of UFD analogs (crosses), it shows how various progenitors influence the final values at $z=0$. For instance, progenitor halos with lower metallicity contribute to lowering the overall average metallicity of the UFD analogs.}
  \label{fig:MZR_minihalo.png}
\end{figure}

Fig.~\ref{fig:Mass_evolution.png} shows the evolution of the virial mass (solid lines) and stellar mass (dashed lines) of progenitor halos, using the same color scheme as in Fig.~\ref{fig:morph_all.png}. When calculating the stellar mass, we exclude stars that have exploded as SNe, since these stars lose their mass in the explosion and do not significantly contribute to the final stellar mass. Our results indicate that progenitor halos begin their star formation at different times, with more massive halos tending to form stars earlier. The variations in the timing of star formation are primarily driven by differences in the mass growth histories of the progenitor halos, as a progenitor halo needs to exceed a virial mass of $M_{\rm vir} = 10^{5}-10^{6} \Msun$ for gas clouds to collapse and form stars (e.g., \citealp{Bromm2004, Hirano2015}). For instance, the earliest star formation occurs at $z \sim 16$ in {\sc prog3} of {\sc Halo4}, where the virial mass reaches $M_{\rm vir} \sim 6.5 \times 10^{6} \Msun$, the highest mass among the progenitor halos in {\sc Halo4} at this epoch. Here, a Pop III star forms first, followed by a Pop II star at $z \sim 15.7$, after a delay of 6 Myr (e.g., \citealp{Ritter2012,Jeon2014}). On the other hand, {\sc prog1} of {\sc Halo1} begins star formation later at $z \sim 7.8$, with $M_{\rm vir} \sim 6.3 \times 10^{6} \Msun$. The viral masses of progenitor halos at the time of first star formation range from $M_{\rm vir}\approx1.1 \times 10^{6} \Msun$ to $6.6 \times 10^{6} \Msun$, reflecting the varying environment of each progenitor halo. Furthermore, it is important to note that the first-formed star is not always a Pop III star; it could be a Pop II star if the interstellar medium (ISM) within the progenitor halo is externally enriched. In this work, 90\% of progenitor halos have a Pop III star as their first formed star, with only 4 out of 40 progenitor halos experiencing external enrichment at the time of their first star formation.

We find that star formation in the progenitor halo does not occur continuously over long periods but rather in episodic bursts. During these episodic phases, the stellar mass increases rapidly and then pauses until the next burst of star formation begins. This episodic nature is attributed to the shallow potential wells of the progenitor halos, which are unable to sustain star formation for extended periods due to stellar feedback. During a starburst, stellar feedback heats the surrounding ISM and expels gas from the central region of the progenitor halo, temporarily suppressing or even halting further star formation. This delayed star formation is particularly pronounced in {\sc prog2} of {\sc Halo4}, where star formation begins at $z \sim 10$ and resumes at $z \sim 7$ after a delay of about 300 Myr.

Fig.~\ref{fig:stellar_mass_ratio.png} shows the stellar masses of each progenitor halos, with numbering and colors matching those in Fig.~\ref{fig:morph_all.png}. Here, a prominent feature is that the main progenitor\footnote{We define a main progenitor as the most massive one at a given time. Note that, in some cases, the main progenitor halo changes over time according to their mass growth histories.} ({\sc prog1}) at $z\approx6$ does not significantly contribute to the total stellar mass. For {\sc Halo2} to {\sc Halo5}, the stellar mass fractions of the main progenitor at $z=6$ are 31.1\%, 52.3\%, 45.7\%, 15.0\%, and 48.0\%, respectively. This suggests that in the simulated UFD analogs, stars formed in the main progenitor account for less than 50\% of the total stellar mass. Furthermore, in the case of {\sc Halo5}, the most substantial contributor to the stellar mass is {\sc prog5}, which contributes 29\% of the total stellar mass of {\sc Halo5}. This implies that caution is needed when assuming that most stars in the observed UFD originate from a single progenitor, especially when deducing their SFHs or stellar environments.

Fig.~\ref{fig:SMHM_minihalo.png} presents the resulting stellar mass-halo mass (SMHM) relation for all progenitor halos at $z=6$ (circle) and the final UFD analogs at $z=0$ (cross). Progenitor halos that constitute the same UFD analog are represented by the same color. In addition, progenitors contributing less than 5\% of the total stellar mass are also included. We find a positive correlation in the SMHM relation with a large scatter, particularly at $M_{\rm vir} \approx 10^{7}\Msun$, where stellar mass ranges in $M_{\star}\approx10^{2}-10^{4}\Msun$. 

This spread could be attributed to the vulnerability of small-mass progenitor halos and their varying mass growth histories, which means their stellar mass is sensitively affected by nonlinear and stochastic processes during evolution.  For instance, if a progenitor halo of the same mass forms relatively late in a low-density region, it will have a shorter SFH, resulting in a lower stellar mass. This occurs because, after the onset of a late starburst, star formation will be immediately quenched by cosmic reionization, leaving no opportunity for subsequent star formation. Our SMHM relation for the UFD analogs at $z=0$ aligns well with other theoretical studies (e.g., \citealp{Munshi2013, Wheeler2019, Applebaum2021}), which demonstrate that stellar masses from $M_{\star} = 10^{3}\Msun$ to a few $10^{4} \Msun$ are produced within a total halo mass range from $M_{\rm vir} = 10^{8} \Msun$ to a few $10^{9} \Msun$. 

On the other hand, \citet{Revaz2018} predicted $M_{\star}\sim10^5\Msun$ in the halo mass within $[2\times10^8 - 4\times10^9] \Msun$. While showing a similar halo mass range to our UFD analogs, their stellar masses are roughly an order of magnitude higher. Given the similarities in feedback implementations---such as HI self-shielding and a UV background---the difference may partly stem from the delay time treatment. In our case, instantaneous SN feedback could lead to the early suppression of the star-forming region, compared to \citet{Revaz2018}, where the delayed feedback enables continuous star formation.

\begin{figure}
  \centering
  \includegraphics[width=85mm]{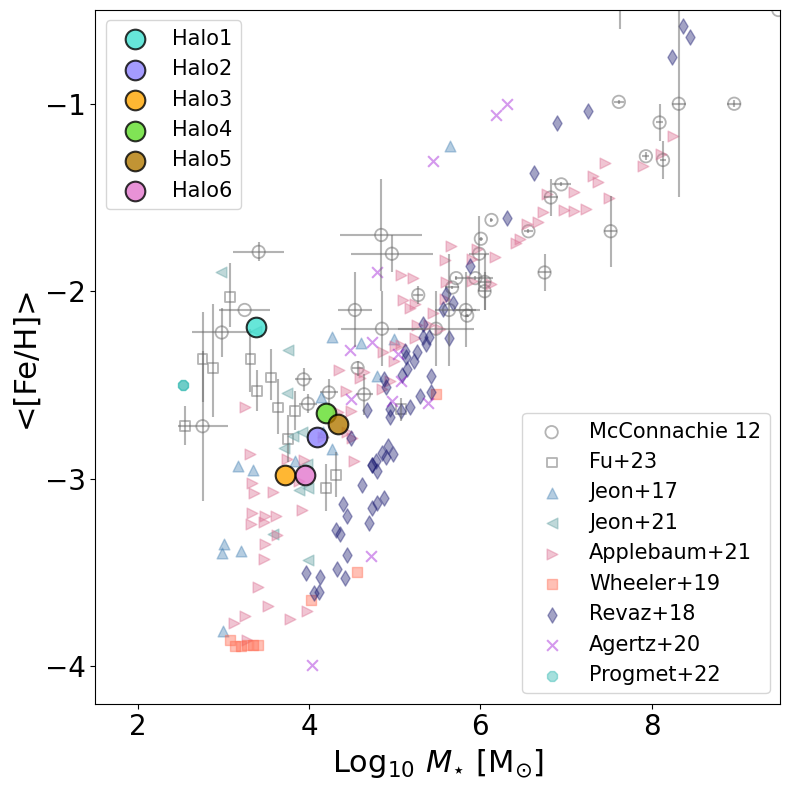}
  \caption{Comparison of the MZR for our simulated UFD analogs with observational data and other theoretical studies, spanning a stellar mass range from $M_{\star}\approx 10^{2.5}\Msun$ to $10^{8.5} \Msun$. Grey symbols represent observed galaxies from \citet{McConaachie2012} and \citet{Fu2023}, while simulation results are shown in various colors (\citealp{Jeon2017, Jeon2021a, Revaz2023, Wheeler2019, Applebaum2021, Agertz2020, Prgomet2022}). Our simulations display generally higher average $\rm \langle[Fe/H]\rangle$ compared to other theoretical works, likely due to the individual IMF sampling method. Nevertheless, the average metallicities in our simulations are 0.5-2.0 dex lower than those observed, except for {\sc Halo1}.  
  }
  \label{fig:mine.png}
\end{figure}

\begin{figure*}
  \centering
  \includegraphics[width=170mm]{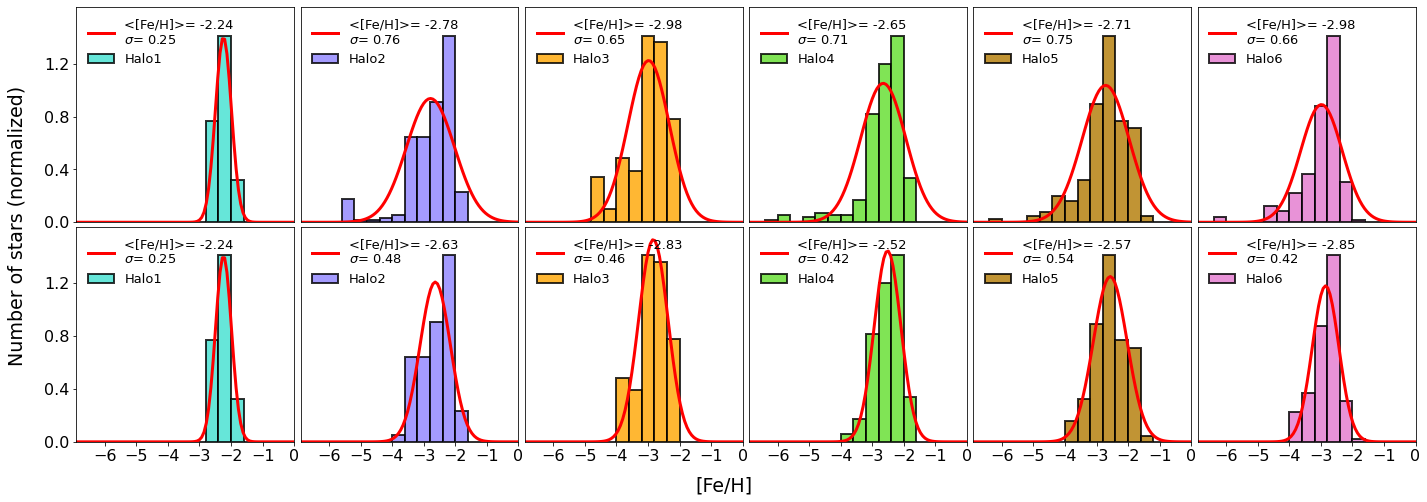}
  \caption{MDFs for our UFD analogs. To determine the average metallicity, $\rm \langle[Fe/H]\rangle$, and metallicity dispersion, $\sigma_{\rm [Fe/H]}$, for Pop II SSP particles consisting of stars below 8$\Msun$ (top panels), we utilize maximum likelihood estimation with a two-dimensional Gaussian likelihood function, similar to the method used for observational data (\citealp{Fu2023}). We also present the MDFs applying the observational limit at $\rm \langle[Fe/H]\rangle= -4$ in the bottom panels. Across all UFDs, the peaks fall within the range of $\rm [Fe/H]-3$ to $\rm [Fe/H]=-2$. Excluding stars with metallicities below $\rm [Fe/H] = -4$ results in an increase in average metallicity and a decrease in metallicity dispersion.}
  \label{fig:MDF_fitting.png}
\end{figure*}

\subsection{Metallicity-related quantities} \label{sec:3.2}

\subsubsection{stellar mass-metallicity relation}

It is well-established that there is a positive correlation between the stellar mass and stellar metallicity of galaxies. This relationship is primarily due to more massive galaxies producing more stars and retaining more metals, which increases the average metallicity of their stars. Fig. \ref{fig:MZR_minihalo.png} illustrates the resulting MZR for progenitor halos at $z\approx6$ (circles) and the corresponding UFD analogs at $z=0$ (crosses). Our results show a positive correlation in the MZR, particularly among progenitor halos. Fig.~\ref{fig:MZR_minihalo.png} also demonstrates how stellar metallicity contributions from multiple progenitor halos can alter the final metallicities of the UFD analogs. For instance, although a progenitor halo with the highest $\rm \langle[Fe/H]\rangle$ value belonging to {\sc Halo5} (depicted in brown) shows $\rm \langle[Fe/H]\rangle=-2.0$ with $M_{\star}\sim10^4\Msun$, the final $\rm \langle[Fe/H]\rangle$ value for {\sc Halo5} is reduced to $\rm \langle[Fe/H]\rangle=-2.71$ due to the contributions from other progenitor halos, lowering the overall $\rm \langle[Fe/H]\rangle$. A similar averaging effect is also observed in {\sc Halo2}, {\sc Halo4}, and {\sc Halo5}. This implies that, similar to the stellar mass assembly process, the metallicity of UFDs is not determined by the single (main) progenitor.  Instead, it is determined by multiple progenitor halos, within which stars exhibit a wide range of metallicities in different environments.

To compare our results with other studies, Fig.~\ref{fig:mine.png} presents the derived MZR for galaxies based on both simulations and observations, covering a stellar mass range from $M_{\star}\approx 10^{2.5}\Msun$ to $10^{8.5} \Msun$. Observed galaxies are depicted in grey: circles denote UFDs and dwarf galaxies in the LG compiled by \cite{McConaachie2012}, from which we use only those with metallicity measurements from spectroscopy, while squares represent MW UFDs reported by \cite{Fu2023}. Simulated galaxies from cosmological simulations are shown in various colors and symbols. Throughout Fig.~\ref{fig:MZR_minihalo.png} and Fig.~\ref{fig:mine.png}, we derive $\rm \langle[Fe/H]\rangle$ values using Pop II SSP star particles by the fitting method introduced in Sec.~\ref{sec:MDF}. 

When comparing the results, an increasing trend with stellar mass is evident in the classical dwarf regime ($M_{\star} \geq \rm 10^{5} \Msun$), showing alignment between observations and simulations. However, this correlation weakens for observed UFD galaxies, exhibiting a plateau rather than an increasing trend. Explaining this observed plateau has been challenging in theoretical work (e.g., \citealp{Jeon2017, Wheeler2019, Agertz2020, Prgomet2022, Jeon2021a, Applebaum2021, Sanati2023}), which shows that smaller-stellar mass UFDs tend to have significantly lower average metallicity ($\rm \langle[Fe/H]\rangle \lesssim -3$).

\begin{figure}
  \centering  \includegraphics[width=85mm]{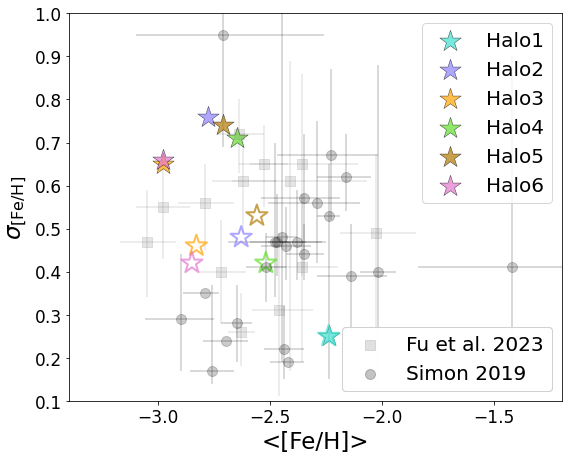}
  \caption{Average metallicity, $\rm \langle[Fe/H]\rangle$, versus metallicity dispersion, $\sigma_{\rm [Fe/H]}$, for our UFD analogs, compared with observed UFDs from \citet{Fu2023} and \citet{Simon2019}. For our simulations, two values are presented for each set: one including all stars across the entire metallicity range (filled stars) and another excluding low-metallicity stars below the observational limit (open stars). With the exception of {\sc Halo1}, the results excluding low-metallicity stars ($\rm [Fe/H] < -4$) show an excellent agreement with the observational data, suggesting that the presence of low-metallicity stars can significantly influence the characteristics of MDFs.}
  \label{fig:disp_feh.png}
\end{figure}

Some possible explanations have been suggested, including a metallicity-dependent IMF (e.g., \citealp{Prgomet2022}) or the consideration of Pop III star formation (e.g., \citealp{Sanati2023}). However, theoretical work still tends to show tension in the UFD regime, with overall $\rm \langle[Fe/H]\rangle$ values being 0.5-2.0 dex lower than observed values. This discrepancy is intriguing given the current ability to simulate UFD galaxies more realistically with high resolutions ($m_{\rm gas}\approx20-100 \Msun$). Meanwhile, the observed plateau for UFDs might be due to tidal effects, as most of them are satellites of massive host galaxies (MW, M31), which could reduce their stellar mass without altering the average metallicity. However, utilizing cosmological simulations, \cite{Applebaum2021} found no significant differences in the metallicity of simulated UFDs when comparing those with and without the host galaxy effect.

In comparison with other simulation studies, the \(\rm \langle[Fe/H]\rangle\) values of \textsc{Halo2} to \textsc{Halo5} range from \(\rm \langle[Fe/H]\rangle \approx -2.98\) to \(-2.65\), which is about 1-1.5 dex higher than those reported in other works (e.g., \citealp{Wheeler2019, Applebaum2021, Sanati2023}). Furthermore, our results exhibit a smaller scatter in \(\rm \langle[Fe/H]\rangle\) values, with the lowest being \(\rm \langle[Fe/H]\rangle \approx -2.98\), whereas other studies yield the lowest values as \(\rm \langle[Fe/H]\rangle \lesssim -3.5\). We attribute these overall higher and narrower values, compared to other simulation studies, to the individual IMF sampling method adopted in this work.

This can be understood as follows: if we use SSP scheme, the combined SNe energy from the SSP stellar particle can exert excessive feedback on the surrounding medium, suppressing subsequent star formation for a period during which metals are too effectively dispersed, leading to the formation of subsequent stars with lower metallicity (e.g., \citealp{Agertz2020, Applebaum2020}). On the other hand, with individual IMF sampling, stars tend to from gas that has been enriched by the previous SN explosion, immediately reflecting the gas metallicity shaped by that event. The impact of the IMF sampling method on the SFHs and resulting stellar metallicities of the simulated UFDs is also demonstrated in \citet{Jeon2024}.

Interestingly, {\sc Halo1} exhibits a higher average metallicity with $\rm \langle[Fe/H]\rangle = -2.19$. This uniqueness arises from two factors: first, Halo1 has a single progenitor, and second, this progenitor halo has high metallicity. At the progenitor halo scale, rather than the galaxy scale, this high metallicity is not particularly unusual. This is demonstrated by Fig.~\ref{fig:MZR_minihalo.png}, where other halos also have progenitor halos with similarly high metallicities. However, for other halos, the average metallicity tends to be reduced as a result of mergers with low-metallicity progenitor halos during growth. In contrast, the average metallicity of Halo1 remains high because of its single high-metallicity progenitor with no further mergers. The properties of the progenitor halo that contribute to the high metallicity are discussed in Sec.~\ref{sec:key-properties}.

\subsubsection{Metallicity distribution function}\label{sec:MDF}

To investigate the detailed distribution of stellar metallicities, Fig.~\ref{fig:MDF_fitting.png} presents MDFs of our simulated UFDs. The top panels display the MDF for all stars, while the bottom panels show the MDF for stars with $\rm [Fe/H] \geq -4$, reflecting the observational limit\footnote{In \citet{Fu2023}, they measured metallicity using the MESA Stellar Isochrones and Tracks (MIST) models (\citealp{Choi2016, Dotter2016}), where the metallicity grid is set from $\rm [Z/H]=-4$ to $\rm [Z/H]=-1$.} (e.g., \citealp{Fu2023}). To determine the average metallicity, $\langle \mathrm{[Fe/H]} \rangle$, and metallicity dispersion, $\sigma_{[\mathrm{Fe}/\mathrm{H}]}$, for Pop II SSP stars with masses below $8~\Msun$ (top panels), we utilize maximum likelihood estimation with a two-dimensional Gaussian likelihood function (e.g., \citealp{Walker2006}), which is the same method used for observational data in \citet{Fu2023}.

As shown in the top panels of Fig.~\ref{fig:MDF_fitting.png}, all the simulated galaxies exhibit a Gaussian-like distribution, with peaks in the range of $\rm -3 \lesssim [Fe/H] \lesssim -2$ and dispersion values between $0.25 \lesssim \sigma_{\rm [Fe/H]} \lesssim 0.76$.
In the case of {\sc Halo1}, the MDF has a narrower distribution than other halos, with the lowest dispersion of $\sigma_{\rm [Fe/H]} \rm = 0.25$. This narrowness arises from having only one progenitor and star formation occurring in a single starburst event. On the other hand, since stars in the other simulated UFD analogs form through bursty star formation in multiple progenitor halos, the metallicity dispersion tends to increase. For example, {\sc Halo5}, where stars formed in a total of 14 different progenitor halos prior to $z=6$, exhibits an increased metallicity dispersion, with $\sigma_{\rm [Fe/H]} \rm = 0.75$.

One notable point in our results is the existence of low-metallicity stars with $\rm [Fe/H] < -4$. We observe that such low-metallicity stars primarily originate from external enrichment, where Pop II stars form first, without prior Pop III star formation, from gas contaminated by metals from nearby halos. The limited amount of metals from the external halo gives rise to low-metallicity stars. In addition, these progenitor halos that experience external enrichment tend to have relatively small virial masses and are often located near more massive progenitor halos, as observed in other works (e.g., \citealp{Smith2015, Hicks2021}). Through all progenitor halos in our sets, few stars form through external metal enrichment, as most progenitor halos underwent internal enrichment for their initial ISM enrichment. For this case, stars with metallicities of $\rm [Fe/H] \geq -4$ form through just one or two SN explosions. For example, in the {\sc prog1} of {\sc Halo4}, a SN from a Pop III star with $m_{\star, \ \rm Pop III} \approx 60 \Msun$ leads to the formation of a Pop II star with $\rm [Fe/H] = -3.5$.

Given the observational limit of $\rm [Fe/H] \sim -4$, we also fit the MDFs excluding stars below this limit. As shown in the bottom panels of Fig.~\ref{fig:MDF_fitting.png}, this exclusion leads to a noticeably narrower distribution, with the dispersion reduced by an average of $\sim$0.2 dex and the average metallicity increasing by about 0.1 dex compared to the MDFs that include all stars. To compare our MDFs to the observational results, Fig.~\ref{fig:disp_feh.png} shows $\rm \langle[Fe/H]\rangle$ versus $\sigma_{\rm [Fe/H]}$ of galaxies, with the observed UFDs depicted in grey. The values for all stars are represented by filled star symbols, whereas those for stars with $\rm [Fe/H] \geq -4$ are depicted by open star symbols. There is a prominent change in values when we exclude those low-metallicity stars with $\rm [Fe/H] < -4$. For all halos except {\sc Halo1}, $\rm \langle[Fe/H]\rangle$ shifts to higher values, and $\sigma_{\rm [Fe/H]}$ shifts to lower values, showing an excellent match with the observational data. This might indicate that if we can discover low-metallicity stars in observed UFDs, the discrepancy in the MZR can partly be mitigated by reducing the average metallicity of the observed data. Also, note that for the observed UFDs with $\rm \sigma_{\rm [Fe/H]} \leq 0.3$, they might have similar SFHs to {\sc Halo1}, where stars form in a single progenitor and/or through a single bursty star formation event.

\begin{figure}
  \centering
  \includegraphics[width=85mm]{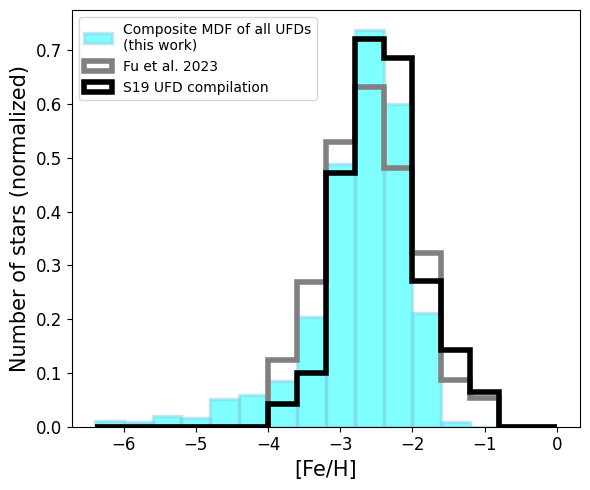}
  \caption{Comparison of the normalized composite MDF between our simulated UFDs (cyan) and the observational data from \citet{Fu2023} (grey) and \citet{Simon2019} (black). Our composite MDF aligns well with the peak location, but it depicts a lower fraction of high-metallicity stars ($\rm [Fe/H] \geq -2$) compared to the observational data. Note that the consistent agreement among observational studies indicates the general presence of such high-metallicity stars.} 
  \label{fig:MDF_integ.png}
\end{figure}

\begin{figure}
  \centering
  \includegraphics[width=80mm]{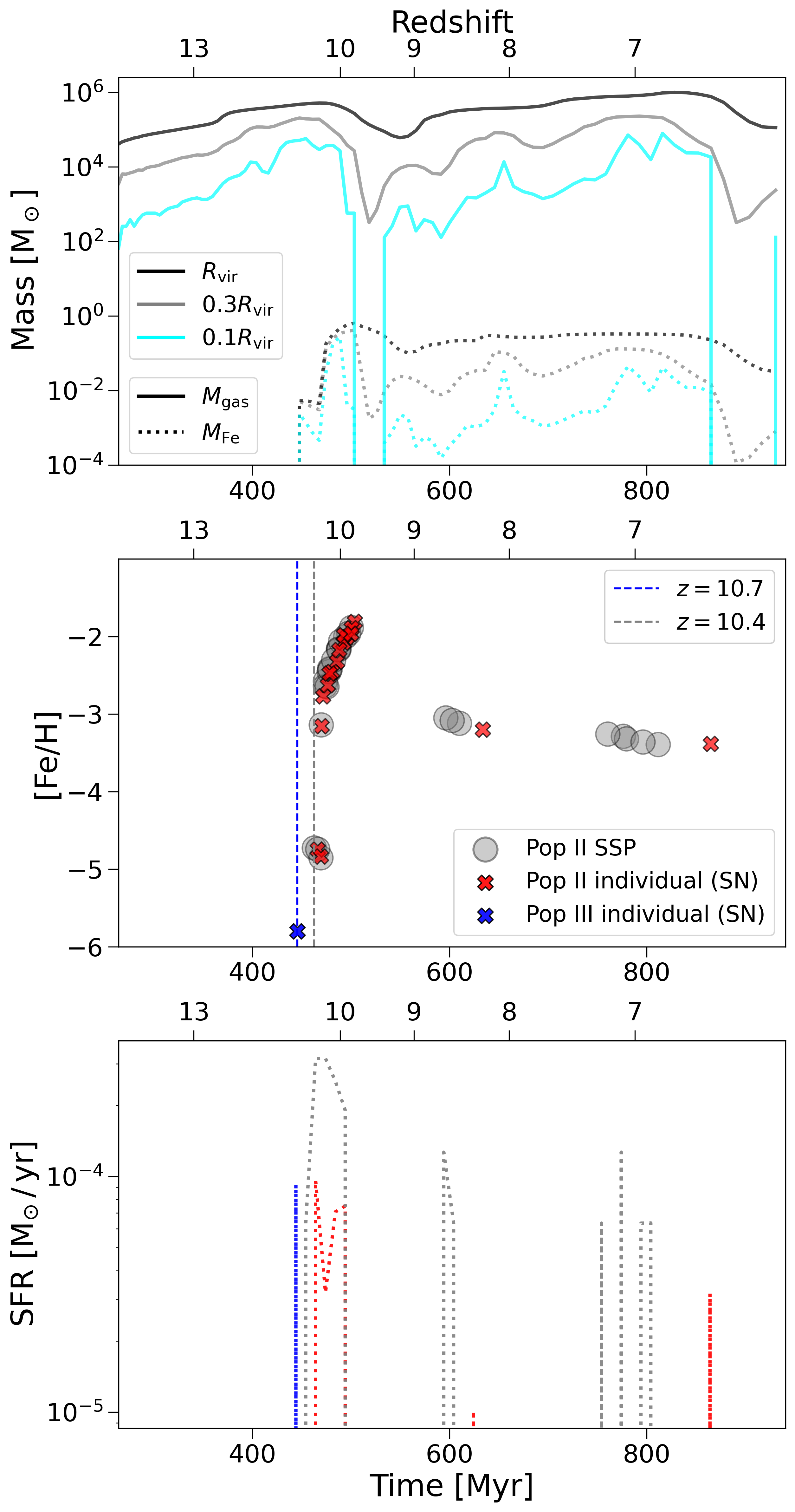}
  \caption{The procedure describing high-metallicity star formation during a starburst in {\sc prog2} of {\sc Halo5}. Top Panel: The evolution of gas (solid lines) and metals (dotted lines) over cosmic time within spheres of radius 0.1 (cyan), 0.3 (grey), and 1 $R_{\rm vir}$ (black). Middle Panel: Stellar metallicity as a function of star formation time is shown, focusing only on in situ stars. Blue crosses represent Pop III stars, Pop II SN stars by red crosses, and Pop II SSP stars that survive to $z=0$ are indicated by grey circles. After the formation of the first Pop II SSP star at $z=10.4$, a period of bursty star formation occurs, leading to the creation of high-metallicity stars from gas enriched by previous continuous star formation. However, subsequent cumulative SNe feedback significantly reduces the metal and gas mass, temporarily evacuating the gas within 0.1 $R_{\rm vir}$. This suppression halts further metal-enriched star formation in the progenitor. Bottom Panel: Star formation rates of each stellar type-Pop III (blue), Pop II individual star (red), and Pop II SSP, which is calculated binning 10 Myr.} 
  \label{fig:Halo5_mh2.png}
\end{figure}

\begin{figure*}
  \centering
  \includegraphics[width=170mm]{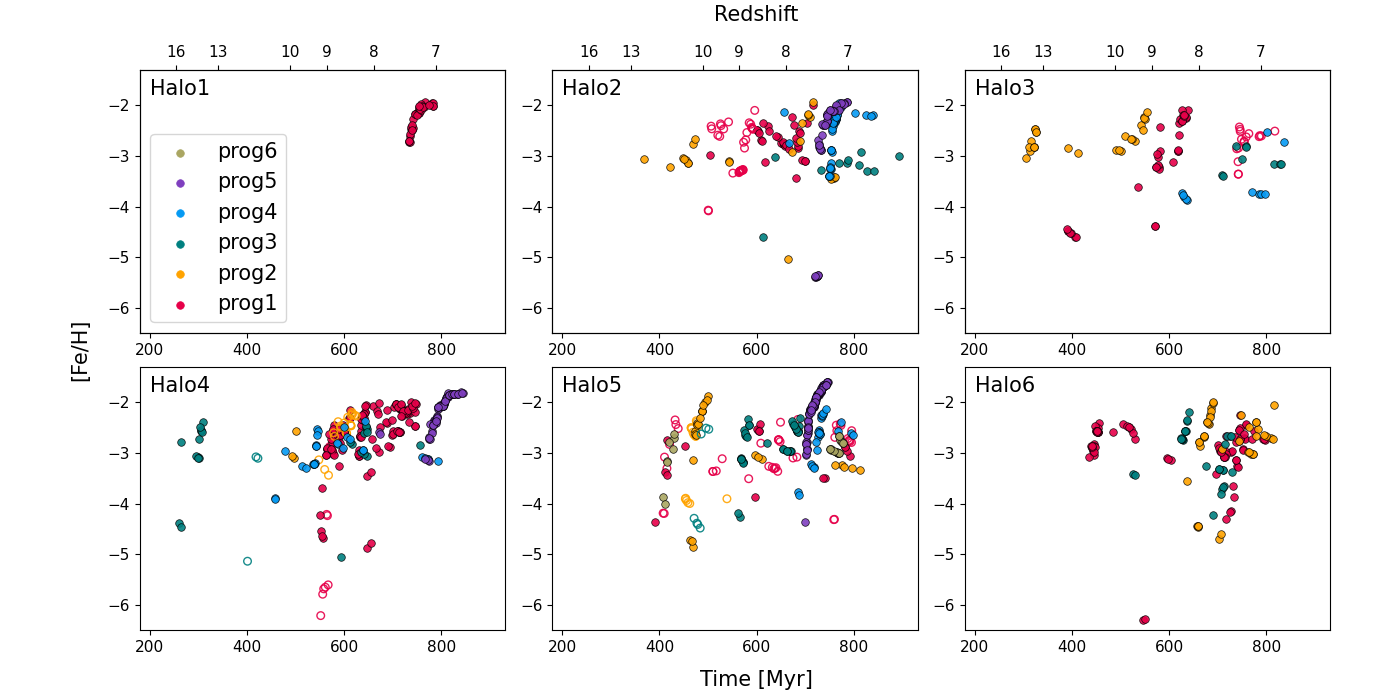}
  \caption{The evolution of stellar metallicity for progenitor halos, with colors matching those in Fig.~\ref{fig:stellar_mass_ratio.png}, generally following a similar starburst pattern to {\sc prog2} of {\sc Halo5} described in Fig.~\ref{fig:Halo5_mh2.png}. In these patterns, stellar metallicity increases during bursty star formation, continuously forming stars with enhanced metallicity until cumulative feedback effects quench further star formation at a certain point.}
  \label{fig:fe_h_2by3.png}
\end{figure*}

To figure out the general trend of MDFs for UFDs, Fig.~\ref{fig:MDF_integ.png} compares the composite MDFs for stars in all our simulated galaxies to those from observational data. The grey histogram displays stars observed in 13 MW UFDs as reported by \cite{Fu2023}, while the black histogram includes stars in observed 26 UFDs in Local Group (LG), as detailed by \cite{Simon2019}. One evident difference in the composite MDFs is that, although both observed composite MDFs show a similar fraction of high-metallicity stars with $\rm [Fe/H] \geq -2.0$, reflecting the typical ratio of these stars in UFDs ($\sim$19\%), our simulated galaxies lack such high-metallicity stars. This absence of high-metallicity stars may explain the discrepancy in the MZR between observations and simulations. In the next section, we discuss the challenges in producing such high-metallicity stars in simulations of UFD analogs. 

\subsubsection{High-metallicity star formation}
Here, we scrutinize the evolution of stellar metallicity by tracking the star-forming environment, gas, and metal reservoirs within a progenitor halo. Since the metallicity evolution in most progenitor halos exhibits a similar monotonically increasing pattern, we focus on {\sc prog2} of {\sc Halo5} as a representative example. In the top panel of Fig.~\ref{fig:Halo5_mh2.png}, we display the mass evolution of gas (solid line) and metal (dotted line) within the progenitor halo as a function of cosmic time. We calculate the mass within 0.1 (cyan), 0.3 (grey), and 1 $R_{\rm vir}$ (black), respectively. This approach is taken because most stars form in the central region ($r_{\rm star} < 0.3 \ R_{\rm vir}$), where star formation is suppressed if there is insufficient gas. In the middle panel, we show the metallicity of stars as a function of their formation time, focusing only on the evolution of in situ stars. Blue crosses represent Pop III stars; red crosses indicate Pop II individual stars that explode as SN; and grey circles denote Pop II SSP stars that survive to $z=0$. Finally, the bottom panel presents the star formation rate of these stars.

We find the initial production of metals at $z \sim 10.7$, triggered by a Pop III SN. About 17 Myr later, a Pop II star forms at $z \sim 10.4$ with a metallicity of $\rm [Fe/H] = -4.7$, marking the onset of the first phase of starburst. During this phase, the stellar metallicity increases monotonically from $\rm [Fe/H] = -3.1$ to $\rm [Fe/H] = -1.8$. 
This increase in metallicity is possible by the localized and short timescale nature of SN feedback. Some dense regions, which have not yet formed stars, can be rapidly enriched by nearby SNe. However, due to their high density, they can withstand the injected energy without being disrupted, allowing star formation to proceed in their enriched state. As shown in the top panel of Fig.~\ref{fig:Halo5_mh2.png}, metal mass within the innermost region ($r < 0.1 \ R_{\rm vir}$) initially increases due to metal ejection from Pop II SNe, and most of metals remain concentrated in this central region, although some diffuse outward. However, as SN energy continues to accumulate over time, it eventually disrupts the dense gas in the core, which causes a rapid outflow of metals from the center to the outer regions. This effective suppression temporarily halts star formation in the halo. Star formation remains suppressed until a high-density region is reestablished at $z \sim 8.6$, about 100 Myr after the first starburst phase.

In the second phase of starburst at $z < 9$, shown in Fig.~\ref{fig:Halo5_mh2.png}, the initially formed in situ star has a metallicity of $\rm [Fe/H] \sim -3$, indicating that it forms from gas enriched by SN feedback of previous starburst. During the period of 100 Myr between first and second starbursts, metals diffuse outward and disperse into extended regions, resulting in a decrease in the metallicity. This explains why the second starburst exhibits lower metallicities than the first starburst. Moreover, in the second starburst, individual stars that explode as SNe are sampled later in the phase, and this absence of early SNe leads to a monotonic decrease in metallicity during the starburst. The third starburst also features a late-forming SN progenitor at $z \sim 6.7$. This final SN feedback, combined with the effects of reionization, suppresses the ISM and completely quenches further star formation. Therefore, unlike in massive galaxies---where stellar metallicity typically increases monatonically over long timescale---high-metallicity stars in the UFD analogs are likely to form over short periods, specifically during each starburst phase.

Fig.~\ref{fig:fe_h_2by3.png} shows the evolution of stellar metallicity in each progenitor halo, whose stellar mass accounts for more than 5\% of the total stellar mass of each simulated UFD analog. Stars formed within the same progenitor halo are represented by the same color. We display only Pop II SSP stars, as they persist until $z = 0$, making them comparable to observational data. Within each progenitor halo, we identify starburst events based on their location, formation time, and density. For instance, there are primarily two starbursts in the progenitor, {\sc prog2} of {\sc Halo3}, and a single starburst in {\sc prog5} of {\sc Halo4} and {\sc Halo5}. All progenitor halos exhibit similar starburst patterns to those described for {\sc prog2} of {\sc Halo5} in Fig.~\ref{fig:Halo5_mh2.png}. However, the maximum metallicity achieved, which typically indicates the last formed star in each starburst, varies among the starbursts. The maximum metallicity of stars ranges from $\rm -3.0 \leq [Fe/H]_{max} \leq -1.5$.

\subsubsection{Key physical properties for high-metallicity star formation} \label{sec:key-properties}
To understand what physical properties determine the maximum metallicity in each starburst, we investigate the correlation between maximum metallicity and other physical properties. We select four physical quantities that could influence the maximum $\rm [Fe/H]$ value during a starburst: $M_{\rm vir}$, $M_{\rm gas}$, $n_{\rm H, max}$, and $N_{\rm SN}$. Here, $M_{\rm vir}$ and $M_{\rm gas}$ denote the virial mass and gas mass within the virial radius of progenitor halos at the onset of each starburst. Meanwhile, $n_{\rm H, max}$ and $N_{\rm SN}$ represent the maximum hydrogen density and the number of SNe during the starburst, respectively. 

\begin{figure}
  \centering 
  \includegraphics[width=85mm]{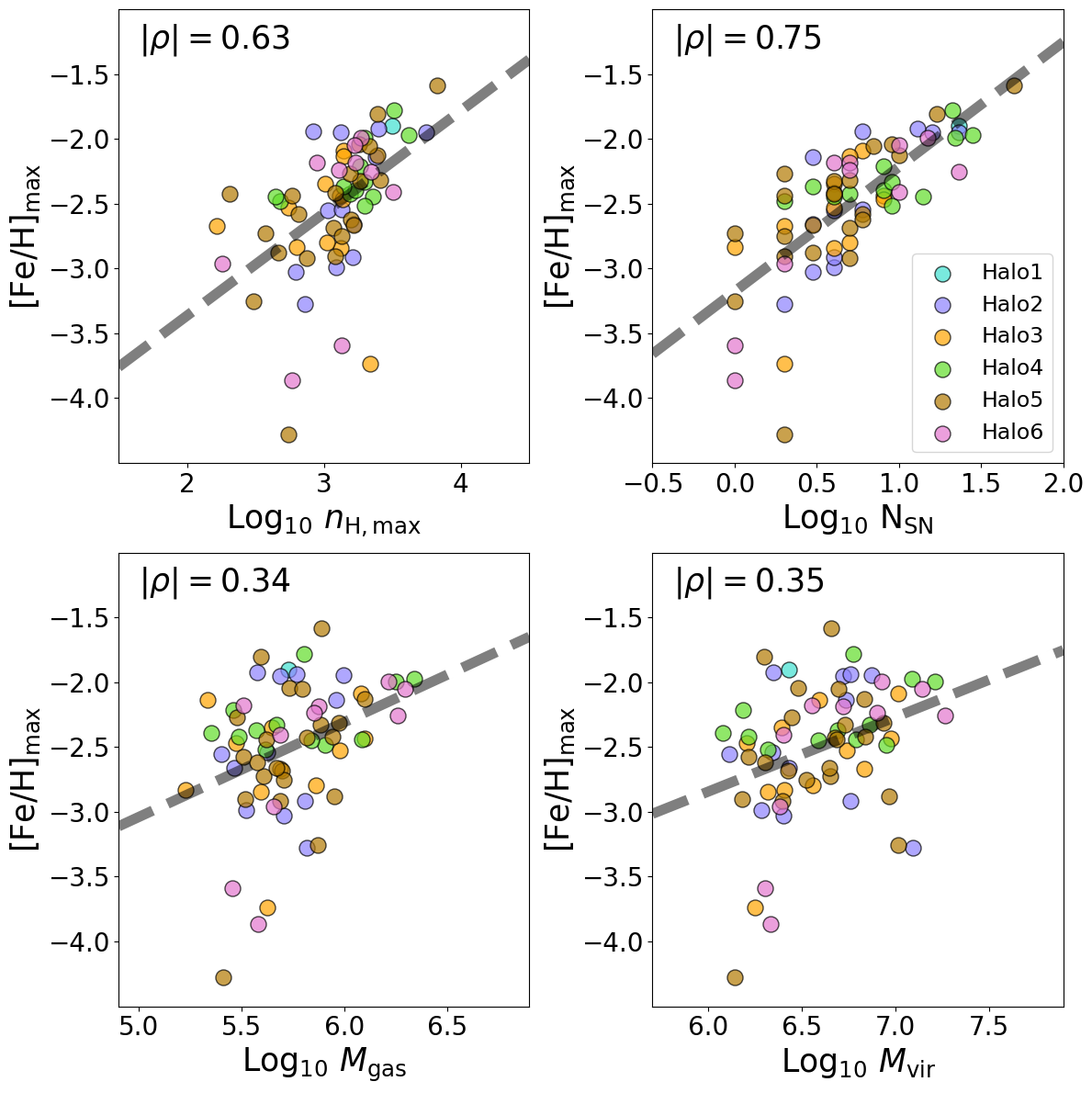}
  \caption{Correlations between maximum metallicity and physical quantities during starbursts are shown clockwise from the upper left: maximum hydrogen number density ($n_{\rm H, max}$), the number of SNe ($N_{\rm SN}$),  
  virial mass ($M_{\rm vir}$), and gas mass ($M_{\rm gas}$). Here, $M_{\rm vir}$ and $M_{\rm gas}$ refer to the virial mass and gas mass within the virial radius of progenitor halos at the onset of each starburst.}
  \label{fig:correlation.png}
\end{figure}

Fig.~\ref{fig:correlation.png} illustrates the relationships between the maximum metallicity, $\rm [Fe/H]_{max}$, and each of these properties. Each data point represents the properties of each starburst, with the color indicating the UFD analog to which the progenitor halo belongs. To quantify the relationship between these selected properties and $\rm [Fe/H]_{max}$, we calculate Spearman's rank correlation coefficient $\vert \rho \vert$, which measures the strength of the monotonic relationship between two variables (\citealp{Spearman1904}). In this test, $\vert \rho \vert \rm = 1$ indicates a perfect monotonic relationship. Our calculations reveal that maximum hydrogen density and the number of SNe exhibit strong correlations with the maximum metallicity, with coefficients of $\vert \rho \vert \rm=0.63$ (p-value $\rm = \ 7\times10^{-9}$) for $n_{\rm H, max}$ and $\vert \rho \vert \rm=0.75$ (p-value$\rm \ = 0$) for $N_{\rm SN}$, respectively. In contrast, gas mass and virial mass show weaker correlations, with coefficients of $\vert \rho \vert \rm = 0.34$ (p-value $=0.0045$) for $M_{\rm gas}$ and $\vert \rho \vert \rm = 0.35$ (p-value $=0.0037$) for $M_{\rm vir}$, respectively.

The strong correlation between maximum hydrogen density ($n_{\rm H, max}$) and maximum metallicity is closely related to the pronounced correlation between the number of SNe ($\rm N_{SN}$) and maximum metallicity. This relationship is expected, as a highly dense region is necessary to trigger a large number of SN events. For instance, the highest metallicity star with $\rm [Fe/H]_{\rm max} = -1.58$, formed in {\sc prog5} of {\sc Halo5}, is achieved through the cumulative effects of 50 SN events. On the other hand, the weak correlations between maximum metallicity and global properties of progenitor halos, such as gas mass ($M_{\rm gas}$) and virial mass ($M_{\rm vir}$), are intriguing. This implies that, unlike relatively massive progenitor halos (e.g., those in classical dwarf galaxies), progenitor halos with small virial masses struggle to retain gas within their virial radius due to their vulnerability. Consequently, simply having a large virial mass does not ensure the formation of high-metallicity stars. Instead, the local properties of the star-forming region play a more crucial role in determining whether high-metallicity stars are formed. 

As the number of SNe increases, while higher metallicity stars may form, the resulting cumulative SN feedback inhibits further star formation. This effect is particularly evident in low-mass progenitor halos existing before $z\approx6$, where shallow potential wells make it unlikely for stars with metallicities higher than $\rm [Fe/H] = -1.5$ to form in our simulation. Certainly, as the virial mass of the progenitor halo increases, these halos can experience metal enrichment over a more extended period, potentially leading to the formation of high-metallicity stars. However, ultimately, around $z\approx6$, the heating effect from reionization halts nearly all star formation, making the formation of stars with metallicity $\rm [Fe/H]\geq-1.5$ extremely challenging, as observed.

\subsubsection{MDF with different SN energy} \label{sec:MDF_SN_energy}
The energy of SN feedback plays a crucial role in shaping MDFs as it directly affects the ISM by expelling gas into the intergalactic medium (IGM) or suppressing the formation of high-density regions. For example, \cite{Agertz2020} demonstrated that increasing SN energy by a factor of 100 results in about a tenfold decrease in stellar mass and a reduction in metallicity by more than 1 dex. Their findings emphasize that metal enrichment in the ISM is highly sensitive to different feedback models. While many simulations have explored the effects of SNe on average metallicity, their impact on MDFs has not yet been studied for the UFD regime. In this context, we further investigate how varying SN energy affects MDFs, particularly focusing on the possibility of forming high-metallicity stars as observed.

\begin{figure*}
  \centering 
  \includegraphics[width=185mm]{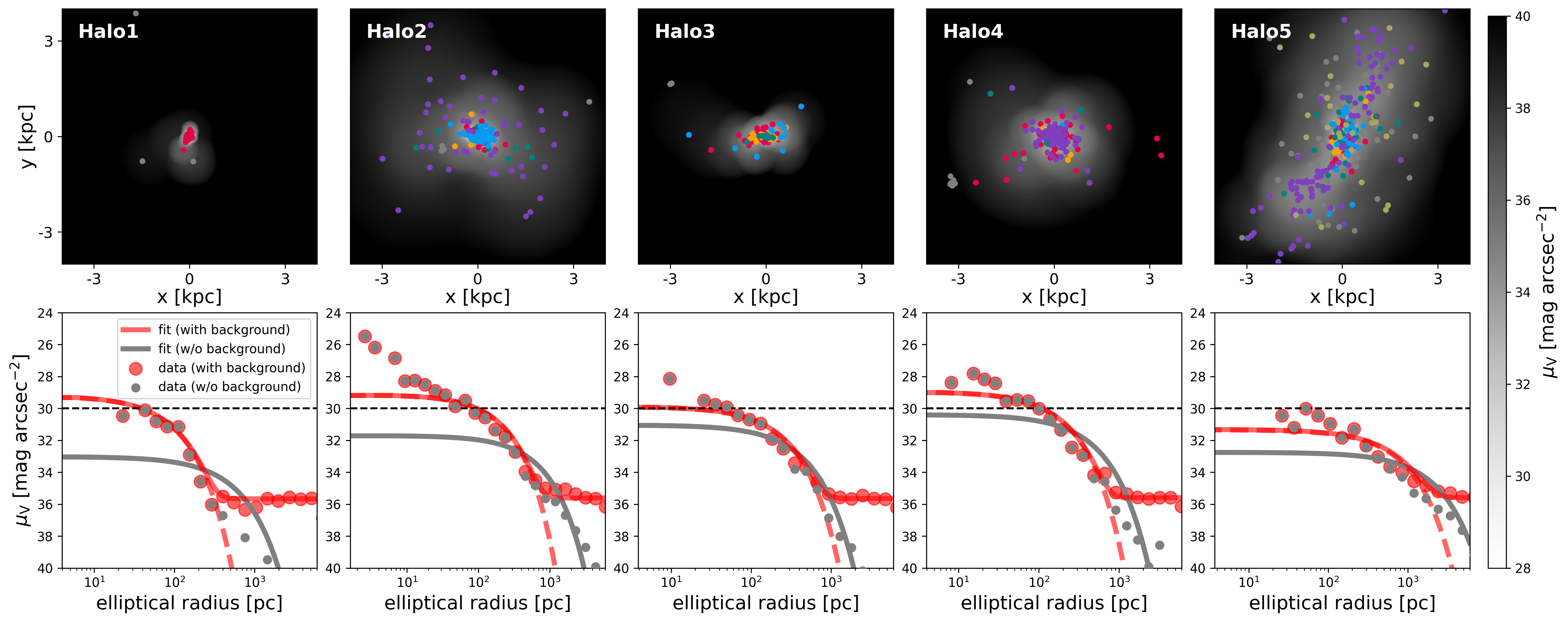}
  \caption{Top panel: Surface brightness map of the UFD analog, with star particles overlaid and color-coded by their progenitor halo. Artificial background stars used for fitting are excluded from this plot. Bottom panel: The surface brightness density profile, calculated by binning the elliptical radius from the center of the simulated UFD analog. Data are shown for both cases, including background stars (red circles) and excluding background stars (grey circles). The solid red line represents the fit with the background, while the grey line represents the fit without the background. The latter one, which considers all stars within the virial radius of the UFD analog as the member stars, results in a poor fit with an exceptionally large radius.}
  \label{fig:desity_profile}
\end{figure*}

For this purpose, we conduct additional simulations on UFD analogs using the same initial conditions of {\sc Halo4}, varying the energy levels for each SN event, with 0.5 $E_{\rm SN}$, 1.0 $E_{\rm SN}$, and 2.0 $E_{\rm SN}$ where $E_{\rm SN} \ = 10^{51} \rm ergs$. Fig.~\ref{fig:MDF_SNe} presents the resulting MDF of the stars within the simulated UFD at $z=0$. As expected, the simulation with the reduced energy value of 0.5 $E_{\rm SN}$ results in higher stellar mass and average metallicity, with $M_{\star} = 2.46 \times 10^{4} \Msun$ and $\rm \langle[Fe/H]\rangle \approx -2.38$. Compared to the fiducial set with 1.0 $E_{\rm SN}$, the stellar mass increases by a factor of 1.4, and $\rm \langle[Fe/H]\rangle$ increases by 0.27 dex (a factor of 1.9). Weaker SN feedback allows more stars to form within the UFD analog, enhancing associated metal production due to the increased number of SNe. Conversely, as SN energy increases, the overall MDF shifts towards a lower metallicity range. This occurs because gas eligible for star formation is more likely to be expelled along with the metals, resulting in reductions in both stellar mass and the average metallicity of the simulated UFD.

When comparing the simulated MDFs to observational data, the fiducial set (1.0 $E_{\rm SN}$) aligns most closely with observations despite lacking stars with metallicity $\rm [Fe/H] \geq -1.5$. On the contrary, the reduced SN energy run (0.5 $E_{\rm SN}$) yields the highest metallicity star with $\rm [Fe/H] \approx -1.2$. However, although stars with metallicities, up to $\rm [Fe/H] \approx -1.2$, form in the reduced SN energy run (0.5 $E_{\rm SN}$), this model overproduces stars with metallicity $\rm [Fe/H] \geq -2.0$ and underproduces those with $\rm [Fe/H] < -2.8$ compared to observations. Moreover, while lower SN energy could facilitate the formation of stars with metallicity $\rm [Fe/H] \geq -1.5$, it leads to an increase in stellar mass as well by a factor of 1.4. This highlights the challenge of forming metal-rich stars in the UFD regime while maintaining the low stellar mass observed in UFDs. The difference in maximum metallicity according to SN energy implies that the accurate measurements of metal-rich stars in UFDs can constrain the sub-grid model, thus contributing to the refinement of the galaxy evolution model (e.g., \citealp{Sandford2024}).

\begin{table*}
  \centering
  \setlength{\tabcolsep}{8pt}
  \begin{tabular}{c c c c c c c c}
    \hline%\toprule
    UFD & $M_{\star}$ & $\rm \langle[Fe/H]\rangle$ & $r_{\rm h, dir}$ & $r_{\rm h, fit \ (w/)}$ & $\epsilon_{\rm h, fit \ (w/)}$ & $r_{\rm h, fit \ (w/o)}$ & $\epsilon_{\rm h, fit \ (w/o)}$ \\
    analog & $[10^4 \rm \Msun]$ & & [pc] & [pc] & & [pc] & \\
    \hline%\midrule
    {\sc Halo1} & 0.24 & -2.19 & 89 & $87\pm16$ & $0.34\pm0.16$ & $516\pm71$ & $0.25\pm0.12$ \\[0.1cm]
    {\sc Halo2} & 1.22 & -2.78 & 228 & $182\pm17$ & $0.44\pm0.06$  & $667\pm39.5$ & $0.31\pm0.05$ \\[0.1cm]
    {\sc Halo3} & 0.52 & -2.98 & 175 & $251\pm33$ & $0.58\pm0.06$  & $390\pm36$ & $0.53\pm0.05$ \\[0.1cm]
    {\sc Halo4} & 1.58 & -2.65 & 176 & $180\pm13$ & $0.23\pm0.07$  & $445\pm23.5$ & $0.40\pm0.04$ \\[0.1cm]
    {\sc Halo5} & 2.14 & -2.71 & 1115 & $754\pm201$ & $0.57\pm0.11$  & $1926\pm86$ & $0.62\pm0.02$ \\[0.1cm]
    \hline%\bottomrule
  \end{tabular}
  \caption{ Structural parameters of the simulated UFDs at $z=0$. The columns represent the following parameters: (1) UFD analog name. (2) $M_{\star}$: Stellar mass. (3) $\rm \langle[Fe/H]\rangle$: Average stellar iron-to-hydrogen ratio for all stars within $r_{\rm vir}$. (4) $r_{\rm h, \ dir}$: 2D half-light radius based on the direct method. (5)–(8): Results from the fitting method: half-light radius and ellipticity derived \textit{with} background stars ($r_{\rm h, \ fit \ (w/)}$, $\epsilon_{\rm h, \ fit \ (w/)}$) and \textit{without} background stars ($r_{\rm h, \ fit \ (w/o)}$, $\epsilon_{\rm h, \ fit \ (w/o)}$).}\label{tab:one_column_table}
\end{table*}

\subsection{Galaxy-size related quantities} \label{sec:3.3}

\subsubsection{Half-light radius} \label{sec:Size-lum}

In addition to MZR, both observations and simulations of UFDs have shown discrepancies in the size-luminosity relation, with cosmological simulations struggling to reproduce the observed compactness ($r_{\rm h}\lesssim 50 \rm \ pc$). Simulated UFDs typically exhibit larger sizes, spanning several hundred parsecs (e.g., \citealp{Jeon2017, Prgomet2022, Sanati2023}), and in some cases, sizes approach nearly 1 kpc (e.g., \citealp{Wheeler2019, Jeon2021a}). A possible cause for this size discrepancy may be differences in the methods used to derive the half-light radius. For simulated UFDs, the half-light radius is directly measured by finding the radius where the stellar mass within the virial radius is halved while applying a constant mass-to-light ratio to the stellar particles. On the other hand, observed UFDs are modeled using specific functional forms, such as an exponential profile, and maximum likelihood fits are performed on the probability density functions with free parameters to describe their spatial structures. For convenience, we refer to the former as the direct method and the latter as the fitting method. To address this methodological difference between simulation and observation, we apply the same method commonly used in observations to derive the structural parameters of stellar distributions, utilizing maximum likelihood estimation techniques. For a detailed mathematical explanation, we refer readers to \citet{Martin2008}.

In this method, stars are assumed to follow an elliptical exponential density profile, defined as,
\begin{equation}
\label{eq_profile1}
\Sigma(r) = \Sigma_{0} e^{-r/r_{\rm e}} + \Sigma_{\rm b},
\end{equation}
where $r$ is the elliptical radius and $r_{\rm e}$ is the exponential scale radius, related to the half-light radius by $r_{\rm h}=1.68 \ r_{\rm e}$. The parameters $\Sigma_{0}$ and $\Sigma_{\rm b}$ represent the central and background stellar densities, respectively, with $\Sigma_{0}$ given by $\Sigma_{0} = N_{\star}/(2\pi r_{\rm e}^{2}(1-\epsilon))$. Here, $N_{\star}$ is the total number of galaxy member stars, and $\epsilon$ is the ellipticity, defined as $\epsilon = 1 - b/a$, with $b/a$ being the minor-to-major axis ratio of the galaxy. The elliptical radius is calculated as
\begin{equation}
r = \bigg\{\Big[\frac{1}{1-\epsilon} (X \cos\theta - Y \sin\theta)\Big]^{2} + (X \sin\theta - Y \cos\theta)^{2}\bigg\}^{1/2},
\end{equation}
where $X$ and $Y$ are the stellar position\footnote{Here, $X$ and $Y$ represent the Cartesian coordinates of stellar positions, with stars projected onto a plane.} with respect to the centroid ($X_{0}, \ Y_{0}$), and $\theta$ is the angular offset of the ellipse from north to east. In total, six parameters ($X_{0}, \ Y_{0}, \ \epsilon, \ r_{\rm e}, \ \theta, \ N_{\star}$) are derived to maximize the likelihood. The likelihood is computed by summing the probabilities of finding each star particle within the distribution, i.e., $l_{i} = \Sigma(r_{i})$. The likelihood function, $\mathcal{L}$, is given by,
\begin{equation}
\centering
\log \mathcal{L}  = \sum_{i} \log l_{i}.
\end{equation}
We determine the best-fitting structural parameters using the {\sc emcee} code (\citealp{Foreman2013}), running 50 walkers over a total of 10,000 steps, with the first 1,000 steps discarded to ensure the convergence of the parameter values.

Here, we should highlight the importance of including artificial background stars and their impact on fitting results. Artificial background stars\footnote{To avoid confusion among observers, we will refer to "artificial background stars," used exclusively in simulated UFDs, simply as "background stars". We also note that the term "background stars" in this paper encompasses both background and foreground stars in real observations of UFDs.} refer to stars added to simulated galaxies to replicate the observational conditions of UFDs. Without accounting for background stars, we find that when there are extended stars in UFDs, maximum likelihood estimation produces an improper fit due to the lack of a suitable exponential stellar density profile that adequately captures both the inner, compact distribution and the outer, extended structure. Consequently, the estimated half-light radius is significantly larger than observed values typically seen in UFDs.

For instance, {\sc Halo1} contains seven extremely extended star particles distributed over a few kiloparsecs. Including these stars and fitting only the star particles within the virial radius of the simulated galaxies, without accounting for background stars, results in a half-light radius of $r_{\rm h, \ fit} = 516 \ \rm pc$, making it difficult to accurately capture both the inner and outer profiles simultaneously. This estimated half-light radius is excessively larger than that obtained via the direct method, which yields a much smaller radius of $r_{\rm h, \ dir} = 89 \ \rm pc$. In actual observations, these seven extended stars would likely be misidentified as background stars due to their low surface brightness and thus not included as members of the galaxy. To take an approach similar to observations, we artificially generate background stars using Poisson random sampling, assuming a background stellar density of $\rm 0.1 \ arcmin^{-2}$, as an initial value ($\Sigma_{\rm b,0}$) during the fitting process. The value is arbitrary but representative of the background density of observed dwarf galaxies such as Draco (e.g., \citealp{Segall2007}). This approach reduces the half-light radius when applying the fitting method, as the extended stars are less likely to be included as members during the fitting.

The estimated structural parameters of the simulated UFD analogs are summarized in Table~\ref{tab:one_column_table}. We compare the half-light radii derived by fitting the best-fitting parameters from equation (\ref{eq_profile1}), both with and without background stars. Additionally, we calculated $r_{\rm h, \ dir}$ to assess the extent to which the fitting results deviate from the directly calculated radius. The comparison between $r_{\rm h, fit \ (w/o)}$ and $r_{\rm h, fit \ (w/)}$ clearly shows that excluding background stars in the fit leads to a systematically higher value, on average, three times larger. The most significant difference is observed in {\sc Halo1}, where $r_{\rm h, fit \ (w/)} = 516$ pc is approximately six times larger than $r_{\rm h, fit \ (w/o)} = 87$ pc. Comparing $r_{\rm h, fit \ (w/)}$ and $r_{\rm h, \ dir}$, we find similar values, with any deviations likely arising from the actual distribution not perfectly following an exponential profile. 

\begin{table}
  \centering
  \hspace*{-1.0cm}
  %\resizebox{\textwitdth}{!}{
  \makebox[0.5\textwidth][c]{
  \begin{tabular}{c c c c c}
    \hline%
    UFD analog & $r_{\rm h} [\rm pc]$ & $r_{s} [\rm pc] $ & $r_{\rm trans} [\rm pc]$ &  $B$  \\
    \hline%
    {\sc Halo1} & $84\pm9$ & $964\pm76$ & 391 & 0.001   \\[0.1cm]
    {\sc Halo2} & $117\pm4$ & $1302\pm38$ & 467 & 0.008  \\[0.1cm]
    {\sc Halo3} & $109\pm8$ & $587\pm26$ & 283 & 0.028  \\[0.1cm]
    {\sc Halo4} & $117\pm3$ & $850\pm19$ & 382 & 0.009  \\[0.1cm]
    {\sc Halo5} & $145\pm13$ & $1676\pm38$ & 309 & 0.042 \\[0.1cm]
    \hline
  \end{tabular}
  }
  \caption{The estimated best-fit parameters for a two-component density profile. The columns provide: (1) $r_{\rm h}$: the half-stellar mass radius, calculated as $r_{\rm h}=1.68 r_{\rm e}$, representing the inner density profile; (2) $r_{\rm s}$: the scale radius for the extended structure; and (3) $r_{\rm trans}$: the transition radius at which the outer density profile becomes dominant over the inner profile.}\label{tab:one_column_table2}
\end{table}

\begin{figure}
  \centering 
  \includegraphics[width=85mm]{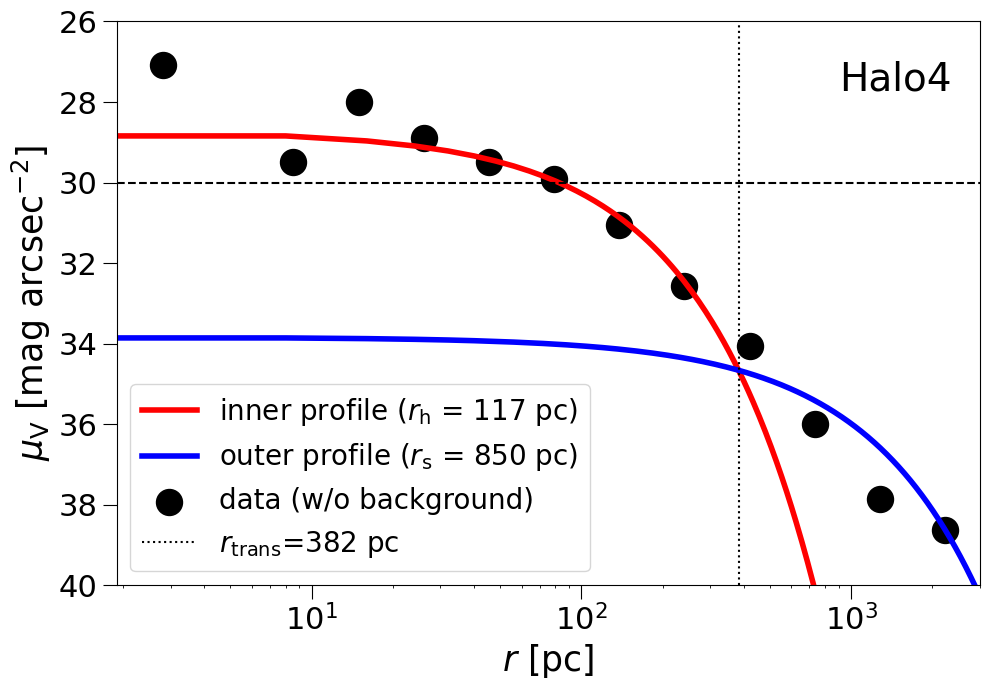}
  \caption{The representative surface brightness profile of {\sc Halo4} is modeled using a two-component approach. The actual density profile is shown as black-filled circles at binned radii. The inner component is depicted in red, while the outer component, characterized by a scale radius of $r_{\rm s}=850$ pc, is illustrated in blue. The transition between the inner and outer components occurs at a radius of $r_{\rm trans}=382$ pc.}
  \label{fig:two-comp}
\end{figure}

\begin{figure*}
  \centering 
  \includegraphics[width=170mm]{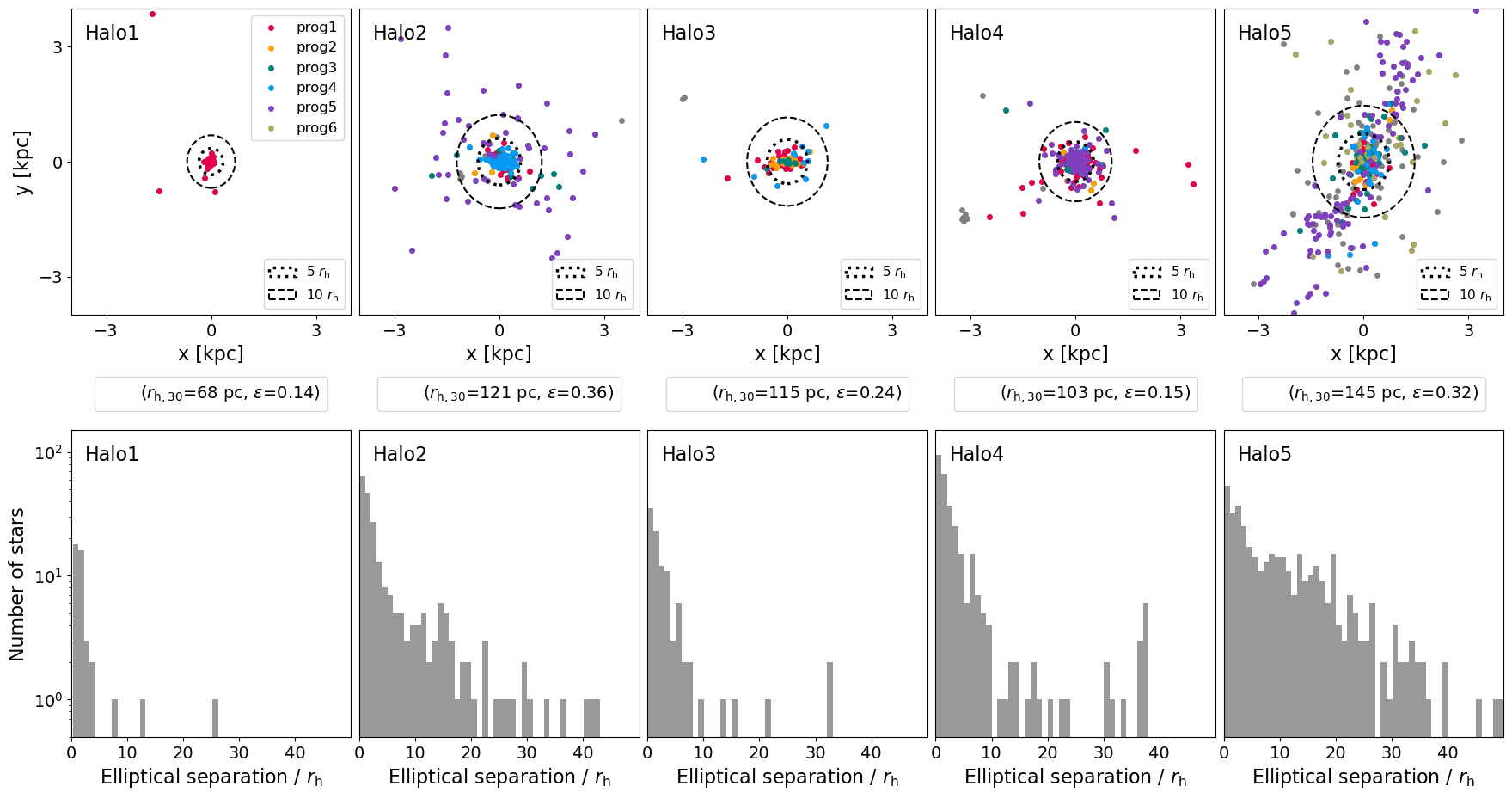}
  \caption{Top panels: Projections of stellar distributions for UFD analogs at $z=0$. Star particles from different progenitors are color-coded. Circles in the central region represent 5 and 10 times the half-light radii, $r_{\rm h}$. This demonstrates that the extended stars likely originate from the same progenitor halo. For example, stars formed in progenitors {\sc prog5} of {\sc Halo2}, {\sc prog1} of {\sc Halo4}, and {\sc prog5} of {\sc Halo5} predominantly contribute to the outer regions of these halos. Bottom panels: Distribution of stars as a function of elliptical radius from the center of each galaxy, normalized by $r_{\rm h}$. The extended stellar structures relative to the half-light radius suggest that stars situated at the outskirts could potentially be overlooked in observational studies.}
  \label{fig:MLE_Size}
\end{figure*}

In the top panels of Fig.~\ref{fig:desity_profile}, we illustrate the spatial distribution of our UFD analogs, with star particles from different progenitor halos represented in distinct colors overlaid on the surface brightness maps. The background stars used for fitting are excluded from this plot. Among the simulated UFD analogs, {\sc Halo1} exhibits the most compact features, with a half-light radius of $r_{\rm h, \ fit} \approx 87$ pc and an ellipticity of $\epsilon_{\rm fit} = 0.34$, indicating a spherical distribution of stellar particles. Conversely, stars in {\sc Halo5} are the most extended, with $r_{\rm h, \ fit} \approx 754$ pc, and tend to deviate from a spherical shape, displaying a more elongated distribution with an ellipticity of $\epsilon_{\rm fit} = 0.57$. Note that {\sc Halo6}, which has a halo mass similar to that of {\sc Halo5}, is excluded from the size relation analysis to avoid redundancy.

The bottom panels of Fig.~\ref{fig:desity_profile} display the surface brightness profile as a function of the elliptical radius from the center of the simulated UFD analogs. The red circles (including background stars) and grey circles (excluding background stars) represent the surface brightness directly computed from stars in each UFD analog. The solid red line depicts the exponential fit derived using maximum likelihood estimation, including background stars, while the grey line represents the fit that excludes background stars. We verify that accurately estimating the spatial structures of simulated UFD analogs using observational methodologies necessitates the inclusion of background stars. Due to the presence of extended stars at large radii (indicated by grey circles), the fit without background stars (grey line) fails to capture the extended structure. This suggests that a different density profile, one that simultaneously accounts for both the inner, compact structure and the outer, extended structure, would be more appropriate (e.g., \citealp{Kang2019, Ricotti2022, Jensen2024}).

Recently, \citet{Jensen2024} discovered an outer component in the stellar distribution of MW dwarf galaxies by utilizing an updated algorithm to find member stars of dwarf galaxies using {\it Gaia} data. To identify member stars in the outer regions, they assumed that the stellar structure of each dwarf galaxy comprises two spatial components: an inner component and an extended component. This method led to the identification of nine dwarf galaxies that exhibit a low-density outer profile. Motivated by this, we employ the two-component profile suggested by \citet{Jensen2024} as follows,
\begin{equation}
\label{eq_profile2}
\Sigma(r) \propto e^{-r/r_{\rm e}} + B e^{-r/r_{\rm s}},
\end{equation}
where the $B$ is the normalization and $r_{\rm s}$ indicates the scale radius for the outer exponential component. We find three unknown parameters ($r_{\rm h} \ = \ 1.68   r_{\rm e}$, $r_{\rm s}$, and $B$) by fitting the profile in equation (\ref{eq_profile2}) using the maximum likelihood estimation method, and the resulting estimates are summarized in Table~\ref{tab:one_column_table2}. It is noteworthy that, unlike the previous fitting method, which involves constructing a two-dimensional map by co-adding multiple realizations of an exponential profile to approximate the dwarf galaxy's spatial distribution, this method determines the best parameters to fit the two-component density profile for the three unknowns using star particles within the virial radius of the simulated galaxies, without including background stars.

Fig.~\ref{fig:two-comp} depicts the representative density profile of {\sc Halo4}, which consists of the inner (red line) and outer (blue line) components. For other halos, we represent the fitting results in Appendix.~\ref{appendix_c}. The actual density profile is represented by black-filled circles at binned radii. In comparison to the $r_{\rm h,\ fit}$ value derived from a single exponential component with background stars, the $r_{\rm h}$ value is approximately 35\% smaller while also effectively capturing the outer structures with $r_{\rm s} = 850$ pc. Notably, the pronounced outer stellar distributions observed in {\sc Halo2} and {\sc Halo5} are accurately represented by the large $r_{\rm s}$ values of $r_{\rm s} = \ 1302 \ \rm pc$ and $r_{\rm s} = 1676 \ \rm pc$, respectively, along with smaller half-light radii of $r_{\rm h} = 117 \ \rm pc$ and $r_{\rm h} = 145 \ \rm  pc$. Specifically, the smaller half-light radius in {\sc Halo5} is reduced by a factor of 5 compared to $r_{\rm h, \ fit} = 754 \ \rm pc$, which is based on the one-component profile with background stars. 

We find that when performing a single-component fitting that includes background stars, the derived half-light radius can vary depending on the background stellar density adopted. For instance, if the background stellar density increases, more extended stars become less distinguishable from the background stars, resulting in a decrease in the estimated half-light radius. On the other hand, the two-component profile model captures both the inner and outer structures without being influenced by the background stellar density. Therefore, in the forthcoming analysis, we utilize the half-light radius of the inner profile as the representative size of our UFD analogs.

\begin{figure*}
  \centering 
  \includegraphics[width=150mm]{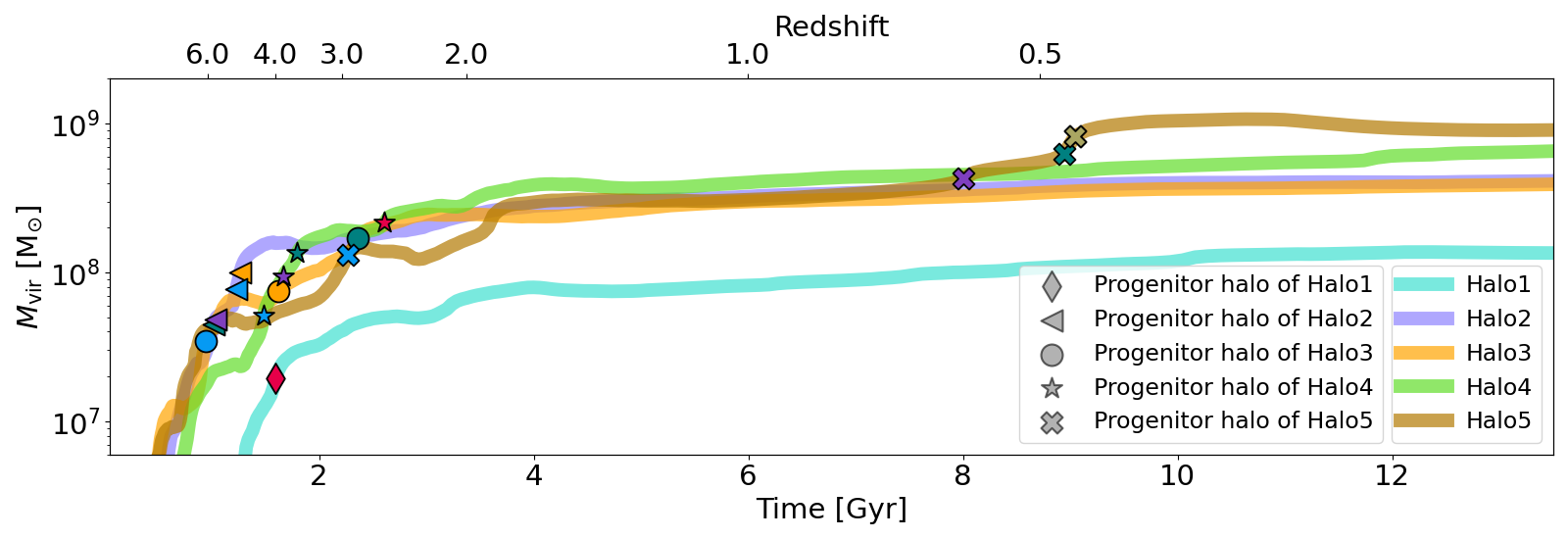}
  \caption{ 
  Merger histories along the virial mass growth of the UFD analogs, illustrating the merger times when progenitor halos accreted onto the traced progenitor. Each progenitor halo within the same UFD analog is represented by the same symbol, with different colors used to distinguish between the progenitor halos. Most mergers occur at early times ($z \gtrsim 3$), but {\sc Halo5} exhibits unique late mergers at $z \lesssim 0.7$, contributing to the formation of the most prominent extended structures in {\sc Halo5}. Similarly, a relatively later merger at $z \sim 2.7$ with {\sc prog1} of {\sc Halo4} leads to a more extensive distribution of stars in {\sc Halo4}.}
  \label{fig:merge}
\end{figure*}

\subsubsection{Extended structure in UFDs} \label{sec:extended_star}

The existence of stars in the outskirts of UFDs has also been confirmed by recent observational data (e.g., \citealp{Chiti2021, Jensen2024, Tau2024}). Then, what physical mechanisms cause small systems like UFDs to have extended structures? Three potential pathways have been proposed for the origin of these extended stars in UFDs: tidal stripping, mergers, and energetic SN feedback. While the predominant scenario is still under debate, some cases do not seem to be explained by tidal stripping or SN feedback. For example, Tucana II is a notable case where an extended star located at $\sim$ 9 $r_{\rm h}$ ($\sim$1 kpc) was discovered by \citet{Chiti2021}, showing no signs of tidal stripping. Furthermore, a recent study by \citet{Chiti2023} argued that it is unlikely that energetic SNe push stars into extended regions, as such extended structures are not frequently observed in other UFDs.

Given that our simulated UFD analogs are chosen as isolated galaxies, we can dismiss tidal stripping as the cause of the extended structures. Instead, our simulations reveal significant details about the origins of these extended stars (see the top panel of Fig.~\ref{fig:MLE_Size}). Firstly, we find that a few stars, shown in silver in the top panel of Fig.~\ref{fig:MLE_Size}, formed in small progenitor halos that contribute less than 5\% of the total stellar content and are located in the outskirts of the simulated UFDs. This trend is particularly pronounced in {\sc Halo5}, where the majority of outer stars are composed of stars from the progenitor {\sc prog5} (denoted as purple) and other miscellaneous halos (silver). Interestingly, these stars have different morphologies; for example, stars from {\sc prog5} are distributed in a highly elongated manner ($\epsilon_{\rm prog5} = 0.74$).
Moreover, as illustrated in Fig.~\ref{fig:morph_all.png}, we observe that these miscellaneous halos were distinctly located even at high redshift ($z \gtrsim 6$), later assembled into other progenitor halos, but still remain in the outskirts of the simulated UFDs.

Secondly, the extended stars likely originate from the same progenitor halo. For instance, stars formed in progenitors {\sc prog5} of {\sc Halo2}, {\sc prog1} of {\sc Halo4}, and {\sc prog5} of {\sc Halo5} predominantly contribute to the outer regions of these halos. To investigate the role of mergers between progenitor halos in forming extended structures, we examine the merger times of progenitor halos along the virial mass growth of the UFD analogs, as shown in Fig.~\ref{fig:merge}. In this figure, progenitor halos belonging to the same UFD analog are represented by the same symbol. This merger history is derived by tracking DM particles backward from $z=0$ to $z=16$, identifying the progenitor with the most shared DM as the ancestor. Note that this "ancestor progenitor" is not always the same as the "main progenitor" used in this study.

The merger history reveals that most mergers between progenitor halos occur at an early epoch ($z \gtrsim 3$). Meanwhile, {\sc Halo5} exhibits a distinct merger history, with some progenitors, including {\sc prog5}, which contributes the most to the outer population of {\sc Halo5}, assembling later at $z \lesssim 0.7$, leading to its extended stellar distribution. Similarly, {\sc Halo4} undergoes a later merger at $z \sim 2.7$ with {\sc prog1} (red), resulting in the broader distribution of stars. On the contrary, {\sc Halo1}, where all stars formed in a single progenitor {\sc prog1}, exhibits the highest compactness. Our finding on the role of mergers in the formation of extended stars is supported by other simulation studies (e.g., \citealp{Rey2019, Tarumi2021, Deason2022, Revaz2023, Goater2024}). In particular, \citet{Goater2024} suggested that stars accreted from late mergers are mainly responsible for the extended structures of simulated galaxies. However, our results do not emphasize the importance of late mergers in creating extended structures, as {\sc Halo2}, which experiences early mergers prior to $z \approx 6$, still shows the second largest extended structures among our UFD analogs, with $r_{\rm s}$ = 689 pc.

We find that the progenitor halo {\sc prog5}, which significantly contributes to the extended structures of {\sc Halo2}, initially merges with other progenitor halos at $z=7.21$. Unlike other progenitors that settle smoothly into the center of the potential well formed by the merged halos, {\sc prog5} of {\sc Halo2} follows a uniquely distinct orbit. After its first infall, some stars of {\sc prog5} move back towards the virial radius and then re-infall into the central region at $z=5.88$, $z=3.43$, and $z=2.24$, with all stars eventually settling within the virial radius by $z \lesssim 1$. This behavior suggests that, unlike other progenitors, {\sc prog5} of {\sc Halo2} is less gravitationally bound and traverses through the center of the emerging galaxy multiple times, thereby contributing to the formation of extended structures.

\begin{figure}
  \centering 
  \includegraphics[width=85mm]{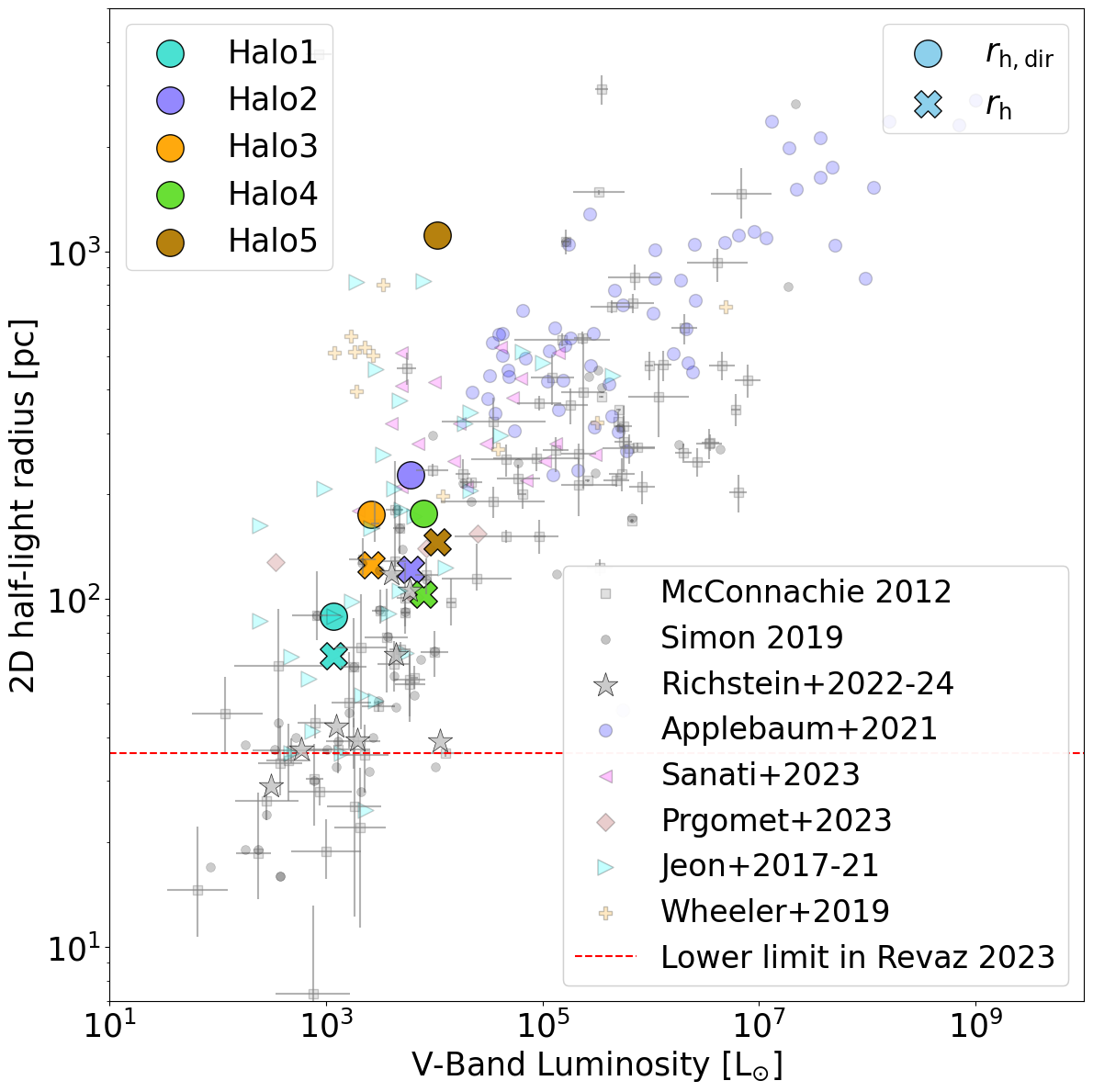}
  \caption{V-Band luminosity versus 2D projected half-light radius for simulated and observed galaxies. We compare our results with updated catalogs of observed UFD and dwarf galaxies from \citet{McConaachie2012}, \citet{Simon2019}, and \citet{Richstein2022, Richstein2024}. Simulated galaxies from various studies (\citealp{Jeon2017,Jeon2021a, Jeon2021b, Wheeler2019, Applebaum2021, Prgomet2022, Sanati2023}) are represented by different colors. We derive the two half-light radii for each UFD analog using the direct method ($r_{\rm h, \ dir}$, denoted as circles) and two-component fitting ($r_{\rm h}$, denoted as crosses). A uniform mass-to-light ratio of 2 is applied to our data and other simulated galaxies. Half-light radius using two-component fitting results in closer values to those of observations.}
  \label{fig:size_luminosity}
\end{figure}

\subsubsection{Size-luminosity relation}\label{sec:Size-luminosity relation}
 
We compare the half-light radii of our simulated UFD analogs with both observational data and other theoretical studies in Fig.~\ref{fig:size_luminosity}. The observational results include an updated catalog of dwarf galaxies from \citet{McConaachie2012} and \citet{Simon2019}, as well as measurements of UFDs reported by \citet{Richstein2022, Richstein2024}. While \citet{Richstein2022, Richstein2024} use azimuthally averaged half-light radii, \textcolor{red}{we adopt the elliptical half-light radius, denoted as $a_{\rm h}$ in their work}, to maintain consistency with other works using this definition. The simulated galaxies are represented in different colors (\citealp{Jeon2017, Jeon2021a, Wheeler2019, Applebaum2021, Prgomet2022, Sanati2023}). For the V-band luminosity, we uniformly apply a mass-to-light ratio of 2 to our results and the other simulated galaxies.

As illustrated in Fig.~\ref{fig:size_luminosity}, we estimate the half-light radius of each UFD analog using two different methods: the direct method ($r_{\rm h, \ dir}$, shown as circles) and the two-component fitting method ($r_{\rm h}$, shown as crosses). While the $r_{\rm h}$ values are consistently smaller than the $r_{\rm h, \ dir}$ values for all UFD analogs, the former closely match the observational data. This consistency suggests that the large half-light radii reported in other simulations (e.g., \citealp{Wheeler2019, Sanati2023}) could be reduced by applying the two-component fitting method if extended structures are present. 

Although our derived values using the two-component approach successfully match the observational data at the upper range, the average values remain larger ($r_h \sim$ a few $100\ \rm pc$) compared to the typical observed values ($r_h \lesssim 100\ \rm pc$). These larger sizes in the UFD regime predicted by theoretical work imply that observed UFDs may possess more extended yet unresolved structures. This possibility is further supported by recent observational studies revealing the presence of extended structures in UFDs (e.g., \citealp{Tau2024, Jensen2024}).

In particular, our simulations struggle to reproduce the highly compact sizes with $r_{\rm h} \lesssim 50$ pc. This discrepancy is particularly notable for UFDs with $M_{\star} \sim 10^3 \Msun$, where observational studies consistently suggest sizes around $r_{\rm h} \sim 30$ pc. To explain the absence of compact sizes in the simulations, numerical heating has been suggested as a possible cause. This effect, which arises from the difference in mass resolution between dark matter and star particles, could contribute to the larger stellar sizes (e.g., \citealp{Revaz2018, Binney2002, Wilkinson2023}). 

Specifically, \citet{Ludlow2019b} showed that the heavier DM particles tend to sink to the center of simulated galaxies, causing the less massive stellar particles to gradually diffuse outward. This process results in a systematic increase in the half-light radius over time until it reaches the convergence radius, $r_{\rm conv} \approx 0.055 \times l(z)$ (\citealp{Ludlow2019a}), where $l(z)$ represents the mean interparticle spacing, defined as $l = L_{\rm box, \ phy} / N_{\rm part}^{1/3}$. Here, $L_{\rm box, \ phy}$ is the physical size of the simulation box, and $N_{\rm part}$ is the number of particles. In our study, the zoom-in region, with a resolution of a $2048^3$ grid, yields a convergence radius of $r_{\rm covn}\approx59$ pc, including baryonic particles. Given that our estimated $r_{\rm h}$ values are larger than this limit, the impact of numerical heating may not be significant in our work. Furthermore, \citet{Wilkinson2023} show that the disk height in a MW-like galaxy can artificially increase due to spurious dynamical heating when the halo is resolved with fewer than $10^6$ DM particles. In this regard, our UFD analogs are unlikely to be affected by such spurious dynamical heating, as all halos except Halo1 are resolved with more than $10^6$ DM particles.

An additional point worth considering is the influence of collisional dynamics. As simulations of galaxy evolution achieve increasingly higher resolution, the resulting decrease in interstellar distances naturally raises the question of whether the collisionless treatment of stars remains valid. To date, collisional effects have primarily been studied in the context of star cluster formation. For example, \citet{Lahen2025} modeled cluster formation in an isolated disc galaxy using collisional dynamics and showed that this approach increases the effective radius of clusters by up to a factor of 10, bringing their sizes into better agreement with observations. However, due to the sub-parsec inter-particle distances involved in such clusters, it remains uncertain how collisional effects might impact stellar distributions in UFDs.

We also mention that \citet{Jeon2021b,Jeon2021a}, where a simulation setup similar to ours ($m_{\rm DM} = 500 \Msun$ and $m_{\rm gas} = 63 \Msun$), demonstrated the existence of simulated UFDs with smaller sizes ($r_{\rm h} < 50$ pc). In these studies, numerical heating caused by differing mass ratios between DM and star particles was not a concern, as each star particle was modeled as a SSP with an initial mass of $m_{\star} \approx 500 \Msun$, formed by accreting gas particles, giving rise to the same mass as a gas particle. However, the total stellar masses in these UFD analogs consisted of fewer than 10 particles for a stellar mass of $M_{\star} \lesssim 2 \times 10^3 \Msun$. This small number of particles representing the total stellar mass may contribute to the lower $r_{\rm h}$ values.

\begin{figure*}
  \centering 
  \includegraphics[width=175mm]{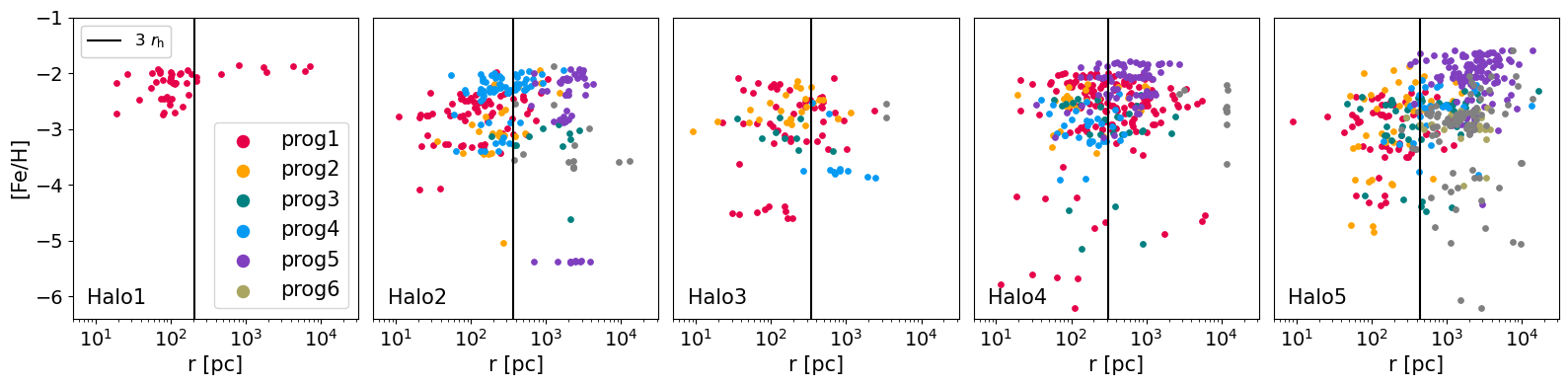}
  \caption{ 
  Stellar metallicity as a function of 2D elliptical radius from the center at $z=0$. Here, the metallicity gradient is not observed in any simulated UFDs; rather, these galaxies exhibit a wide range of metallicity distributions, reflecting the diverse origins of stars from different progenitor halos. Specifically, in {\sc Halo5}, stars originating from {\sc prog5} (indicated in purple) are found dispersed across vast regions extending from 1 to 10 kpc.}
  \label{fig:metallicity_distance}
\end{figure*}

\begin{figure*}
  \centering 
  \includegraphics[width=175mm]{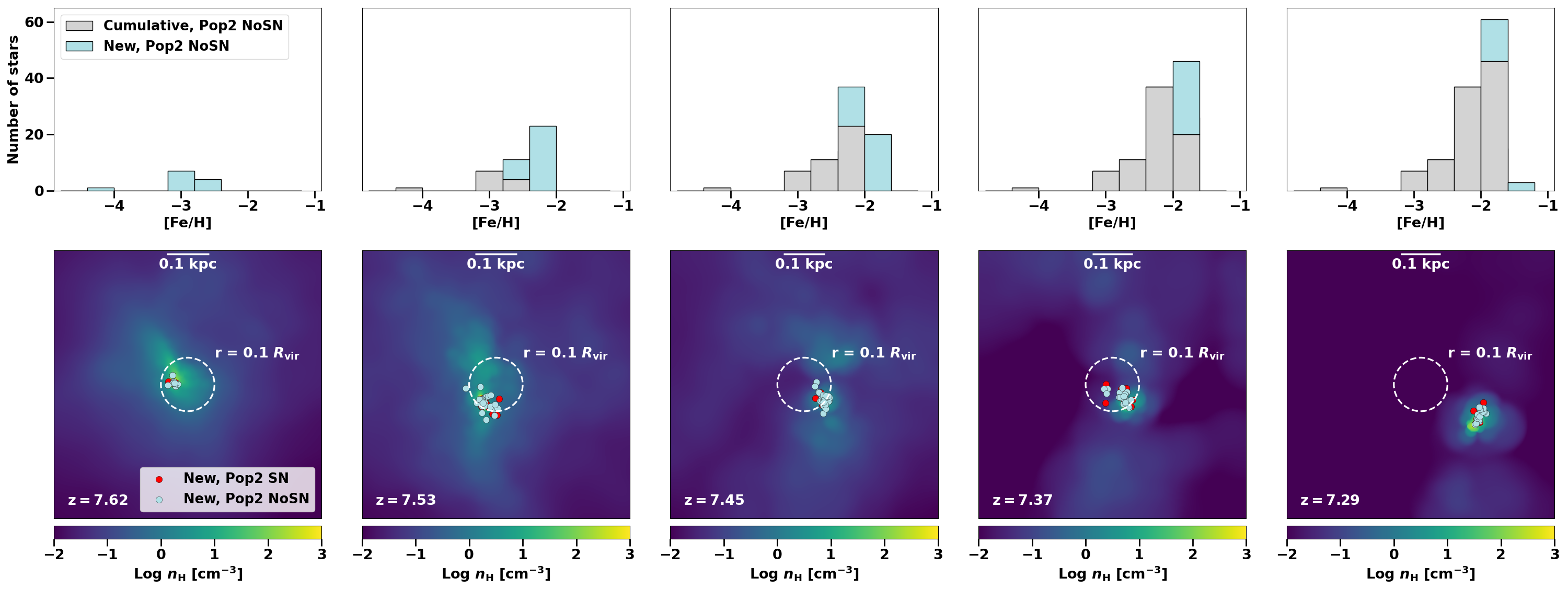}
  \caption{The time sequences from left to right depicting the distribution of stars over cosmic time in {\sc prog5} of {\sc Halo5}. Top Panel: MDFs of stars formed within the progenitor. Newly formed stars are represented in blue, while the cumulative MDF of all stars is shown in grey. Over time, there is a noticeable increase in the metallicity of newly formed stars. Bottom Panel: This shows the hydrogen density projection, revealing the movement of star-forming clouds away from the center due to SN feedback. Consequently, high-metallicity stars are relocated to regions far from the progenitor halo's center.}
  \label{fig:MDF_evolution}
\end{figure*}

\subsubsection{Gradient in stellar metallicity}
\label{sec:gradient}

The discovery of member stars in the outskirts of observed UFDs raises an intriguing question: Is the metallicity gradient, where metal-poor stars are found farther from the center than metal-rich stars, also present in UFDs, similar to what is observed in more massive dwarf galaxies (e.g., \citealp{Harbeck2001, Kirby2011, Fu2024})? \citet{Chiti2021} proposed the presence of a metallicity gradient in Tucana II, noting the detection of distant stars with $\rm [Fe/H]\approx -3.02$ at distances greater than $\gtrsim 9 \ r_{\rm h}$. Furthermore, \citet{Tarumi2021} demonstrated through simulations that a metallicity gradient in a galaxy with $M_{\rm vir}=2.5\times10^8\Msun$ may arise from mergers between dwarf galaxies with different metallicities. In their scenario, the metal-rich, central galaxy remains intact while the metal-poor, infalling galaxy is disrupted, spreading its stars into the outer regions and thereby producing the gradient. Also, induced star formation during and after the merger further enhances the gradient.

On the other hand, not all simulated UFDs in our study show such a metallicity gradient. Fig.~\ref{fig:metallicity_distance} illustrates the stellar metallicities as a function of 2D elliptical radius from the center of each UFD analog at $z=0$. Stars are color-coded according to their progenitor halos to identify their origins. In the case of {\sc Halo1}, for example, there are seven extended stars lying in $r \gtrsim 3 \ r_{\rm h}$, which have higher metallicities than those lying in $r < 3 \ r_{\rm h}$, resulting in no clear metallicity gradient. Furthermore, as discussed in Section~\ref{sec:MDF}, low-metallicity stars  with $\rm [Fe/H] < -4$ shown in {\sc Halo2}$-${\sc Halo5}, are primarily result of external metal enrichment. However, these stars are not predominantly located in the outer regions and make up only a small fraction of the total stellar population; thus, they do not significantly contribute to the overall metallicity gradient.

Notably, {\sc Halo5} exhibits high-metallicity stars at larger radii ($\gtrsim 1$ kpc), most of which originate from a single progenitor halo, {\sc prog5} (purple), as shown in Fig.~\ref{fig:metallicity_distance}. Star formation in the progenitor {\sc prog5} of {\sc Halo5} began relatively late, around $z \lesssim 8$ (see also Fig.~\ref{fig:fe_h_2by3.png}), resulting in stars with metallicities ranging between $-3 \lesssim \rm [Fe/H] \lesssim -1.5$. These stars, spanning a wide metallicity range, are distributed over an extensive spatial range, up to $\sim$10 kpc at $z=0$. This dispersion is a consequence of a late merger at $z \approx 0.7$, leading to the formation of extended structures. We point out that if current observational capabilities are limited to the central regions ($\lesssim 3 \ r_{\rm h}$), as indicated by the vertical line in Fig.~\ref{fig:metallicity_distance}, our simulations suggest that the metallicity properties will be primarily determined by stars formed from $1-3$ progenitor halos located in the inner region at $z=0$. For instance, in the case of {\sc Halo2} within this region, stars from three different progenitors ({\sc prog1, prog2}, and {\sc prog4}) contribute to the metallicity analysis.

\begin{figure*}
  \centering 
  \includegraphics[width=170mm]{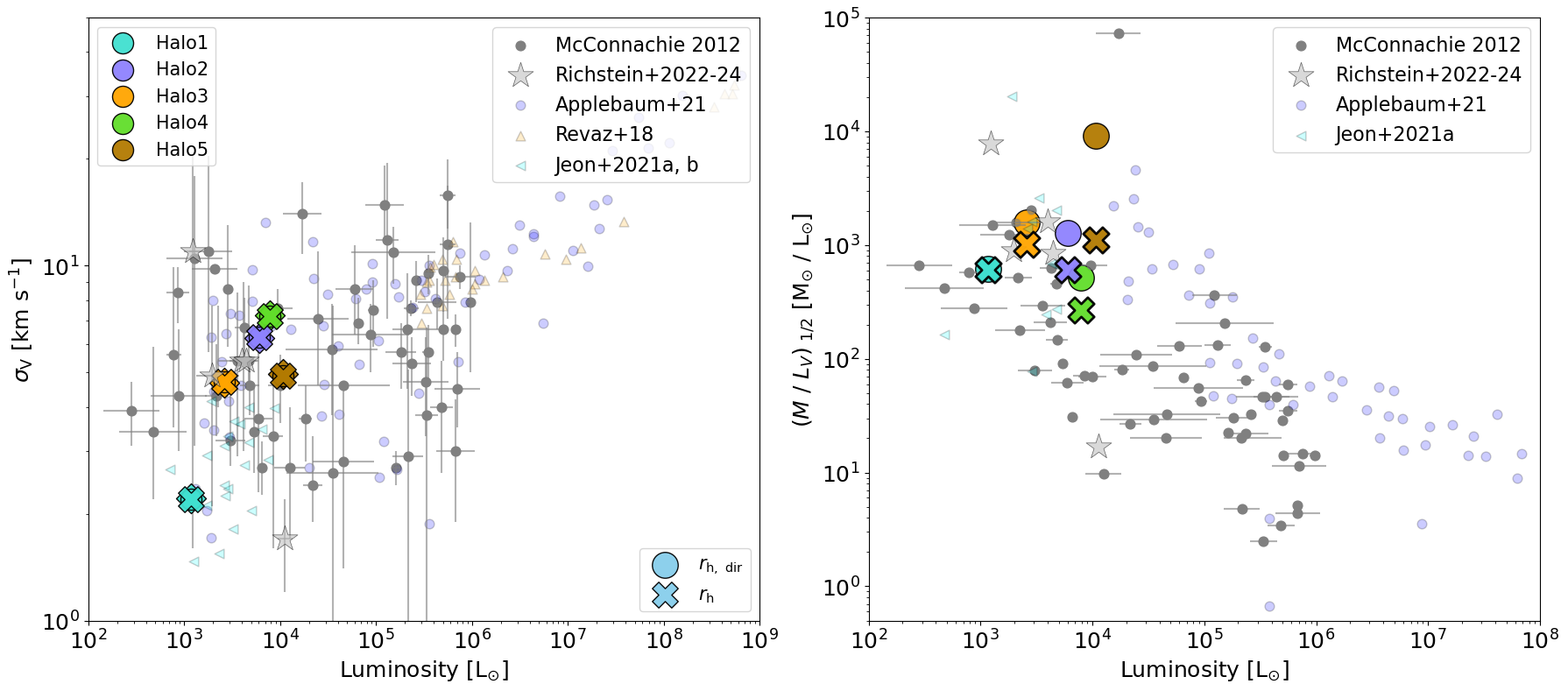}
  \caption{
  Global properties of the simulated UFD analogs at $z=0$, displaying luminosity versus line-of-sight velocity dispersion (left panel), and luminosity versus mass-to-light ratio within the half-light radius (right panel). These properties are derived by adopting two different half-light radii ($r_{\rm h, dir}$, $r_{\rm h}$), with symbols and colors consistent with Fig.~\ref{fig:size_luminosity}. The results include comparisons with observational data (\citealp{McConaachie2012, Kirby2015, Kirby2017, Simon2020, Richstein2022, Richstein2024}) and simulations (\citealp{Jeon2017, Revaz2018, Jeon2021a, Applebaum2021}). Our simulated galaxies tend to be intrinsically cold systems, exhibiting low $\sigma_v = 2-9 \ \rm km \ s^{-1}$ values, and show good agreement with observed UFDs. Also, the mass-to-light ratios align with observed values, confirming that our UFD analogs are dark matter-supported systems.}
  \label{fig:compare2}
\end{figure*}

Furthermore, the lack of a metallicity gradient in this work is in part attributed to the displacement of the star-forming region to the outer areas. Initially, we expected that stars would form in the central region and then relocate to the outer region due to SN feedback from newly formed stars. However, during star formation, it is not the stars that expand outward but rather the high-density gas blobs. For a detailed explanation, Fig.~\ref{fig:MDF_evolution} illustrates the time sequence from left to right, showing the distribution of stars over cosmic time in {\sc prog5} of {\sc Halo5}. The bottom panels display a density projection, with a circle at the central region of the progenitor halo, having a radius of $0.1R_{\rm vir}$. Note that each projection displays only the stars that formed after the previous snapshot, highlighting regions where star formation was actively occurring. The top panel shows the cumulative MDF denoted as grey, whereas newly formed stars are marked in blue.

The generation of the metallicity distribution as a function of distance in our simulated UFDs involves several steps: (1) Stars form from the central star-forming gas cloud, with the MDF depicted in the top panel of Fig.~\ref{fig:MDF_evolution}. (2) SNe feedback from these stars clears the central region of the progenitor halo, causing the shock front to move outward beyond $0.1R_{\rm vir}$. (3) Although gas within $0.1R_{\rm vir}$ is entirely ejected, stars continue to form from the gas compressed by the shock front, which has been enriched by previous SNe explosions, resulting in relatively high-metallicity stars ($\rm [Fe/H] \geq -2$). This process leads to the formation of relatively high-metallicity stars far from the center. In addition, due to the shallow gravitational potential well of the simulated UFDs, the central star-forming region is highly susceptible to stellar feedback, which makes sustained star formation in this region difficult. Consequently, this leads to a lack of a metallicity gradient.
Observational efforts to investigate the existence of a metallicity gradient over large distances are an ongoing area of research using high-resolution spectroscopy (e.g., \citealp{Chiti2021, Sestito2023, Waller2023}). Recently, \citet{Sestito2023} discovered five new member stars on the outskirts of Ursa Minor (one star at $\sim12 \  r_{\rm h}$ and four stars at $\sim5.2-6.3 \ r_{\rm h}$) using an updated identification technique proposed by \citet{Jensen2024}. They reported that the metallicity of the most distant stars at $\sim12 \ r_{\rm h}$ is $\rm [Fe/H] \approx -2.09$, which is higher than the average value of $\rm [Fe/H] \approx -2.65$ for the four stars at $\sim5.2-6.3 \ r_{\rm h}$, although still lower than metallicity of stars in the inner region ($\lesssim3 \ r_{\rm h}$). Furthermore, \citet{Waller2023} reported the metallicity of newly discovered stars in three UFD galaxies: Coma Berenices, Bootes I, and Ursa Major I. Among them, they introduced two new stars located at $2.5 \ r_{\rm h}$ of Coma Berenices, which exhibit the same metallicity range as the previously studied inner stars, with $-3 < \rm [Fe/H] < -2$.

\subsubsection{Velocity dispersion and mass-to-light ratio} \label{sec:vel_disp}

In Fig.~\ref{fig:compare2}, we present a comparison of the velocity dispersion (left panel) and mass-to-light ratio (right panel) between our simulated galaxies and observed galaxies (e.g., \citealp{Kirby2015, Kirby2017, Simon2020, McConaachie2012, Richstein2022, Richstein2024}). For our UFD analogs, we compute the line-of-sight velocity dispersion, $\sigma_{v}$, in the z-direction and determine the dynamical mass within the 2D half-light radius. The (dynamic) mass-to-light ratio calculations are performed using both $r_{\rm h, \ dir}$ and $r_{\rm h}$, consistent with Fig.~\ref{fig:size_luminosity}. For the observed UFDs, the dynamical mass within the half-light radius is determined based on the velocity dispersion and half-light radius as suggested by \citet{Wolf2010}, with the relation given by
\begin{equation}
\label{eq_wolf}
M_{\rm 1/2} \approx 930 \bigg(\frac{\sigma_{v}^{2}}{\rm km^{2} \ s^{-2}}\bigg) \bigg(\frac{r_{\rm h}}{\rm pc}\bigg) \Msun.
\end{equation}

Our results indicate that the velocity dispersions of our UFD analogs are in good agreement with observational data, displaying values around $\sigma_{v} \sim \rm 2-8 \ km \ s^{-1}$. These values are the same for both $r_{\rm h, \ dir}$ and $r_{\rm h}$, since $\sigma_{v}$ is calculated for all stars within the virial radius. Given that our simulated UFD analogs represent isolated field galaxies, these low-mass systems are inherently dynamically cold, not requiring mass loss through tidal stripping interactions with host galaxies to achieve such low $\sigma_v$ values (e.g., \citealp{Penarrubia2008, Buck2019}). Specifically, {\sc Halo1} demonstrates the lowest $\sigma_v\sim \rm 2 \ km \ s^{-1}$ values, with the majority of stars ($M_{\star}\approx2\times10^3\msun$) predominantly formed in a single progenitor halo. Compared to \citet{Jeon2021a}, stars from {\sc Halo2} to {\sc Halo5} tend to show higher dispersion values ($\sigma_v\sim 4-9 \rm \ km \ s^{-1}$) within a similar luminosity range. As mentioned in Section~\ref{sec:Size-luminosity relation}, the stellar particles in \citet{Jeon2021a} have approximately eight times the mass of those in this work, which leads to these quantities being determined by a smaller number of stellar particles. 

Meanwhile, the derived mass-to-light ratios vary based on the selected half-light radius, unlike the consistent velocity dispersions. Specifically, the mass-to-light ratio is higher when the direct method is employed. This increase is due to the rise in dynamical mass associated with a larger half-light radius. For example, in the case of {\sc Halo5}, there is nearly one order of magnitude increase in the mass-to-light ratio when using the direct method compared to the values obtained from fitting a two-component stellar density profile. Consequently, except for the case of {\sc Halo5}, in which $r_{\rm h, \ dir}$ is significantly increased due to the presence of extended stars, our estimated mass-to-light ratios ($200 < M/L_{V} < 2\times 10^3$) are consistent with observed values, supporting the idea that UFD galaxies are primarily dark matter-dominated systems.

\section{DISCUSSION AND CONCLUSION}

In this study, we examine the physical properties, including stellar metallicity and galaxy size (i.e., half-light radius), of Ultra-Faint Dwarfs (UFDs), the faintest galaxy systems in the Universe ($L_{V} \gtrsim -7.7$) (e.g., \citealp{Simon2019}). Despite achieving high mass resolution ($m_{\rm gas} \lesssim 100 \msun$), recent theoretical work predicts lower average stellar metallicity ($\rm -4 \lesssim \langle[Fe/H]\rangle \lesssim -3$) compared to observed values in UFDs ($\rm -3 \lesssim \langle[Fe/H]\rangle \lesssim -2$) and larger half-light radii ($r_{\rm h} > 100$ pc) than those observed. To address this discrepancy in stellar metallicity and galaxy size between observations and theoretical predictions, we conduct a series of high-resolution hydrodynamic simulations ($m_{\rm gas}\approx60\msun, m_{\rm DM}\approx300\msun$) on UFD analogs with virial masses $M_{\rm vir} \approx 2\times10^8 \msun - 1.3\times 10^9 \msun$ and stellar masses ranging from $M_{\star} \sim 10^3 \msun$ to a few $10^4 \msun$. In addition, we adopt an individual sampling method to extract massive stars for supernovae (SNe) explosions from a given initial mass function (IMF), allowing us to address the discrete nature of SNe.

To understand the origin of the metallicity in the stellar components of UFD analogs, we focus on their star-forming environments at high redshift ($z \gtrsim 6$). Unlike more massive classical dwarf galaxies, the shallow potential wells of our UFD analogs lead to the quenching of star formation before $z \approx 6$, when their progenitor halos are still physically separated and have not yet merged. As these progenitor halos form stars in their unique environments, we investigate them in detail to understand how the stellar properties of UFD analogs are determined. For our UFD analogs from {\sc Halo1} to {\sc Halo6}, at $z \approx 6$, the number of progenitor halos contributing more than 5\% of the total stellar mass of the final UFD analogs are 1, 5, 4, 5, 6, and 3, respectively. Furthermore, to investigate the resultant structural properties of the simulated UFD analogs at $z=0$, we study the role of mergers between progenitor halos and examine the observational effects of the stellar components when deriving structural properties.

The main findings of this study are as follows.

\par
%\begin{enumerate}
\begin{itemize}
 
\item[$\circ$] \text{Stellar metalliticy findings}
\begin{itemize}  

    \item[$\bullet$] 
    In our UFD analogs, the shallow potential wells result in star formation occurring in episodic bursts rather than continuously. Notably, the most massive progenitor at $z \sim 6$ contributes less than 50\% of the total stellar mass. This suggests that caution is needed when analyzing the properties of observed UFDs, as it is often assumed that most observed stars formed in a single predominant progenitor. 
    
    \item[$\bullet$]    
    Our UFD analogs at $z=0$ exhibit a linearly correlated stellar mass-metallicity relation (MZR), with the lowest average metallicity being $\rm \langle[Fe/H]\rangle \approx -2.98$. This value is generally higher than those reported in other simulation studies, where the lowest values are $\rm \langle[Fe/H]\rangle \lesssim -3.5$. We attribute our higher $\rm \langle[Fe/H]\rangle$ values to the individual IMF sampling method, which allows stars to form immediately from gas enriched by previous SN explosions. Nonetheless, our results still show average metallicity values that are 0.5-1 dex lower than those observed.

    \item[$\bullet$] 
    The metallicity distribution functions (MDFs) of our UFD analogs show a Gaussian-like distribution, peaking at $\rm -3 \lesssim [Fe/H] \lesssim -2$ with dispersion values ranging from $0.25 \lesssim \sigma_{\rm [Fe/H]} \lesssim 0.76$. Low-metallicity stars ($\rm [Fe/H] < -4$) mainly result from external enrichment, where Pop II stars form first in gas contaminated by nearby halos, without prior Pop III star formation. Meanwhile, stars formed through internal enrichment achieve higher metallicity ($\rm [Fe/H] \gtrsim -4$) with just one or two SN explosions.    
    \item[$\bullet$]   
    To compare our MDFs with observations, we derive $\rm \langle[Fe/H]\rangle$ and $\sigma_{\rm [Fe/H]}$ values, excluding stars with $\rm [Fe/H] < -4$ to align with observational limits. When these low-metallicity stars are excluded, the $\rm \langle[Fe/H]\rangle$ values for all halos, except for {\sc Halo1}, shift to higher values, while the $\sigma_{\rm [Fe/H]}$ values shift to lower values. These adjusted values show an excellent match with observational data.
    
    \item[$\bullet$] 
    When comparing the composite MDFs of observed UFDs, our simulations exhibit similar distributions but reveal a significant paucity of stars with $\rm [Fe/H] \geq -2.0$. This indicates that achieving such high metallicities within progenitor halo environments at $z \gtrsim 6$ is highly challenging. Although metallicity increases monotonically during starbursts, the process is constrained as SN feedback accumulates over time, progressively suppressing star formation by expelling gas from the central region ($r < 0.3 \ R_{\rm vir}$). Consequently, the formation of stars with $\rm [Fe/H] \geq -2.0$ becomes particularly difficult, though not entirely precluded.
    
    \item[$\bullet$]   
    To understand the physical properties that determine the maximum metallicity in each starburst, we investigate the correlation between maximum metallicity and various physical properties. We find that the maximum hydrogen density ($n_{\rm H, max}$) and the number of SNe ($N_{\rm SN}$) during a starburst are strongly correlated with the maximum metallicity. This suggests that a highly dense region is essential for triggering numerous SN events, which provide a substantial supply of metals to the remaining dense regions. However, as the number of SNe increases, the cumulative feedback from these events begins to inhibit further star formation in the progenitor halo, making it challenging to form stars with $\rm [Fe/H] \geq -1.5$.
    
    \item[$\bullet$]   
    We investigate how varying SN energy affects $\rm \langle[Fe/H]\rangle$ and MDFs through additional simulations. We vary the energy levels for each SN event (0.5 $E_{\rm SN}$, 1.0 $E_{\rm SN}$, and 2.0 $E_{\rm SN}$ for $E_{\rm SN} \ = \ 1.0 \times 10^{51} \rm ergs$). The weakest SN energy ($0.5 \ E_{\rm SN}$) results in higher stellar mass and average metallicity compared to the fiducial set ($1.0 \ E_{\rm SN}$), producing stars with an exceptionally high metallicity of $\rm [Fe/H] \approx -1.2$. However, this scenario overproduces stars with $\rm [Fe/H] \geq -2.0$ and underproduces stars with $\rm [Fe/H] < -2.8$ compared to the composite MDFs of observed UFDs. Additionally, the increase in stellar mass by a factor of 1.7 when applying a $0.5 \ E_{\rm SN}$ makes it difficult to reconcile the formation of high-metallicity stars with the low stellar masses observed in UFDs.
    
\end{itemize}

\item[$\circ$] \text{Galaxy size findings}
\begin{itemize}

    \item[$\bullet$]  
    For size measurements of UFD analogs, to ensure consistency between simulations and observations, we use the most commonly adopted observational approach, which fits a two-dimensional single exponential surface density profile to derive spatial structural parameters. We also compute the half-light radius using the direct method commonly used in simulations, where the radius is defined as the distance enclosing half of the total stellar mass based on star particles.
    
    When using the fitting method, it is important to account for artificial background stars because the presence of extended star particles in UFD analogs can distort the surface density profile. This distortion may lead to an overestimation of the half-light radius if background stars are not properly considered. In fact, for UFD analogs, the half-light radius derived without background stars ($r_{\rm fit,\ (w/o)}$) is, on average, three times larger than that obtained with background stars ($r_{\rm fit,\ (w/)}$). Furthermore, $r_{\rm fit,\ (w/)}$ shows similar values compared to the those from the direct method ($r_{\rm dir}$). This highlights the necessity of accounting for artificial background stars to recover the true stellar distribution.

    \item[$\bullet$]
    We find extended structures of varying degrees in all halos of our UFD analogs. Single exponential component fitting methods with background stars effectively describe the inner profile but still struggle with extended outer structures. To address this issue, we adopt the recently proposed two-component stellar density profile (\citealp{Jensen2024}), which successfully captures both the inner and outer components in our simulated UFDs, allowing for more accurate estimation of structural parameters.

    \item[$\bullet$]
    We find extended structures of varying degrees in all halos of our UFD analogs. Single exponential component fitting methods with background stars effectively describe the inner profile but still struggle with extended outer structures. To address this issue, we adopt the recently proposed two-component stellar density profile (\citealp{Jensen2024}), which successfully captures both the inner and outer components in our simulated UFDs, allowing for more accurate estimation of structural parameters.
    
    \item[$\bullet$]
    The physical mechanism responsible for the extended structures in our UFD analogs is the result of dry mergers between progenitor halos. Specifically, {\sc Halo4} and {\sc Halo5}, which exhibit the most prominent extended structures, experience late dry mergers with particular progenitor halos, {\sc prog2} in {\sc Halo4} and {\sc prog5} in {\sc Halo5}. This indicates that mergers between progenitor halos could play a significant role in producing extended structures in UFD analogs without tidal effects from their host halo.
    
    \item[$\bullet$]
    Our simulated UFDs do not exhibit a significant metallicity gradient. Unlike the typical scenario where a metallicity gradient develops as stars form in the central region and are then relocated to the outer regions due to SN feedback from newly formed stars, we observe that it is the high-density gas blobs that migrate outward, not the stars. Consequently, high-metallicity stars can also form from the displaced gas blobs, leading to the absence of a metallicity gradient.
    
    \item[$\bullet$]  
    In the size-luminosity relation, our findings closely match observational data when the two-component stellar density profile is applied. However, when using the half-light radius derived from the direct method, the values tend to be larger, especially for {\sc Halo5}, where they exceed 1 kpc due to the presence of extended stars. Despite our simulated galaxies having small sizes with $r_{\rm h}$, our UFD analogs still struggle to reproduce the highly compact sizes with $r_{\rm h} \lesssim 50$ pc observed in some UFDs.
    
%\end{enumerate}
\end{itemize}
\end{itemize}

%High metallicity regime에서 관측과의 gap을 설명하고자, patchy을 사용할 경우 extended star formation이 higher metallicity star를 만들 수 있음을 말하고자 함.
In our simulations, all UFD analogs experience complete quenching of star formation at $z \approx 6$ due to the uniformly applied UV background that activates from $z = 7$. However, for observed UFDs, the effect of the UV background at high redshift can vary depending on the distance between the UFD and its host galaxy, reflecting the patchy reionization effect (e.g., \citealp{Sacchi2021, Kim2023}). For instance, if a galaxy was distant from its host during the epoch of reionization, the reduced intensity of the UV background could allow for extended star formation after $z \approx 6$, possibly leading to the formation of stars with higher metallicity compared to our UFD analogs.

We emphasize that we do not account for delay time for stars dying as SNe, as a means to compensate for the absence of RT. Otherwise, considering delay time without RT would artificially increase star formation and lead to overestimating both stellar mass and metallicity. A more accurate approach should consider delay time with radiative feedback, which will be addressed in future work.  The work by \citet{Smith2021} found that incorporating RT could suppress the clustering of SNe, resulting in a less effective outflow rate compared to scenarios without RT, while maintaining the total number of SNe. Given this effect, accounting for RT would lead to a more dispersed distribution of SNe, potentially reducing the overall stellar metallicity. Consequently, a more realistic cosmological simulation with advanced feedback mechanisms would likely yield a lower average metallicity than our current results, suggesting that the stellar metallicity observed in this study may represent an upper limit. However, intriguingly, in such a case, the simulated UFD analogs would exhibit lower metallicity compared to the observed UFDs, therefore further widening the gap between observational data and theoretical models.

As mentioned earlier, the presence and frequency of metal-rich stars provide vital insights into the environments of star-forming regions, a point also highlighted in other studies. In a recent study by \citet{Sandford2024}, the authors offer an analytical prediction of the galactic evolution of the UFD galaxy Eridanus II (Eri II), suggesting that the abundance of metal-rich stars is a key indicator of various physical processes. They specifically assessed the influence of Type Ia SNe, which promote the formation of metal-rich stars by increasing iron yields, examining their effects on both the MDF and the $\rm [Mg/Fe]-[Fe/H]$ space. Similarly, \citet{Ting2024} showed that analyzing the $\rm [\alpha/Fe]-[Fe/H]$ space can reveal whether dwarf galaxies experienced bursty star formation. However, in our simulations, Type Ia SNe are so rare that they had little effect, and the “knee” feature in the $\rm [\alpha/Fe]-[Fe/H]$ plane was not observed. This low frequency of Type Ia SNe is mainly due to the small stellar mass of our UFD analogs. Nevertheless, given that Type Ia SNe are not well constrained (e.g., \citealp{Dubay2024, Alexander2024}), it would be interesting to examine whether frequent Type Ia SNe, depending on the delay time distribution, alter the MDF and the abundance of high-metallicity stars in UFDs.

During the preparation of this paper, recent work by \citet{Andersson2024} reported results similar to ours, suggesting that the differences in physical properties predicted by simulations and observations could be due to methodological differences in deriving structural parameters. Specifically, they showed that using the methods commonly employed in observations (the same as our one-component fitting method with background stars) could result in lower luminosity estimates compared to the direct method. This discrepancy causes the simulated galaxies to be positioned to the left of the size-luminosity relation. They emphasized that this reduction in luminosity is due to the exclusion of extended star particles from the simulation data during the fitting process, rather than the omission of stars below the observational limit. Our simulations also exhibit a similar trend, where the number of particles included in the fitting process is, on average, 22\% lower than the total number of particles within the virial radius of the simulated UFDs.

As demonstrated in this study, employing consistent methodologies for deriving physical quantities in simulated galaxies is crucial for making fair comparisons with observational data. Simulations may also account for factors not yet considered in observations. For instance, metal-poor stars ($\rm [Fe/H] \lesssim -4$) are likely underrepresented in observational data, while simulations predict a higher abundance, leading to expected discrepancies in average metallicity and the MDF. Furthermore, simulations tend to suggest extended structures in the outer regions of UFD analogs. However, observationally, if galaxy sizes are derived solely from observed stars, missing potential outer member stars may result in an underestimation of galaxy size.

Discovering the diffuse outskirts of small UFDs is challenging due to their low surface brightness. However, next-generation surveys, such as the Vera Rubin Observatory's Legacy Survey of Space and Time (LSST; \citealp{LSST2009}), will provide deeper photometric data, enabling the study of dwarf galaxies at greater distances. Additionally, analyzing resolved stellar populations within our Local Group shows promise. As the kinematic data from {\it Gaia} improves with each release, we expect to identify new member stars in the outer regions of UFDs. Furthermore, spectroscopic efforts to measure detailed metal abundances in individual stars will aid in the identification and classification of member stars in the outer regions of dwarf galaxies.

\section*{acknowledgements}
We appreciate the insightful discussions with Mordecai-Mark Mac Low and Eric P. Andersson. We extend our gratitude to Volker Springel, Joop Schaye, and Claudio Dalla Vecchia for granting us permission to use their versions of \textsc{gadget}. We also thank Seongjun Hyung for his assistance in improving the Fig.~\ref{fig:desity_profile}. M.~J. and M.~K. acknowledge support from the National Research Foundation (NRF) grants funded by the Korean government (MSIT) (No. 2021R1A2C109491713, No. 2022M3K3A1093827).

\section*{DATA AVAILABILITY}
The simulation data and results of this paper may be available upon
request.

%%%%%%%%%%%%%%%%%%%% REFERENCES %%%%%%%%%%%%%%%%%%

% The best way to enter references is to use BibTeX:

\bibliographystyle{aasjournal}
\bibliography{ref}

\begin{thebibliography}{}
\expandafter\ifx\csname natexlab\endcsname\relax\def\natexlab#1{#1}\fi
\providecommand{\url}[1]{\href{#1}{#1}}
\providecommand{\dodoi}[1]{doi:~\href{http://doi.org/#1}{\nolinkurl{#1}}}
\providecommand{\doeprint}[1]{\href{http://ascl.net/#1}{\nolinkurl{http://ascl.net/#1}}}
\providecommand{\doarXiv}[1]{\href{https://arxiv.org/abs/#1}{\nolinkurl{https://arxiv.org/abs/#1}}}

\bibitem[{{Abdurro'uf} {et~al.}(2022){Abdurro'uf}, {Accetta}, {Aerts}, {Silva Aguirre}, {Ahumada}, {Ajgaonkar}, {Filiz Ak}, {Alam}, {Allende Prieto}, {Almeida}, {Anders}, {Anderson}, {Andrews}, {Anguiano}, {Aquino-Ort{\'\i}z}, {Arag{\'o}n-Salamanca}, {Argudo-Fern{\'a}ndez}, {Ata}, {Aubert}, {Avila-Reese}, {Badenes}, {Barb{\'a}}, {Barger}, {Barrera-Ballesteros}, {Beaton}, {Beers}, {Belfiore}, {Bender}, {Bernardi}, {Bershady}, {Beutler}, {Bidin}, {Bird}, {Bizyaev}, {Blanc}, {Blanton}, {Boardman}, {Bolton}, {Boquien}, {Borissova}, {Bovy}, {Brandt}, {Brown}, {Brownstein}, {Brusa}, {Buchner}, {Bundy}, {Burchett}, {Bureau}, {Burgasser}, {Cabang}, {Campbell}, {Cappellari}, {Carlberg}, {Wanderley}, {Carrera}, {Cash}, {Chen}, {Chen}, {Cherinka}, {Chiappini}, {Choi}, {Chojnowski}, {Chung}, {Clerc}, {Cohen}, {Comerford}, {Comparat}, {da Costa}, {Covey}, {Crane}, {Cruz-Gonzalez}, {Culhane}, {Cunha}, {Dai}, {Damke}, {Darling}, {Davidson}, {Davies}, {Dawson}, {De Lee}, {Diamond-Stanic}, {Cano-D{\'\i}az}, {S{\'a}nchez},
  {Donor}, {Duckworth}, {Dwelly}, {Eisenstein}, {Elsworth}, {Emsellem}, {Eracleous}, {Escoffier}, {Fan}, {Farr}, {Feng}, {Fern{\'a}ndez-Trincado}, {Feuillet}, {Filipp}, {Fillingham}, {Frinchaboy}, {Fromenteau}, {Galbany}, {Garc{\'\i}a}, {Garc{\'\i}a-Hern{\'a}ndez}, {Ge}, {Geisler}, {Gelfand}, {G{\'e}ron}, {Gibson}, {Goddy}, {Godoy-Rivera}, {Grabowski}, {Green}, {Greener}, {Grier}, {Griffith}, {Guo}, {Guy}, {Hadjara}, {Harding}, {Hasselquist}, {Hayes}, {Hearty}, {Hern{\'a}ndez}, {Hill}, {Hogg}, {Holtzman}, {Horta}, {Hsieh}, {Hsu}, {Hsu}, {Huber}, {Huertas-Company}, {Hutchinson}, {Hwang}, {Ibarra-Medel}, {Chitham}, {Ilha}, {Imig}, {Jaekle}, {Jayasinghe}, {Ji}, {Johnson}, {Jones}, {J{\"o}nsson}, {Katkov}, {Khalatyan}, {Kinemuchi}, {Kisku}, {Knapen}, {Kneib}, {Kollmeier}, {Kong}, {Kounkel}, {Kreckel}, {Krishnarao}, {Lacerna}, {Lane}, {Langgin}, {Lavender}, {Law}, {Lazarz}, {Leung}, {Leung}, {Lewis}, {Li}, {Li}, {Lian}, {Liang}, {Lin}, {Lin}, {Lin}, {Lintott}, {Long}, {Longa-Pe{\~n}a}, {L{\'o}pez-Cob{\'a}}, {Lu},
  {Lundgren}, {Luo}, {Mackereth}, {de la Macorra}, {Mahadevan}, {Majewski}, {Manchado}, {Mandeville}, {Maraston}, {Margalef-Bentabol}, {Masseron}, {Masters}, {Mathur}, {McDermid}, {Mckay}, {Merloni}, {Merrifield}, {Meszaros}, {Miglio}, {Di Mille}, {Minniti}, {Minsley}, {Monachesi}, {Moon}, {Mosser}, {Mulchaey}, {Muna}, {Mu{\~n}oz}, {Myers}, {Myers}, {Nadathur}, {Nair}, {Nandra}, {Neumann}, {Newman}, {Nidever}, {Nikakhtar}, {Nitschelm}, {O'Connell}, {Garma-Oehmichen}, {Luan Souza de Oliveira}, {Olney}, {Oravetz}, {Ortigoza-Urdaneta}, {Osorio}, {Otter}, {Pace}, {Padilla}, {Pan}, {Pan}, {Parikh}, {Parker}, {Peirani}, {Pe{\~n}a Ram{\'\i}rez}, {Penny}, {Percival}, {Perez-Fournon}, {Pinsonneault}, {Poidevin}, {Poovelil}, {Price-Whelan}, {B{\'a}rbara de Andrade Queiroz}, {Raddick}, {Ray}, {Rembold}, {Riddle}, {Riffel}, {Riffel}, {Rix}, {Robin}, {Rodr{\'\i}guez-Puebla}, {Roman-Lopes}, {Rom{\'a}n-Z{\'u}{\~n}iga}, {Rose}, {Ross}, {Rossi}, {Rubin}, {Salvato}, {S{\'a}nchez}, {S{\'a}nchez-Gallego}, {Sanderson}, {Santana
  Rojas}, {Sarceno}, {Sarmiento}, {Sayres}, {Sazonova}, {Schaefer}, {Schiavon}, {Schlegel}, {Schneider}, {Schultheis}, {Schwope}, {Serenelli}, {Serna}, {Shao}, {Shapiro}, {Sharma}, {Shen}, {Shetrone}, {Shu}, {Simon}, {Skrutskie}, {Smethurst}, {Smith}, {Sobeck}, {Spoo}, {Sprague}, {Stark}, {Stassun}, {Steinmetz}, {Stello}, {Stone-Martinez}, {Storchi-Bergmann}, {Stringfellow}, {Stutz}, {Su}, {Taghizadeh-Popp}, {Talbot}, {Tayar}, {Telles}, {Teske}, {Thakar}, {Theissen}, {Tkachenko}, {Thomas}, {Tojeiro}, {Hernandez Toledo}, {Troup}, {Trump}, {Trussler}, {Turner}, {Tuttle}, {Unda-Sanzana}, {V{\'a}zquez-Mata}, {Valentini}, {Valenzuela}, {Vargas-Gonz{\'a}lez}, {Vargas-Maga{\~n}a}, {Alfaro}, {Villanova}, {Vincenzo}, {Wake}, {Warfield}, {Washington}, {Weaver}, {Weijmans}, {Weinberg}, {Weiss}, {Westfall}, {Wild}, {Wilde}, {Wilson}, {Wilson}, {Wilson}, {Wolf}, {Wood-Vasey}, {Yan}, {Zamora}, {Zasowski}, {Zhang}, {Zhao}, {Zheng}, {Zheng}, \& {Zhu}}]{Abdurrouf2022}
{Abdurro'uf}, {Accetta}, K., {Aerts}, C., {et~al.} 2022, \apjs, 259, 35, \dodoi{10.3847/1538-4365/ac4414}

\bibitem[{{Agertz} {et~al.}(2020){Agertz}, {Pontzen}, {Read}, {Rey}, {Orkney}, {Rosdahl}, {Teyssier}, {Verbeke}, {Kretschmer}, \& {Nickerson}}]{Agertz2020}
{Agertz}, O., {Pontzen}, A., {Read}, J.~I., {et~al.} 2020, \mnras, 491, 1656, \dodoi{10.1093/mnras/stz3053}

\bibitem[{{Alexander} \& {Vincenzo}(2024)}]{Alexander2024}
{Alexander}, R.~K., \& {Vincenzo}, F. 2024, arXiv e-prints, arXiv:2408.07443, \dodoi{10.48550/arXiv.2408.07443}

\bibitem[{{Andersson} {et~al.}(2024){Andersson}, {Rey}, {Pontzen}, {Cadiou}, {Agertz}, {Read}, \& {Martin}}]{Andersson2024}
{Andersson}, E.~P., {Rey}, M.~P., {Pontzen}, A., {et~al.} 2024, arXiv e-prints, arXiv:2409.08073, \dodoi{10.48550/arXiv.2409.08073}

\bibitem[{{Applebaum} {et~al.}(2021){Applebaum}, {Brooks}, {Christensen}, {Munshi}, {Quinn}, {Shen}, \& {Tremmel}}]{Applebaum2021}
{Applebaum}, E., {Brooks}, A.~M., {Christensen}, C.~R., {et~al.} 2021, \apj, 906, 96, \dodoi{10.3847/1538-4357/abcafa}

\bibitem[{{Applebaum} {et~al.}(2020){Applebaum}, {Brooks}, {Quinn}, \& {Christensen}}]{Applebaum2020}
{Applebaum}, E., {Brooks}, A.~M., {Quinn}, T.~R., \& {Christensen}, C.~R. 2020, \mnras, 492, 8, \dodoi{10.1093/mnras/stz3331}

\bibitem[{{Azartash-Namin} {et~al.}(2024){Azartash-Namin}, {Engelhardt}, {Munshi}, {Keller}, {Brooks}, {Van Nest}, {Christensen}, {Quinn}, \& {Wadsley}}]{Azartash-Namin2024}
{Azartash-Namin}, B., {Engelhardt}, A., {Munshi}, F., {et~al.} 2024, arXiv e-prints, arXiv:2401.06041, \dodoi{10.48550/arXiv.2401.06041}

\bibitem[{{Bate}(2019)}]{Bate2019}
{Bate}, M.~R. 2019, \mnras, 484, 2341, \dodoi{10.1093/mnras/stz103}

\bibitem[{{Behroozi} {et~al.}(2010){Behroozi}, {Conroy}, \& {Wechsler}}]{Behroozi2010}
{Behroozi}, P.~S., {Conroy}, C., \& {Wechsler}, R.~H. 2010, \apj, 717, 379, \dodoi{10.1088/0004-637X/717/1/379}

\bibitem[{{Behroozi} {et~al.}(2013){Behroozi}, {Wechsler}, \& {Conroy}}]{Behroozi2013}
{Behroozi}, P.~S., {Wechsler}, R.~H., \& {Conroy}, C. 2013, ApJ, 770, 57, \dodoi{10.1088/0004-637X/770/1/57}

\bibitem[{{Binney} \& {Knebe}(2002)}]{Binney2002}
{Binney}, J., \& {Knebe}, A. 2002, \mnras, 333, 378, \dodoi{10.1046/j.1365-8711.2002.05400.x}

\bibitem[{{Bovill} \& {Ricotti}(2009)}]{Bovill2009}
{Bovill}, M.~S., \& {Ricotti}, M. 2009, ApJ, 693, 1859, \dodoi{10.1088/0004-637X/693/2/1859}

\bibitem[{{Bromm} \& {Larson}(2004)}]{Bromm2004}
{Bromm}, V., \& {Larson}, R.~B. 2004, ARA\&A, 42, 79, \dodoi{10.1146/annurev.astro.42.053102.134034}

\bibitem[{{Brown} {et~al.}(2014){Brown}, {Tumlinson}, {Geha}, {Simon}, {Vargas}, {VandenBerg}, {Kirby}, {Kalirai}, {Avila}, {Gennaro}, {Ferguson}, {Mu{\~n}oz}, {Guhathakurta}, \& {Renzini}}]{Brown2014}
{Brown}, T.~M., {Tumlinson}, J., {Geha}, M., {et~al.} 2014, \apj, 796, 91, \dodoi{10.1088/0004-637X/796/2/91}

\bibitem[{{Buck} {et~al.}(2019){Buck}, {Macci{\`o}}, {Dutton}, {Obreja}, \& {Frings}}]{Buck2019}
{Buck}, T., {Macci{\`o}}, A.~V., {Dutton}, A.~A., {Obreja}, A., \& {Frings}, J. 2019, \mnras, 483, 1314, \dodoi{10.1093/mnras/sty2913}

\bibitem[{{Cerny} {et~al.}(2023){Cerny}, {Mart{\'\i}nez-V{\'a}zquez}, {Drlica-Wagner}, {Pace}, {Mutlu-Pakdil}, {Li}, {Riley}, {Crnojevi{\'c}}, {Bom}, {Carballo-Bello}, {Carlin}, {Chiti}, {Choi}, {Collins}, {Darragh-Ford}, {Ferguson}, {Geha}, {Mart{\'\i}nez-Delgado}, {Massana}, {Mau}, {Medina}, {Mu{\~n}oz}, {Nadler}, {No{\"e}l}, {Olsen}, {Pieres}, {Sakowska}, {Simon}, {Stringfellow}, {Tollerud}, {Vivas}, {Walker}, {Wechsler}, \& {Delve Collaboration}}]{Cerny2023}
{Cerny}, W., {Mart{\'\i}nez-V{\'a}zquez}, C.~E., {Drlica-Wagner}, A., {et~al.} 2023, \apj, 953, 1, \dodoi{10.3847/1538-4357/acdd78}

\bibitem[{{Cerny} {et~al.}(2024){Cerny}, {Chiti}, {Geha}, {Mutlu-Pakdil}, {Drlica-Wagner}, {Tan}, {Adam{\'o}w}, {Pace}, {Simon}, {Sand}, {Ji}, {Li}, {Vivas}, {Bell}, {Carlin}, {Carballo-Bello}, {Chaturvedi}, {Choi}, {Doliva-Dolinsky}, {Gnedin}, {Limberg}, {Mart{\'\i}nez-V{\'a}zquez}, {Mau}, {Medina}, {Navabi}, {No{\"e}l}, {Placco}, {Riley}, {Roederer}, {Stringfellow}, {Bom}, {Ferguson}, {James}, {Mart{\'\i}nez-Delgado}, {Massana}, {Nidever}, {Sakowska}, {Santana-Silva}, {Sherman}, \& {Tollerud}}]{Cerny2024}
{Cerny}, W., {Chiti}, A., {Geha}, M., {et~al.} 2024, arXiv e-prints, arXiv:2410.00981, \dodoi{10.48550/arXiv.2410.00981}

\bibitem[{{Chiti} {et~al.}(2018){Chiti}, {Frebel}, {Ji}, {Jerjen}, {Kim}, \& {Norris}}]{Chiti2018}
{Chiti}, A., {Frebel}, A., {Ji}, A.~P., {et~al.} 2018, \apj, 857, 74, \dodoi{10.3847/1538-4357/aab4fc}

\bibitem[{{Chiti} {et~al.}(2021){Chiti}, {Frebel}, {Simon}, {Erkal}, {Chang}, {Necib}, {Ji}, {Jerjen}, {Kim}, \& {Norris}}]{Chiti2021}
{Chiti}, A., {Frebel}, A., {Simon}, J.~D., {et~al.} 2021, Nature Astronomy, 5, 392, \dodoi{10.1038/s41550-020-01285-w}

\bibitem[{{Chiti} {et~al.}(2023){Chiti}, {Frebel}, {Ji}, {Mardini}, {Ou}, {Simon}, {Jerjen}, {Kim}, \& {Norris}}]{Chiti2023}
{Chiti}, A., {Frebel}, A., {Ji}, A.~P., {et~al.} 2023, \aj, 165, 55, \dodoi{10.3847/1538-3881/aca416}

\bibitem[{{Choi} {et~al.}(2016){Choi}, {Dotter}, {Conroy}, {Cantiello}, {Paxton}, \& {Johnson}}]{Choi2016}
{Choi}, J., {Dotter}, A., {Conroy}, C., {et~al.} 2016, \apj, 823, 102, \dodoi{10.3847/0004-637X/823/2/102}

\bibitem[{{Dalla Vecchia} \& {Schaye}(2012)}]{Vecchia2012}
{Dalla Vecchia}, C., \& {Schaye}, J. 2012, MNRAS, 426, 140, \dodoi{10.1111/j.1365-2966.2012.21704.x}

\bibitem[{{Deason} {et~al.}(2022){Deason}, {Bose}, {Fattahi}, {Amorisco}, {Hellwing}, \& {Frenk}}]{Deason2022}
{Deason}, A.~J., {Bose}, S., {Fattahi}, A., {et~al.} 2022, \mnras, 511, 4044, \dodoi{10.1093/mnras/stab3524}

\bibitem[{{Dekel} \& {Silk}(1986)}]{Dekel1986}
{Dekel}, A., \& {Silk}, J. 1986, \apj, 303, 39, \dodoi{10.1086/164050}

\bibitem[{{Dotter}(2016)}]{Dotter2016}
{Dotter}, A. 2016, \apjs, 222, 8, \dodoi{10.3847/0067-0049/222/1/8}

\bibitem[{{Dubay} {et~al.}(2024){Dubay}, {Johnson}, \& {Johnson}}]{Dubay2024}
{Dubay}, L.~O., {Johnson}, J.~A., \& {Johnson}, J.~W. 2024, \apj, 973, 55, \dodoi{10.3847/1538-4357/ad61df}

\bibitem[{{Escala} {et~al.}(2018){Escala}, {Wetzel}, {Kirby}, {Hopkins}, {Ma}, {Wheeler}, {Kere{\v{s}}}, {Faucher-Gigu{\`e}re}, \& {Quataert}}]{Escala2018}
{Escala}, I., {Wetzel}, A., {Kirby}, E.~N., {et~al.} 2018, \mnras, 474, 2194, \dodoi{10.1093/mnras/stx2858}

\bibitem[{{Fan et al.}(2006)}]{Fan2006}
{Fan et al.} 2006, AJ, 131, 1203, \dodoi{10.1086/500296}

\bibitem[{{Ferland} {et~al.}(1998){Ferland}, {Korista}, {Verner}, {Ferguson}, {Kingdon}, \& {Verner}}]{Ferland1998}
{Ferland}, G.~J., {Korista}, K.~T., {Verner}, D.~A., {et~al.} 1998, PASP, 110, 761, \dodoi{10.1086/316190}

\bibitem[{{Fitts} {et~al.}(2018){Fitts}, {Boylan-Kolchin}, {Bullock}, {Weisz}, {El-Badry}, {Wheeler}, {Faucher-Gigu{\`e}re}, {Quataert}, {Hopkins}, {Kere{\v{s}}}, {Wetzel}, \& {Hayward}}]{Fitts2018}
{Fitts}, A., {Boylan-Kolchin}, M., {Bullock}, J.~S., {et~al.} 2018, \mnras, 479, 319, \dodoi{10.1093/mnras/sty1488}

\bibitem[{{Foreman-Mackey} {et~al.}(2013){Foreman-Mackey}, {Hogg}, {Lang}, \& {Goodman}}]{Foreman2013}
{Foreman-Mackey}, D., {Hogg}, D.~W., {Lang}, D., \& {Goodman}, J. 2013, \pasp, 125, 306, \dodoi{10.1086/670067}

\bibitem[{{F{\"o}rster} {et~al.}(2006){F{\"o}rster}, {Wolf}, {Podsiadlowski}, \& {Han}}]{Forster2006}
{F{\"o}rster}, F., {Wolf}, C., {Podsiadlowski}, P., \& {Han}, Z. 2006, \mnras, 368, 1893, \dodoi{10.1111/j.1365-2966.2006.10258.x}

\bibitem[{{Fran{\c{c}}ois} {et~al.}(2016){Fran{\c{c}}ois}, {Monaco}, {Bonifacio}, {Moni Bidin}, {Geisler}, \& {Sbordone}}]{Francois2016}
{Fran{\c{c}}ois}, P., {Monaco}, L., {Bonifacio}, P., {et~al.} 2016, \aap, 588, A7, \dodoi{10.1051/0004-6361/201527181}

\bibitem[{{Frebel} \& {Norris}(2015)}]{Frebel2015}
{Frebel}, A., \& {Norris}, J.~E. 2015, ARA\&A, 53, 631, \dodoi{10.1146/annurev-astro-082214-122423}

\bibitem[{{Frebel} {et~al.}(2010){Frebel}, {Simon}, {Geha}, \& {Willman}}]{Frebel2010}
{Frebel}, A., {Simon}, J.~D., {Geha}, M., \& {Willman}, B. 2010, \apj, 708, 560, \dodoi{10.1088/0004-637X/708/1/560}

\bibitem[{{Frebel} {et~al.}(2014){Frebel}, {Simon}, \& {Kirby}}]{Frebel2014}
{Frebel}, A., {Simon}, J.~D., \& {Kirby}, E.~N. 2014, \apj, 786, 74, \dodoi{10.1088/0004-637X/786/1/74}

\bibitem[{{Fu} {et~al.}(2023){Fu}, {Weisz}, {Starkenburg}, {Martin}, {Savino}, {Boylan-Kolchin}, {C{\^o}t{\'e}}, {Dolphin}, {Ji}, {Longeard}, {Mateo}, {Patel}, \& {Sandford}}]{Fu2023}
{Fu}, S.~W., {Weisz}, D.~R., {Starkenburg}, E., {et~al.} 2023, \apj, 958, 167, \dodoi{10.3847/1538-4357/ad0030}

\bibitem[{{Fu} {et~al.}(2024){Fu}, {Weisz}, {Starkenburg}, {Martin}, {Mercado}, {Savino}, {Boylan-Kolchin}, {C{\^o}t{\'e}}, {Dolphin}, {Longeard}, {Mateo}, {Samuel}, \& {Sandford}}]{Fu2024}
---. 2024, \apj, 965, 36, \dodoi{10.3847/1538-4357/ad25ed}

\bibitem[{{Geha} {et~al.}(2013){Geha}, {Brown}, {Tumlinson}, {Kalirai}, {Simon}, {Kirby}, {VandenBerg}, {Mu{\~n}oz}, {Avila}, {Guhathakurta}, \& {Ferguson}}]{Geha2013}
{Geha}, M., {Brown}, T.~M., {Tumlinson}, J., {et~al.} 2013, \apj, 771, 29, \dodoi{10.1088/0004-637X/771/1/29}

\bibitem[{{Gilmore} {et~al.}(2013){Gilmore}, {Norris}, {Monaco}, {Yong}, {Wyse}, \& {Geisler}}]{Gilmore2013}
{Gilmore}, G., {Norris}, J.~E., {Monaco}, L., {et~al.} 2013, \apj, 763, 61, \dodoi{10.1088/0004-637X/763/1/61}

\bibitem[{{Goater} {et~al.}(2024){Goater}, {Read}, {No{\"e}l}, {Orkney}, {Kim}, {Rey}, {Andersson}, {Agertz}, {Pontzen}, {Vieliute}, {Kataria}, \& {Jeneway}}]{Goater2024}
{Goater}, A., {Read}, J.~I., {No{\"e}l}, N. E.~D., {et~al.} 2024, \mnras, 527, 2403, \dodoi{10.1093/mnras/stad3354}

\bibitem[{{Greif} {et~al.}(2009){Greif}, {Glover}, {Bromm}, \& {Klessen}}]{Greif2009a}
{Greif}, T.~H., {Glover}, S.~C.~O., {Bromm}, V., \& {Klessen}, R.~S. 2009, MNRAS, 392, 1381, \dodoi{10.1111/j.1365-2966.2008.14169.x}

\bibitem[{{Gunn} \& {Peterson}(1965)}]{Gunn1965}
{Gunn}, J.~E., \& {Peterson}, B.~A. 1965, ApJ, 142, 1633, \dodoi{10.1086/148444}

\bibitem[{{Guszejnov} {et~al.}(2022){Guszejnov}, {Grudi{\'c}}, {Offner}, {Faucher-Gigu{\`e}re}, {Hopkins}, \& {Rosen}}]{Guszejnov2022}
{Guszejnov}, D., {Grudi{\'c}}, M.~Y., {Offner}, S. S.~R., {et~al.} 2022, \mnras, 515, 4929, \dodoi{10.1093/mnras/stac2060}

\bibitem[{{Gutcke} {et~al.}(2021){Gutcke}, {Pakmor}, {Naab}, \& {Springel}}]{Gutcke2021}
{Gutcke}, T.~A., {Pakmor}, R., {Naab}, T., \& {Springel}, V. 2021, \mnras, 501, 5597, \dodoi{10.1093/mnras/staa3875}

\bibitem[{{Haardt} \& {Madau}(2012)}]{Haardt2012}
{Haardt}, F., \& {Madau}, P. 2012, ApJ, 746, 125, \dodoi{10.1088/0004-637X/746/2/125}

\bibitem[{{Hahn} \& {Abel}(2011)}]{Hahn2011}
{Hahn}, O., \& {Abel}, T. 2011, MNRAS, 415, 2101, \dodoi{10.1111/j.1365-2966.2011.18820.x}

\bibitem[{{Hansen} {et~al.}(2024){Hansen}, {Simon}, {Li}, {Sharkey}, {Ji}, {Thompson}, {Reggiani}, \& {Galarza}}]{Hansen2024}
{Hansen}, T.~T., {Simon}, J.~D., {Li}, T.~S., {et~al.} 2024, \apj, 968, 21, \dodoi{10.3847/1538-4357/ad3a52}

\bibitem[{{Hansen} {et~al.}(2017){Hansen}, {Simon}, {Marshall}, {Li}, {Carollo}, {DePoy}, {Nagasawa}, {Bernstein}, {Drlica-Wagner}, {Abdalla}, {Allam}, {Annis}, {Bechtol}, {Benoit-L{\'e}vy}, {Brooks}, {Buckley-Geer}, {Carnero Rosell}, {Carrasco Kind}, {Carretero}, {Cunha}, {da Costa}, {Desai}, {Eifler}, {Fausti Neto}, {Flaugher}, {Frieman}, {Garc{\'\i}a-Bellido}, {Gaztanaga}, {Gerdes}, {Gruen}, {Gruendl}, {Gschwend}, {Gutierrez}, {James}, {Krause}, {Kuehn}, {Kuropatkin}, {Lahav}, {Miquel}, {Plazas}, {Romer}, {Sanchez}, {Santiago}, {Scarpine}, {Smith}, {Soares-Santos}, {Sobreira}, {Suchyta}, {Swanson}, {Tarle}, {Walker}, \& {DES Collaboration}}]{Hansen2017}
{Hansen}, T.~T., {Simon}, J.~D., {Marshall}, J.~L., {et~al.} 2017, \apj, 838, 44, \dodoi{10.3847/1538-4357/aa634a}

\bibitem[{{Harbeck} {et~al.}(2001){Harbeck}, {Grebel}, {Holtzman}, {Guhathakurta}, {Brandner}, {Geisler}, {Sarajedini}, {Dolphin}, {Hurley-Keller}, \& {Mateo}}]{Harbeck2001}
{Harbeck}, D., {Grebel}, E.~K., {Holtzman}, J., {et~al.} 2001, \aj, 122, 3092, \dodoi{10.1086/324232}

\bibitem[{{Heger} {et~al.}(2003){Heger}, {Fryer}, {Woosley}, {Langer}, \& {Hartmann}}]{Heger2003}
{Heger}, A., {Fryer}, C.~L., {Woosley}, S.~E., {Langer}, N., \& {Hartmann}, D.~H. 2003, ApJ, 591, 288, \dodoi{10.1086/375341}

\bibitem[{{Heger} \& {Woosley}(2002)}]{Heger2002}
{Heger}, A., \& {Woosley}, S.~E. 2002, ApJ, 567, 532, \dodoi{10.1086/338487}

\bibitem[{{Heger} \& {Woosley}(2010)}]{Heger2010}
---. 2010, ApJ, 724, 341, \dodoi{10.1088/0004-637X/724/1/341}

\bibitem[{{Hicks} {et~al.}(2021){Hicks}, {Wells}, {Norman}, {Wise}, {Smith}, \& {O'Shea}}]{Hicks2021}
{Hicks}, W.~M., {Wells}, A., {Norman}, M.~L., {et~al.} 2021, \apj, 909, 70, \dodoi{10.3847/1538-4357/abda3a}

\bibitem[{{Hirano} {et~al.}(2015){Hirano}, {Hosokawa}, {Yoshida}, {Omukai}, \& {Yorke}}]{Hirano2015}
{Hirano}, S., {Hosokawa}, T., {Yoshida}, N., {Omukai}, K., \& {Yorke}, H.~W. 2015, MNRAS, 448, 568, \dodoi{10.1093/MNRAS/stv044}

\bibitem[{{Hopkins} {et~al.}(2014){Hopkins}, {Kere{\v{s}}}, {O{\~n}orbe}, {Faucher-Gigu{\`e}re}, {Quataert}, {Murray}, \& {Bullock}}]{hopkins2014}
{Hopkins}, P.~F., {Kere{\v{s}}}, D., {O{\~n}orbe}, J., {et~al.} 2014, \mnras, 445, 581, \dodoi{10.1093/mnras/stu1738}

\bibitem[{{Ishigaki} {et~al.}(2014){Ishigaki}, {Aoki}, {Arimoto}, \& {Okamoto}}]{Ishigaki2014}
{Ishigaki}, M.~N., {Aoki}, W., {Arimoto}, N., \& {Okamoto}, S. 2014, \aap, 562, A146, \dodoi{10.1051/0004-6361/201322796}

\bibitem[{{Jensen} {et~al.}(2024){Jensen}, {Hayes}, {Sestito}, {McConnachie}, {Waller}, {Smith}, {Navarro}, \& {Venn}}]{Jensen2024}
{Jensen}, J., {Hayes}, C.~R., {Sestito}, F., {et~al.} 2024, \mnras, 527, 4209, \dodoi{10.1093/mnras/stad3322}

\bibitem[{{Jeon} {et~al.}(2017){Jeon}, {Besla}, \& {Bromm}}]{Jeon2017}
{Jeon}, M., {Besla}, G., \& {Bromm}, V. 2017, \apj, 848, 85, \dodoi{10.3847/1538-4357/aa8c80}

\bibitem[{{Jeon} {et~al.}(2021{\natexlab{a}}){Jeon}, {Besla}, \& {Bromm}}]{Jeon2021b}
---. 2021{\natexlab{a}}, \mnras, 506, 1850, \dodoi{10.1093/mnras/stab1771}

\bibitem[{{Jeon} {et~al.}(2021{\natexlab{b}}){Jeon}, {Bromm}, {Besla}, {Yoon}, \& {Choi}}]{Jeon2021a}
{Jeon}, M., {Bromm}, V., {Besla}, G., {Yoon}, J., \& {Choi}, Y. 2021{\natexlab{b}}, \mnras, 502, 1, \dodoi{10.1093/mnras/staa4017}

\bibitem[{{Jeon} \& {Ko}(2024)}]{Jeon2024}
{Jeon}, M., \& {Ko}, M. 2024, arXiv e-prints, arXiv:2411.17862, \dodoi{10.48550/arXiv.2411.17862}

\bibitem[{{Jeon} {et~al.}(2014){Jeon}, {Pawlik}, {Bromm}, \& {Milosavljevi{\'c}}}]{Jeon2014}
{Jeon}, M., {Pawlik}, A.~H., {Bromm}, V., \& {Milosavljevi{\'c}}, M. 2014, \mnras, 444, 3288, \dodoi{10.1093/mnras/stu1980}

\bibitem[{{Ji} \& {Frebel}(2018)}]{Ji2018}
{Ji}, A.~P., \& {Frebel}, A. 2018, \apj, 856, 138, \dodoi{10.3847/1538-4357/aab14a}

\bibitem[{{Ji} {et~al.}(2016){Ji}, {Frebel}, {Simon}, \& {Geha}}]{Ji2016a}
{Ji}, A.~P., {Frebel}, A., {Simon}, J.~D., \& {Geha}, M. 2016, \apj, 817, 41, \dodoi{10.3847/0004-637X/817/1/41}

\bibitem[{{Ji} {et~al.}(2019){Ji}, {Simon}, {Frebel}, {Venn}, \& {Hansen}}]{Ji2019}
{Ji}, A.~P., {Simon}, J.~D., {Frebel}, A., {Venn}, K.~A., \& {Hansen}, T.~T. 2019, \apj, 870, 83, \dodoi{10.3847/1538-4357/aaf3bb}

\bibitem[{{Kang} \& {Ricotti}(2019)}]{Kang2019}
{Kang}, H.~D., \& {Ricotti}, M. 2019, \mnras, 488, 2673, \dodoi{10.1093/mnras/stz1886}

\bibitem[{{Kim} {et~al.}(2023){Kim}, {Jeon}, {Choi}, {Richstein}, {Sacchi}, \& {Kallivayalil}}]{Kim2023}
{Kim}, J., {Jeon}, M., {Choi}, Y., {et~al.} 2023, \apj, 959, 31, \dodoi{10.3847/1538-4357/acfe08}

\bibitem[{{Kirby} {et~al.}(2013){Kirby}, {Cohen}, {Guhathakurta}, {Cheng}, {Bullock}, \& {Gallazzi}}]{Kirby2013}
{Kirby}, E.~N., {Cohen}, J.~G., {Guhathakurta}, P., {et~al.} 2013, \apj, 779, 102, \dodoi{10.1088/0004-637X/779/2/102}

\bibitem[{{Kirby} {et~al.}(2017){Kirby}, {Cohen}, {Simon}, {Guhathakurta}, {Thygesen}, \& {Duggan}}]{Kirby2017}
{Kirby}, E.~N., {Cohen}, J.~G., {Simon}, J.~D., {et~al.} 2017, \apj, 838, 83, \dodoi{10.3847/1538-4357/aa6570}

\bibitem[{{Kirby} {et~al.}(2011){Kirby}, {Lanfranchi}, {Simon}, {Cohen}, \& {Guhathakurta}}]{Kirby2011}
{Kirby}, E.~N., {Lanfranchi}, G.~A., {Simon}, J.~D., {Cohen}, J.~G., \& {Guhathakurta}, P. 2011, \apj, 727, 78, \dodoi{10.1088/0004-637X/727/2/78}

\bibitem[{{Kirby} {et~al.}(2015){Kirby}, {Simon}, \& {Cohen}}]{Kirby2015}
{Kirby}, E.~N., {Simon}, J.~D., \& {Cohen}, J.~G. 2015, \apj, 810, 56, \dodoi{10.1088/0004-637X/810/1/56}

\bibitem[{{Komatsu} {et~al.}(2011){Komatsu}, {Smith}, {Dunkley}, {Bennett}, {Gold}, {Hinshaw}, {Jarosik}, {Larson}, {Nolta}, {Page}, \& {Spergel}}]{Komatsu2011}
{Komatsu}, E., {Smith}, K.~M., {Dunkley}, J., {et~al.} 2011, ApJS, 192, 18, \dodoi{10.1088/0067-0049/192/2/18}

\bibitem[{{Lah{\'e}n} {et~al.}(2025){Lah{\'e}n}, {Rantala}, {Naab}, {Partmann}, {Johansson}, \& {Hislop}}]{Lahen2025}
{Lah{\'e}n}, N., {Rantala}, A., {Naab}, T., {et~al.} 2025, \mnras, 538, 2129, \dodoi{10.1093/mnras/staf350}

\bibitem[{{Lee} {et~al.}(2024){Lee}, {Jeon}, \& {Bromm}}]{Lee2024}
{Lee}, T., {Jeon}, M., \& {Bromm}, V. 2024, \mnras, 527, 1257, \dodoi{10.1093/mnras/stad3198}

\bibitem[{{Leroy} {et~al.}(2008){Leroy}, {Walter}, {Brinks}, {Bigiel}, {de Blok}, {Madore}, \& {Thornley}}]{Leroy2008}
{Leroy}, A.~K., {Walter}, F., {Brinks}, E., {et~al.} 2008, \aj, 136, 2782, \dodoi{10.1088/0004-6256/136/6/2782}

\bibitem[{{Longeard} {et~al.}(2022){Longeard}, {Jablonka}, {Arentsen}, {Thomas}, {Aguado}, {Carlberg}, {Lucchesi}, {Malhan}, {Martin}, {McConnachie}, {Navarro}, {S{\'a}nchez-Janssen}, {Sestito}, {Starkenburg}, \& {Yuan}}]{Longeard2022}
{Longeard}, N., {Jablonka}, P., {Arentsen}, A., {et~al.} 2022, \mnras, 516, 2348, \dodoi{10.1093/mnras/stac1827}

\bibitem[{{LSST Science Collaboration} {et~al.}(2009){LSST Science Collaboration}, {Abell}, {Allison}, {Anderson}, {Andrew}, {Angel}, {Armus}, {Arnett}, {Asztalos}, {Axelrod}, {Bailey}, {Ballantyne}, {Bankert}, {Barkhouse}, {Barr}, {Barrientos}, {Barth}, {Bartlett}, {Becker}, {Becla}, {Beers}, {Bernstein}, {Biswas}, {Blanton}, {Bloom}, {Bochanski}, {Boeshaar}, {Borne}, {Bradac}, {Brandt}, {Bridge}, {Brown}, {Brunner}, {Bullock}, {Burgasser}, {Burge}, {Burke}, {Cargile}, {Chandrasekharan}, {Chartas}, {Chesley}, {Chu}, {Cinabro}, {Claire}, {Claver}, {Clowe}, {Connolly}, {Cook}, {Cooke}, {Cooray}, {Covey}, {Culliton}, {de Jong}, {de Vries}, {Debattista}, {Delgado}, {Dell'Antonio}, {Dhital}, {Di Stefano}, {Dickinson}, {Dilday}, {Djorgovski}, {Dobler}, {Donalek}, {Dubois-Felsmann}, {Durech}, {Eliasdottir}, {Eracleous}, {Eyer}, {Falco}, {Fan}, {Fassnacht}, {Ferguson}, {Fernandez}, {Fields}, {Finkbeiner}, {Figueroa}, {Fox}, {Francke}, {Frank}, {Frieman}, {Fromenteau}, {Furqan}, {Galaz}, {Gal-Yam}, {Garnavich},
  {Gawiser}, {Geary}, {Gee}, {Gibson}, {Gilmore}, {Grace}, {Green}, {Gressler}, {Grillmair}, {Habib}, {Haggerty}, {Hamuy}, {Harris}, {Hawley}, {Heavens}, {Hebb}, {Henry}, {Hileman}, {Hilton}, {Hoadley}, {Holberg}, {Holman}, {Howell}, {Infante}, {Ivezic}, {Jacoby}, {Jain}, {R}, {Jedicke}, {Jee}, {Garrett Jernigan}, {Jha}, {Johnston}, {Jones}, {Juric}, {Kaasalainen}, {Styliani}, {Kafka}, {Kahn}, {Kaib}, {Kalirai}, {Kantor}, {Kasliwal}, {Keeton}, {Kessler}, {Knezevic}, {Kowalski}, {Krabbendam}, {Krughoff}, {Kulkarni}, {Kuhlman}, {Lacy}, {Lepine}, {Liang}, {Lien}, {Lira}, {Long}, {Lorenz}, {Lotz}, {Lupton}, {Lutz}, {Macri}, {Mahabal}, {Mandelbaum}, {Marshall}, {May}, {McGehee}, {Meadows}, {Meert}, {Milani}, {Miller}, {Miller}, {Mills}, {Minniti}, {Monet}, {Mukadam}, {Nakar}, {Neill}, {Newman}, {Nikolaev}, {Nordby}, {O'Connor}, {Oguri}, {Oliver}, {Olivier}, {Olsen}, {Olsen}, {Olszewski}, {Oluseyi}, {Padilla}, {Parker}, {Pepper}, {Peterson}, {Petry}, {Pinto}, {Pizagno}, {Popescu}, {Prsa}, {Radcka}, {Raddick},
  {Rasmussen}, {Rau}, {Rho}, {Rhoads}, {Richards}, {Ridgway}, {Robertson}, {Roskar}, {Saha}, {Sarajedini}, {Scannapieco}, {Schalk}, {Schindler}, {Schmidt}, {Schmidt}, {Schneider}, {Schumacher}, {Scranton}, {Sebag}, {Seppala}, {Shemmer}, {Simon}, {Sivertz}, {Smith}, {Allyn Smith}, {Smith}, {Spitz}, {Stanford}, {Stassun}, {Strader}, {Strauss}, {Stubbs}, {Sweeney}, {Szalay}, {Szkody}, {Takada}, {Thorman}, {Trilling}, {Trimble}, {Tyson}, {Van Berg}, {Vanden Berk}, {VanderPlas}, {Verde}, {Vrsnak}, {Walkowicz}, {Wandelt}, {Wang}, {Wang}, {Warner}, {Wechsler}, {West}, {Wiecha}, {Williams}, {Willman}, {Wittman}, {Wolff}, {Wood-Vasey}, {Wozniak}, {Young}, {Zentner}, \& {Zhan}}]{LSST2009}
{LSST Science Collaboration}, {Abell}, P.~A., {Allison}, J., {et~al.} 2009, arXiv e-prints, arXiv:0912.0201, \dodoi{10.48550/arXiv.0912.0201}

\bibitem[{{Ludlow} {et~al.}(2019{\natexlab{a}}){Ludlow}, {Schaye}, \& {Bower}}]{Ludlow2019a}
{Ludlow}, A.~D., {Schaye}, J., \& {Bower}, R. 2019{\natexlab{a}}, \mnras, 488, 3663, \dodoi{10.1093/mnras/stz1821}

\bibitem[{{Ludlow} {et~al.}(2019{\natexlab{b}}){Ludlow}, {Schaye}, {Schaller}, \& {Richings}}]{Ludlow2019b}
{Ludlow}, A.~D., {Schaye}, J., {Schaller}, M., \& {Richings}, J. 2019{\natexlab{b}}, \mnras, 488, L123, \dodoi{10.1093/mnrasl/slz110}

\bibitem[{{Marigo}(2001)}]{Marigo2001}
{Marigo}, P. 2001, A\&A, 370, 194, \dodoi{10.1051/0004-6361:20000247}

\bibitem[{{Martin} {et~al.}(2008){Martin}, {de Jong}, \& {Rix}}]{Martin2008}
{Martin}, N.~F., {de Jong}, J. T.~A., \& {Rix}, H.-W. 2008, \apj, 684, 1075, \dodoi{10.1086/590336}

\bibitem[{{McConnachie}(2012)}]{McConaachie2012}
{McConnachie}, A.~W. 2012, \aj, 144, 4, \dodoi{10.1088/0004-6256/144/1/4}

\bibitem[{{Mercado} {et~al.}(2021){Mercado}, {Bullock}, {Boylan-Kolchin}, {Moreno}, {Wetzel}, {El-Badry}, {Graus}, {Fitts}, {Hopkins}, {Faucher-Gigu{\`e}re}, \& {Gurvich}}]{Mercado2021}
{Mercado}, F.~J., {Bullock}, J.~S., {Boylan-Kolchin}, M., {et~al.} 2021, \mnras, 501, 5121, \dodoi{10.1093/mnras/staa3958}

\bibitem[{{Munshi} {et~al.}(2013){Munshi}, {Governato}, {Brooks}, {Christensen}, {Shen}, {Loebman}, {Moster}, {Quinn}, \& {Wadsley}}]{Munshi2013}
{Munshi}, F., {Governato}, F., {Brooks}, A.~M., {et~al.} 2013, ApJ, 766, 56, \dodoi{10.1088/0004-637X/766/1/56}

\bibitem[{{Nagasawa} {et~al.}(2018){Nagasawa}, {Marshall}, {Li}, {Hansen}, {Simon}, {Bernstein}, {Balbinot}, {Drlica-Wagner}, {Pace}, {Strigari}, {Pellegrino}, {DePoy}, {Suntzeff}, {Bechtol}, {Walker}, {Abbott}, {Abdalla}, {Allam}, {Annis}, {Benoit-L{\'e}vy}, {Bertin}, {Brooks}, {Carnero Rosell}, {Carrasco Kind}, {Carretero}, {Cunha}, {D'Andrea}, {da Costa}, {Davis}, {Desai}, {Doel}, {Eifler}, {Flaugher}, {Fosalba}, {Frieman}, {Garc{\'\i}a-Bellido}, {Gaztanaga}, {Gerdes}, {Gruen}, {Gruendl}, {Gschwend}, {Gutierrez}, {Hartley}, {Honscheid}, {James}, {Jeltema}, {Krause}, {Kuehn}, {Kuhlmann}, {Kuropatkin}, {March}, {Miquel}, {Nord}, {Roodman}, {Sanchez}, {Santiago}, {Scarpine}, {Schindler}, {Schubnell}, {Sevilla-Noarbe}, {Smith}, {Smith}, {Soares-Santos}, {Sobreira}, {Suchyta}, {Tarle}, {Thomas}, {Tucker}, {Wechsler}, {Wolf}, \& {Yanny}}]{Nagasawa2010}
{Nagasawa}, D.~Q., {Marshall}, J.~L., {Li}, T.~S., {et~al.} 2018, \apj, 852, 99, \dodoi{10.3847/1538-4357/aaa01d}

\bibitem[{{Omukai}(2000)}]{Omukai2000}
{Omukai}, K. 2000, ApJ, 534, 809, \dodoi{10.1086/308776}

\bibitem[{{Pe{\~n}arrubia} {et~al.}(2008){Pe{\~n}arrubia}, {Navarro}, \& {McConnachie}}]{Penarrubia2008}
{Pe{\~n}arrubia}, J., {Navarro}, J.~F., \& {McConnachie}, A.~W. 2008, \apj, 673, 226, \dodoi{10.1086/523686}

\bibitem[{{Planck Collaboration}(2016)}]{planck2016}
{Planck Collaboration}. 2016, A\&A, 594, A13, \dodoi{10.1051/0004-6361/201525830}

\bibitem[{{Portinari} {et~al.}(1998){Portinari}, {Chiosi}, \& {Bressan}}]{Portinari1998}
{Portinari}, L., {Chiosi}, C., \& {Bressan}, A. 1998, A\&A, 334, 505

\bibitem[{{Prgomet} {et~al.}(2022){Prgomet}, {Rey}, {Andersson}, {Segovia Otero}, {Agertz}, {Renaud}, {Pontzen}, \& {Read}}]{Prgomet2022}
{Prgomet}, M., {Rey}, M.~P., {Andersson}, E.~P., {et~al.} 2022, \mnras, 513, 2326, \dodoi{10.1093/mnras/stac1074}

\bibitem[{{Read} {et~al.}(2017){Read}, {Iorio}, {Agertz}, \& {Fraternali}}]{Read2017}
{Read}, J.~I., {Iorio}, G., {Agertz}, O., \& {Fraternali}, F. 2017, \mnras, 467, 2019, \dodoi{10.1093/mnras/stx147}

\bibitem[{{Revaz}(2023)}]{Revaz2023}
{Revaz}, Y. 2023, \aap, 679, A2, \dodoi{10.1051/0004-6361/202347239}

\bibitem[{{Revaz} {et~al.}(2016){Revaz}, {Arnaudon}, {Nichols}, {Bonvin}, \& {Jablonka}}]{Revaz2016}
{Revaz}, Y., {Arnaudon}, A., {Nichols}, M., {Bonvin}, V., \& {Jablonka}, P. 2016, \aap, 588, A21, \dodoi{10.1051/0004-6361/201526438}

\bibitem[{{Revaz} \& {Jablonka}(2018)}]{Revaz2018}
{Revaz}, Y., \& {Jablonka}, P. 2018, \aap, 616, A96, \dodoi{10.1051/0004-6361/201832669}

\bibitem[{{Rey} {et~al.}(2019){Rey}, {Pontzen}, {Agertz}, {Orkney}, {Read}, {Saintonge}, \& {Pedersen}}]{Rey2019}
{Rey}, M.~P., {Pontzen}, A., {Agertz}, O., {et~al.} 2019, ApJL, 886, L3, \dodoi{10.3847/2041-8213/ab53dd}

\bibitem[{{Richstein} {et~al.}(2022){Richstein}, {Patel}, {Kallivayalil}, {Simon}, {Zivick}, {Tollerud}, {Fritz}, {Warfield}, {Besla}, {van der Marel}, {Wetzel}, {Choi}, {Deason}, {Geha}, {Guhathakurta}, {Jeon}, {Kirby}, {Libralato}, {Sacchi}, \& {Sohn}}]{Richstein2022}
{Richstein}, H., {Patel}, E., {Kallivayalil}, N., {et~al.} 2022, \apj, 933, 217, \dodoi{10.3847/1538-4357/ac7226}

\bibitem[{{Richstein} {et~al.}(2024){Richstein}, {Kallivayalil}, {Simon}, {Garling}, {Wetzel}, {Warfield}, {van der Marel}, {Jeon}, {Rose}, {Torrey}, {Engelhardt}, {Besla}, {Choi}, {Geha}, {Guhathakurta}, {Kirby}, {Patel}, {Sacchi}, \& {Sohn}}]{Richstein2024}
{Richstein}, H., {Kallivayalil}, N., {Simon}, J.~D., {et~al.} 2024, \apj, 967, 72, \dodoi{10.3847/1538-4357/ad393c}

\bibitem[{{Ricotti} {et~al.}(2022){Ricotti}, {Polisensky}, \& {Cleland}}]{Ricotti2022}
{Ricotti}, M., {Polisensky}, E., \& {Cleland}, E. 2022, \mnras, 515, 302, \dodoi{10.1093/mnras/stac1485}

\bibitem[{{Ritter} {et~al.}(2012){Ritter}, {Safranek-Shrader}, {Gnat}, {Milosavljevi{\'c}}, \& {Bromm}}]{Ritter2012}
{Ritter}, J.~S., {Safranek-Shrader}, C., {Gnat}, O., {Milosavljevi{\'c}}, M., \& {Bromm}, V. 2012, \apj, 761, 56, \dodoi{10.1088/0004-637X/761/1/56}

\bibitem[{Sacchi {et~al.}(2021)Sacchi, Richstein, Kallivayalil, van~der Marel, Libralato, Zivick, Besla, Brown, Choi, Deason, Fritz, Geha, Guhathakurta, Jeon, Kirby, Majewski, Patel, Simon, Sohn, Tollerud, \& Wetzel}]{Sacchi2021}
Sacchi, E., Richstein, H., Kallivayalil, N., {et~al.} 2021, The Astrophysical Journal Letters, 920, L19, \dodoi{10.3847/2041-8213/ac2aa3}

\bibitem[{{Safranek-Shrader} {et~al.}(2016){Safranek-Shrader}, {Montgomery}, {Milosavljevi{\'c}}, \& {Bromm}}]{Safranek2016}
{Safranek-Shrader}, C., {Montgomery}, M.~H., {Milosavljevi{\'c}}, M., \& {Bromm}, V. 2016, \mnras, 455, 3288, \dodoi{10.1093/mnras/stv2545}

\bibitem[{{Salpeter}(1955)}]{Salpeter1955}
{Salpeter}, E.~E. 1955, \apj, 121, 161, \dodoi{10.1086/145971}

\bibitem[{{Sanati} {et~al.}(2023){Sanati}, {Jeanquartier}, {Revaz}, \& {Jablonka}}]{Sanati2023}
{Sanati}, M., {Jeanquartier}, F., {Revaz}, Y., \& {Jablonka}, P. 2023, \aap, 669, A94, \dodoi{10.1051/0004-6361/202244309}

\bibitem[{{Sandford} {et~al.}(2024){Sandford}, {Weinberg}, {Weisz}, \& {Fu}}]{Sandford2024}
{Sandford}, N.~R., {Weinberg}, D.~H., {Weisz}, D.~R., \& {Fu}, S.~W. 2024, \mnras, 530, 2315, \dodoi{10.1093/mnras/stae1010}

\bibitem[{{Schaye} {et~al.}(2010){Schaye}, {Dalla Vecchia}, {Booth}, {Wiersma}, {Theuns}, {Haas}, {Bertone}, {Duffy}, {McCarthy}, \& {van de Voort}}]{Schaye2010}
{Schaye}, J., {Dalla Vecchia}, C., {Booth}, C.~M., {et~al.} 2010, MNRAS, 402, 1536, \dodoi{10.1111/j.1365-2966.2009.16029.x}

\bibitem[{{Schmidt}(1959)}]{Schmidt1959}
{Schmidt}, M. 1959, ApJ, 129, 243, \dodoi{10.1086/146614}

\bibitem[{{Schneider} \& {Omukai}(2010)}]{Schneider2010}
{Schneider}, R., \& {Omukai}, K. 2010, MNRAS, 402, 429, \dodoi{10.1111/j.1365-2966.2009.15891.x}

\bibitem[{{S{\'e}gall} {et~al.}(2007){S{\'e}gall}, {Ibata}, {Irwin}, {Martin}, \& {Chapman}}]{Segall2007}
{S{\'e}gall}, M., {Ibata}, R.~A., {Irwin}, M.~J., {Martin}, N.~F., \& {Chapman}, S. 2007, \mnras, 375, 831, \dodoi{10.1111/j.1365-2966.2006.11356.x}

\bibitem[{{Sestito} {et~al.}(2023){Sestito}, {Zaremba}, {Venn}, {D'Aoust}, {Hayes}, {Jensen}, {Navarro}, {Jablonka}, {Fern{\'a}ndez-Alvar}, {Glover}, {McConnachie}, \& {Chen{\'e}}}]{Sestito2023}
{Sestito}, F., {Zaremba}, D., {Venn}, K.~A., {et~al.} 2023, \mnras, 525, 2875, \dodoi{10.1093/mnras/stad2427}

\bibitem[{{Sharda} \& {Krumholz}(2022)}]{Sharda2022}
{Sharda}, P., \& {Krumholz}, M.~R. 2022, \mnras, 509, 1959, \dodoi{10.1093/mnras/stab2921}

\bibitem[{{Simon}(2019)}]{Simon2019}
{Simon}, J.~D. 2019, \araa, 57, 375, \dodoi{10.1146/annurev-astro-091918-104453}

\bibitem[{{Simon} {et~al.}(2010){Simon}, {Frebel}, {McWilliam}, {Kirby}, \& {Thompson}}]{Simon2010}
{Simon}, J.~D., {Frebel}, A., {McWilliam}, A., {Kirby}, E.~N., \& {Thompson}, I.~B. 2010, \apj, 716, 446, \dodoi{10.1088/0004-637X/716/1/446}

\bibitem[{{Simon} {et~al.}(2020){Simon}, {Li}, {Erkal}, {Pace}, {Drlica-Wagner}, {James}, {Marshall}, {Bechtol}, {Hansen}, {Kuehn}, {Lidman}, {Allam}, {Annis}, {Avila}, {Bertin}, {Brooks}, {Burke}, {Rosell}, {Carrasco Kind}, {Carretero}, {da Costa}, {De Vicente}, {Desai}, {Doel}, {Eifler}, {Everett}, {Fosalba}, {Frieman}, {Garc{\'\i}a-Bellido}, {Gaztanaga}, {Gerdes}, {Gruen}, {Gruendl}, {Gschwend}, {Gutierrez}, {Hollowood}, {Honscheid}, {Krause}, {Kuropatkin}, {MacCrann}, {Maia}, {March}, {Miquel}, {Palmese}, {Paz-Chinch{\'o}n}, {Plazas}, {Reil}, {Roodman}, {Sanchez}, {Santiago}, {Scarpine}, {Schubnell}, {Serrano}, {Smith}, {Suchyta}, {Tarle}, {Walker}, \& {DES Collaboration}}]{Simon2020}
{Simon}, J.~D., {Li}, T.~S., {Erkal}, D., {et~al.} 2020, \apj, 892, 137, \dodoi{10.3847/1538-4357/ab7ccb}

\bibitem[{{Smith} {et~al.}(2015){Smith}, {Wise}, {O'Shea}, {Norman}, \& {Khochfar}}]{Smith2015}
{Smith}, B.~D., {Wise}, J.~H., {O'Shea}, B.~W., {Norman}, M.~L., \& {Khochfar}, S. 2015, \mnras, 452, 2822, \dodoi{10.1093/mnras/stv1509}

\bibitem[{{Smith} {et~al.}(2021){Smith}, {Bryan}, {Somerville}, {Hu}, {Teyssier}, {Burkhart}, \& {Hernquist}}]{Smith2021}
{Smith}, M.~C., {Bryan}, G.~L., {Somerville}, R.~S., {et~al.} 2021, \mnras, 506, 3882, \dodoi{10.1093/mnras/stab1896}

\bibitem[{Spearman(1904)}]{Spearman1904}
Spearman, C. 1904, The American Journal of Psychology, 15, 72.
\newblock \url{http://www.jstor.org/stable/1412159}

\bibitem[{{Springel}(2005)}]{Springel2005}
{Springel}, V. 2005, \mnras, 364, 1105, \dodoi{10.1111/j.1365-2966.2005.09655.x}

\bibitem[{{Springel} {et~al.}(2001){Springel}, {White}, {Tormen}, \& {Kauffmann}}]{Springel2001}
{Springel}, V., {White}, S.~D.~M., {Tormen}, G., \& {Kauffmann}, G. 2001, MNRAS, 328, 726, \dodoi{10.1046/j.1365-8711.2001.04912.x}

\bibitem[{{Tarumi} {et~al.}(2021){Tarumi}, {Yoshida}, \& {Frebel}}]{Tarumi2021}
{Tarumi}, Y., {Yoshida}, N., \& {Frebel}, A. 2021, \apjl, 914, L10, \dodoi{10.3847/2041-8213/ac024e}

\bibitem[{{Tau} {et~al.}(2024){Tau}, {Vivas}, \& {Mart{\'\i}nez-V{\'a}zquez}}]{Tau2024}
{Tau}, E.~A., {Vivas}, A.~K., \& {Mart{\'\i}nez-V{\'a}zquez}, C.~E. 2024, \aj, 167, 57, \dodoi{10.3847/1538-3881/ad1509}

\bibitem[{{Ting} \& {Ji}(2024)}]{Ting2024}
{Ting}, Y.-S., \& {Ji}, A.~P. 2024, arXiv e-prints, arXiv:2408.06807, \dodoi{10.48550/arXiv.2408.06807}

\bibitem[{{Tolstoy} {et~al.}(2009){Tolstoy}, {Hill}, \& {Tosi}}]{Tolstoy2009}
{Tolstoy}, E., {Hill}, V., \& {Tosi}, M. 2009, ARA\&A, 47, 371, \dodoi{10.1146/annurev-astro-082708-101650}

\bibitem[{{Vargas} {et~al.}(2013){Vargas}, {Geha}, {Kirby}, \& {Simon}}]{Vargas2013}
{Vargas}, L.~C., {Geha}, M., {Kirby}, E.~N., \& {Simon}, J.~D. 2013, \apj, 767, 134, \dodoi{10.1088/0004-637X/767/2/134}

\bibitem[{{Walker} {et~al.}(2006){Walker}, {Mateo}, {Olszewski}, {Bernstein}, {Wang}, \& {Woodroofe}}]{Walker2006}
{Walker}, M.~G., {Mateo}, M., {Olszewski}, E.~W., {et~al.} 2006, \aj, 131, 2114, \dodoi{10.1086/500193}

\bibitem[{{Waller} {et~al.}(2023){Waller}, {Venn}, {Sestito}, {Jensen}, {Kielty}, {Borukhovetskaya}, {Hayes}, {McConnachie}, \& {Navarro}}]{Waller2023}
{Waller}, F., {Venn}, K.~A., {Sestito}, F., {et~al.} 2023, \mnras, 519, 1349, \dodoi{10.1093/mnras/stac3563}

\bibitem[{{Weinberg} {et~al.}(2017){Weinberg}, {Andrews}, \& {Freudenburg}}]{Weinberg2017}
{Weinberg}, D.~H., {Andrews}, B.~H., \& {Freudenburg}, J. 2017, \apj, 837, 183, \dodoi{10.3847/1538-4357/837/2/183}

\bibitem[{{Weisz} {et~al.}(2014){Weisz}, {Dolphin}, {Skillman}, {Holtzman}, {Gilbert}, {Dalcanton}, \& {Williams}}]{Weisz2014}
{Weisz}, D.~R., {Dolphin}, A.~E., {Skillman}, E.~D., {et~al.} 2014, \apj, 789, 147, \dodoi{10.1088/0004-637X/789/2/147}

\bibitem[{{Wheeler} {et~al.}(2019){Wheeler}, {Hopkins}, {Pace}, {Garrison-Kimmel}, {Boylan-Kolchin}, {Wetzel}, {Bullock}, {Kere{\v{s}}}, {Faucher-Gigu{\`e}re}, \& {Quataert}}]{Wheeler2019}
{Wheeler}, C., {Hopkins}, P.~F., {Pace}, A.~B., {et~al.} 2019, \mnras, 490, 4447, \dodoi{10.1093/mnras/stz2887}

\bibitem[{{Wiersma} {et~al.}(2009){Wiersma}, {Schaye}, {Theuns}, {Dalla Vecchia}, \& {Tornatore}}]{Wiersma2009}
{Wiersma}, R.~P.~C., {Schaye}, J., {Theuns}, T., {Dalla Vecchia}, C., \& {Tornatore}, L. 2009, MNRAS, 399, 574, \dodoi{10.1111/j.1365-2966.2009.15331.x}

\bibitem[{{Wilkinson} {et~al.}(2023){Wilkinson}, {Ludlow}, {Lagos}, {Fall}, {Schaye}, \& {Obreschkow}}]{Wilkinson2023}
{Wilkinson}, M.~J., {Ludlow}, A.~D., {Lagos}, C. d.~P., {et~al.} 2023, \mnras, 519, 5942, \dodoi{10.1093/mnras/stad055}

\bibitem[{{Wise} {et~al.}(2012){Wise}, {Turk}, {Norman}, \& {Abel}}]{Wise2012}
{Wise}, J.~H., {Turk}, M.~J., {Norman}, M.~L., \& {Abel}, T. 2012, ApJ, 745, 50, \dodoi{10.1088/0004-637X/745/1/50}

\bibitem[{{Wolf} {et~al.}(2010){Wolf}, {Martinez}, {Bullock}, {Kaplinghat}, {Geha}, {Mu{\~n}oz}, {Simon}, \& {Avedo}}]{Wolf2010}
{Wolf}, J., {Martinez}, G.~D., {Bullock}, J.~S., {et~al.} 2010, \mnras, 406, 1220, \dodoi{10.1111/j.1365-2966.2010.16753.x}

\bibitem[{{Yang} {et~al.}(2022){Yang}, {Hammer}, {Jiao}, \& {Pawlowski}}]{Yang2022}
{Yang}, Y., {Hammer}, F., {Jiao}, Y., \& {Pawlowski}, M.~S. 2022, \mnras, 512, 4171, \dodoi{10.1093/mnras/stac644}

\bibitem[{{Yoon} {et~al.}(2012){Yoon}, {Dierks}, \& {Langer}}]{Yoon2012}
{Yoon}, S.~C., {Dierks}, A., \& {Langer}, N. 2012, \aap, 542, A113, \dodoi{10.1051/0004-6361/201117769}

\bibitem[{{Zaritsky} {et~al.}(2008){Zaritsky}, {Zabludoff}, \& {Gonzalez}}]{Zaritsky2008}
{Zaritsky}, D., {Zabludoff}, A.~I., \& {Gonzalez}, A.~H. 2008, \apj, 682, 68, \dodoi{10.1086/529577}

\end{thebibliography}
%%%%%%%%%%%%%%%%%%%%%%%%%%%%%%%%%%%%%%%%%%%%%%%%%%

%%%%%%%%%%%%%%%%% APPENDICES %%%%%%%%%%%%%%%%%%%%%
\appendix

\section{Resultant MDF of the UFD analogs with different SN energy}\label{appendix_a}

\begin{figure*}
  \centering 
  \includegraphics[width=170mm]{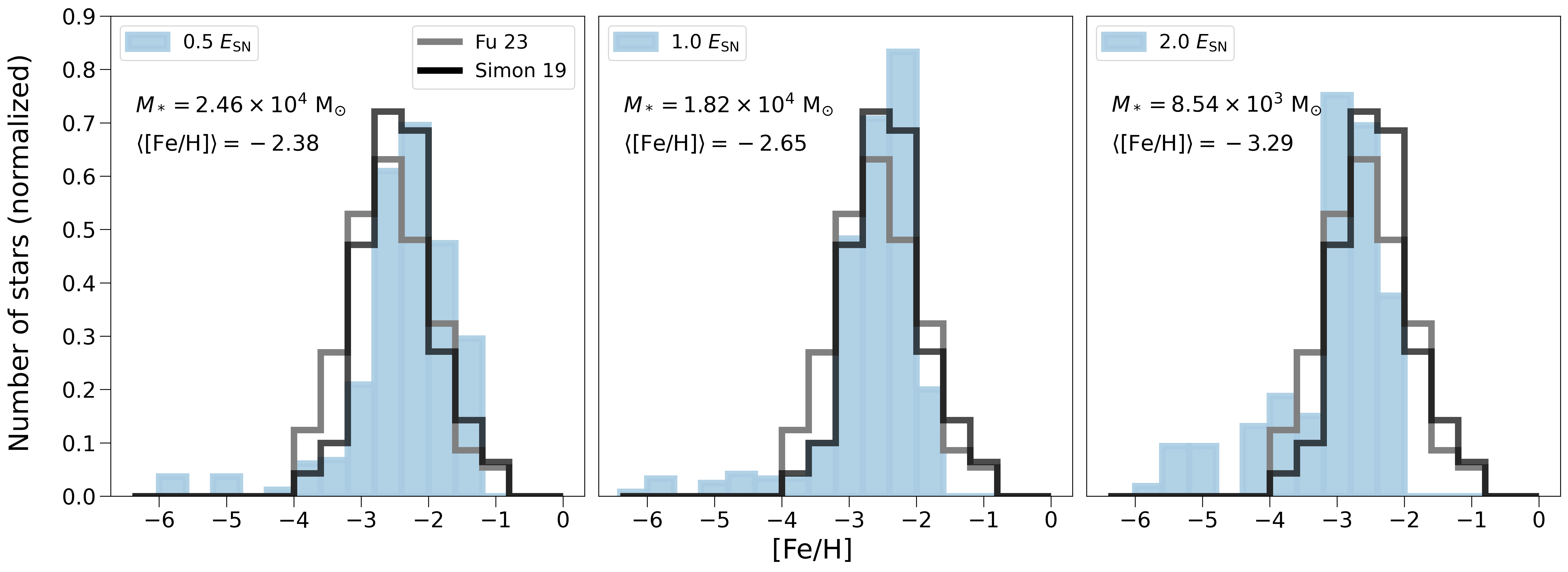}
  \caption{The resultant MDFs of {\sc Halo4} under varying SN energies. From left to right, each panel corresponds to runs with 0.5 $E_{\rm SN}$, 1.0 $E_{\rm SN}$, and 2.0 $E_{\rm SN}$ for $E_{\rm SN} \ = \ 1.0 \times 10^{51} \rm ergs$, respectively, with the middle panel representing the fiducial run of {\sc Halo4}. As SN energy increases, the average stellar mass and metallicity tend to decrease due to the excessively strong SN energy. Furthermore, it becomes more challenging to produce relatively high-metallicity stars ($\rm [Fe/H] \gtrsim -2$) with the increased SN energy of $2.0 \ E_{\rm SN}$}
  \label{fig:MDF_SNe}
\end{figure*}

We conduct additional simulations using {\sc Halo4} to examine the effects of varying SN energies, specifically at 0.5 $E_{\rm SN}$, 1.0 $E_{\rm SN}$, and 2.0 $E_{\rm SN}$ where $E_{\rm SN} = 10^{51} \rm erg$, on the MDF. As depicted in Fig.~\ref{fig:MDF_SNe}, a reduction in SN energy by a factor of two, relative to the fiducial scenario (the middle panel), results in a rightward shift of the MDF, thereby producing a higher fraction of metal-rich stars. On the contrary, increasing the SN energy to $2.0 \ E_{\rm SN}$ induces a leftward shift in the MDF. This indicates that the enhanced SN energy leads to greater metal dispersal, subsequently giving rise to a higher proportion of metal-poor stars.

\section{Alpha element abundances}\label{appendix_b}
We investigate the $\rm [Mg/Fe] \ vs \ [Fe/H]$ distributions of member stars in our UFD analogs. For comparison with observations, we select galaxies with absolute magnitudes fainter than $M_{\rm V}\sim-6.3$, which is similar to that of our most luminous UFD analog {\sc Halo5}. These observed galaxies include Horologium I (\citealp{Nagasawa2010}), Reticulum II (\citealp{Ji2016a, Ji2018}), Bootes II (\citealp{Ji2016a, Francois2016}), Tucana II (\citealp{Chiti2018}), Canes Venatici II (\citealp{Vargas2013, Francois2016}), Coma Berenices (\citealp{Frebel2010, Vargas2013}), Segue II (\citealp{Kirby2013}), Triangulum II (\citealp{Kirby2017, Ji2019}), Tucana III (\citealp{Hansen2017}), Ursa Major II (\citealp{Frebel2010, Vargas2013}), Grus I (\citealp{Ji2019}), Segue I (\citealp{Vargas2013, Frebel2014}), Leo IV (\citealp{Simon2010, Francois2016}), and Bootes I (\citealp{Ishigaki2014, Gilmore2013}). Although our data generally agree well with observations, the low-metallicity stars with $\rm [Fe/H]\lesssim-4$ exhibit extremely high $\rm [Mg/Fe]$ ratios. This is primarily due to the influence of Pop III SN, which yields a significantly higher abundance of alpha-elements than Pop II CCSNe.

\begin{figure}
  \centering 
  \includegraphics[width=70mm]{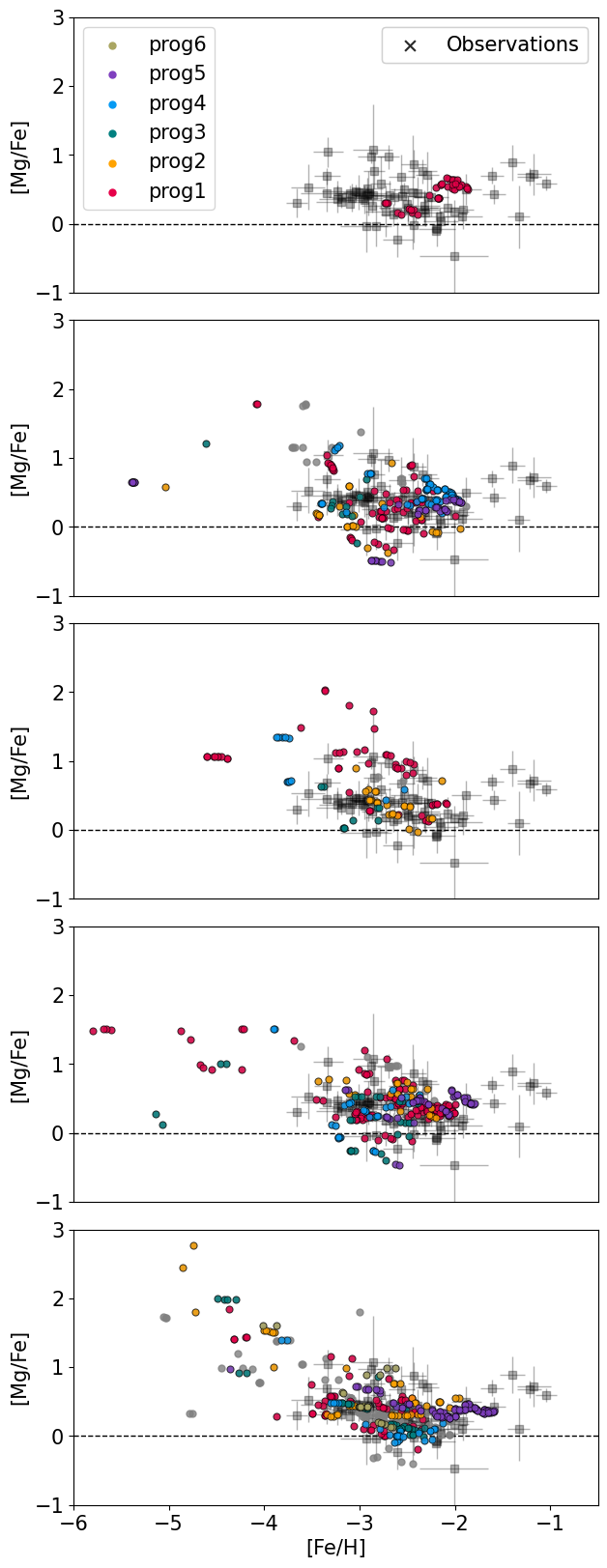}
  \caption{$\rm [Mg/Fe] \ vs. [Fe/H]$ distribution for simulated UFD analogs, with stars represented as circles colored according to their progenitor halos, compared with observed data shown in grey.}
  \label{fig:alpha}
\end{figure}

\section{Two-component fitting results of surface brightness profile}\label{appendix_c}

\begin{figure}
  \centering
  \includegraphics[width=180mm]{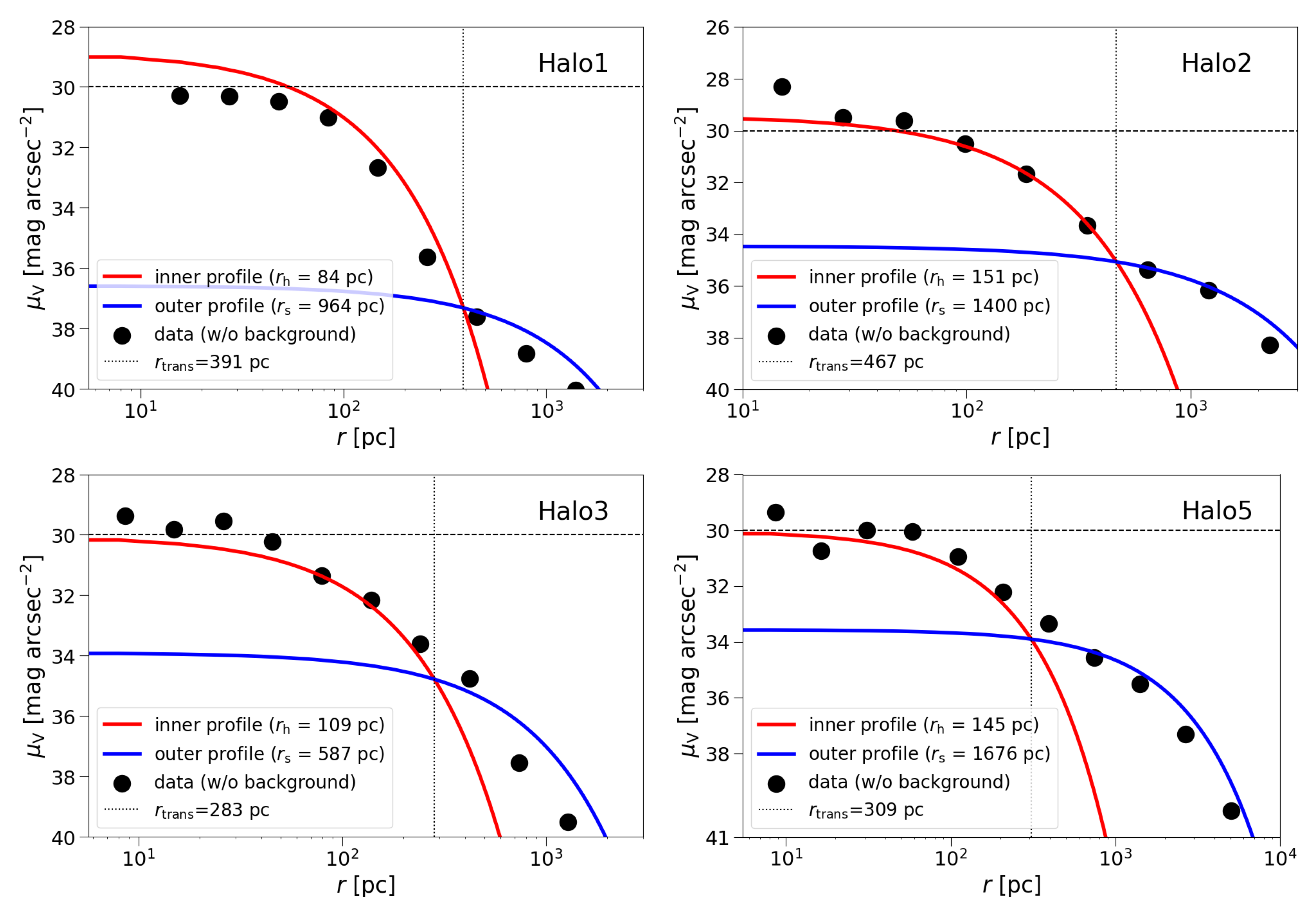}
  \caption{The surface brightness profiles of UFD analogs are modeled using a two-component approach. The actual density profile is shown as black-filled circles at binned radii. The inner component is depicted in red, while the outer component is illustrated in blue. A vertical line indicates the transition radius between the inner and outer components.}
  \label{fig:two-comp2}
\end{figure}

Fig.~\ref{fig:two-comp2} shows results of the two-component surface brightness fitting for other UFD analogs. We perform the same fitting method as represented in Section~\ref{sec:3.3}.

%%%%%%%%%%%%%%%%%%%%%%%%%%%%%%%%%%%%%%%%%%%%%%%%%%
% Don't change these lines
%\bsp	% typesetting comment
%\label{lastpage}
\end{document}